\documentclass[a4paper,12pt]{article}
\usepackage{jheppub,esint,shuffle,psfrag}
\usepackage[utf8]{inputenc}

\usepackage{soul}
\usepackage{tikz}
\usepackage{tikz-3dplot}
\usepackage{mathrsfs}
\usepackage{enumitem}

\usepackage{tikzit}
\usetikzlibrary{calc,arrows.meta,angles,quotes}
\tikzstyle{small circle}=[shape=circle, fill=white, draw=black]
\tikzstyle{medium circle}=[fill=white, draw=black, shape=circle, minimum width=1cm, minimum height=1cm]
\tikzstyle{twoPoint}=[fill=white, draw=black, shape=circle, twopt]
\tikzstyle{threePoint}=[fill=white, draw=black, shape=circle, threept]
\tikzstyle{m vert rect}=[fill=white, draw=black, shape=rectangle, minimum height=2cm, minimum width=.5cm]
\tikzstyle{s box}=[fill=white, draw=black, shape=rectangle]
\tikzstyle{celestialThreePoint}=[fill=white, draw=red, shape=circle, threept]
\tikzstyle{large circle}=[fill=white, draw=black, shape=circle, minimum height=1.5cm, minimum width=1.5cm]
\tikzstyle{green circle of inversion}=[draw={black!20!green}, shape=circle, minimum size=2cm, dashed, tikzit draw={black!20!green}]
\tikzstyle{large green circle}=[draw={black!20!green}, shape=circle, dashed, minimum size=3cm, semithick]
\tikzstyle{small triangle}=[fill=white, draw=black, regular polygon, regular polygon sides=3, inner sep=1.5pt, rotate=90]
\tikzstyle{small blue triangle}=[fill=white, draw={black!20!blue}, regular polygon, regular polygon sides=3, inner sep=1.5pt, rotate=-90, tikzit draw=blue]
\tikzstyle{small green circle}=[draw={black!20!green}, shape=circle, tikzit draw={black!20!green}, dashed, minimum size=.5cm]
\tikzstyle{celestialTwoPoint}=[fill=red, draw=red, shape=circle, inner sep=1pt, minimum size=1pt]

\tikzstyle{arrow}=[->]
\tikzstyle{op}=[-, spinning]
\tikzstyle{dashedOp}=[-, scalar]
\tikzstyle{celestial}=[-, color=red, scalar, tikzit draw=red]
\tikzstyle{minkowski}=[-, spinning, color={black!20!blue}, tikzit draw=blue]

\input{harmonic_analysis.tikzdefs}
\tikzstyle{pole}=[fill=black, draw=black, shape=circle, inner sep=0pt, minimum size=2pt]

\tikzstyle{Arrow}=[->, draw=blue]
\tikzstyle{anti arrow}=[<-, draw=blue]
\tikzstyle{axis}=[->]
\tikzstyle{dashed_arrow}=[->, draw=red]
\tikzstyle{dashed_only}=[-, draw=red]
\tikzstyle{dashed_anti}=[<-, draw=red]

\tikzstyle{point}=[fill=black, draw=black, shape=circle, inner sep=0pt, minimum size=3pt]
\tikzstyle{lr_point}=[fill=red, draw=red, shape=circle, inner sep=0pt, minimum size=2pt]
\tikzstyle{sh_point}=[fill={rgb,255: red,128; green,128; blue,128}, draw={rgb,255: red,128; green,128; blue,128}, shape=circle, inner sep=0pt, minimum size=3pt]
\tikzstyle{sh_lr_point}=[fill={red!40}, draw={red!40}, shape=circle, inner sep=0pt, minimum size=2pt, tikzit fill={rgb,255: red,255; green,128; blue,0}, tikzit draw={rgb,255: red,255; green,128; blue,0}]

\tikzstyle{dashed_st}=[-, dashed]
\tikzstyle{arrow}=[->]
\tikzstyle{dashed_grey}=[-, draw=black, dotted]
\tikzstyle{blue_line}=[draw=blue, ->]
\tikzstyle{energy}=[draw={rgb,255: red,255; green,128; blue,0}, decoration={{snake,amplitude=1pt,segment length=6pt,post length=1pt}}, decorate, ->]
\tikzstyle{initial}=[draw={rgb,255: red,255; green,128; blue,0},  line width= 1pt , decoration={{snake,amplitude=1pt,segment length=6pt,post length=1pt}}, decorate, ->]
\tikzstyle{blue_line_0}=[-, draw=blue]
\tikzstyle{redline}=[-, draw=red]

\usepackage{float}  
\usepackage{esint} 
\usepackage{breqn}
\usepackage{bm}
\usepackage{tikz}
\usepackage{pgfplots}
\pgfplotsset{compat=newest}
\usepackage{hyphenat}

\usepackage[colorlinks=true,
            linkcolor=blue,
            citecolor=blue,
            urlcolor=blue]{hyperref}

\definecolor{mycolor}{rgb}{0.2,0.6,0.5}

\def \Re {\mathop{\rm Re}\nolimits}

\newcommand\lr[1]{{\left({#1}\right)}}
\newcommand \widebar [1] {\overline{#1}}

\newcommand \vev [1] {\langle{#1}\rangle}

\newcommand \ket [1] {|{#1}\rangle}
\newcommand \bra [1] {\langle {#1}|}
\newcommand\re[1]{(\ref{#1})}
\def \qqquad {\qquad\quad}
\def \qqqquad {\qquad\qquad}

\newcommand{\p}[1]{(\ref{#1})}
\newcommand{\ep}{\epsilon}
\newcommand{\la}{\lambda}
\newcommand{\app}{{\alpha'}}
\newcommand{\vp}{\varphi}

\def\be#1\ee{\begin{align}#1\end{align}}
\newcommand{\beq}{\begin{equation}}
\newcommand{\eeq}{\end{equation}}
\newcommand{\beqq}{\begin{equation*}}
\newcommand{\eeqq}{\end{equation*}}
\newcommand\beqa{\begin{eqnarray}}
\newcommand\eeqa{\end{eqnarray}}
\newcommand\beqaa{\begin{eqnarray*}}
\newcommand\eeqaa{\end{eqnarray*}}
\newcommand\bea{\begin{array}}
\newcommand\eea{\end{array}}

\def\numberbysection{\@addtoreset{equation}{section}
                     \def\theequation{\thesection.\arabic{equation}}}

\begin{document}

\begin{flushleft}
 \hfill \parbox[c]{40mm}{CERN-TH-2025-266 \\  LAPTH-060/25}
\end{flushleft}

\title{
Energy correlators in four-dimensional  gravity
}

\author{Dmitry Chicherin$^{a}$, Gregory P. Korchemsky$^{b}$, Emery Sokatchev$^{a}$, and  Alexander Zhiboedov$^{c}$}

\affiliation{$^a$LAPTh-CNRS-USMB, 9 chemin de Bellevue, 74940, Annecy, France }
\affiliation{$^b$Institut de Physique Th\'eorique\footnote{Unit\'e Mixte de Recherche 3681 du CNRS}, Universit\'e Paris Saclay, CNRS, F-91191 Gif-sur-Yvette, France}
\affiliation{$^c$CERN, Theoretical Physics Department, CH-1211 Geneva 23, Switzerland}

\abstract{ We investigate energy correlators in four-dimensional gravitational theories, which provide a simple class of infrared-finite observables. We compute the one- and two-point energy correlators at one loop in $\mathcal{N}=8$ supergravity and in pure Einstein gravity, with particular emphasis on the contact terms arising from the interplay between virtual corrections and real emissions. We explicitly demonstrate the cancellation of infrared divergences and verify the Ward identities associated with energy–momentum conservation. In the back-to-back limit, we derive an all-order expression for the energy-energy correlator, showing that it is governed by universal soft-graviton dynamics. We further introduce a particularly simple beam-averaged energy–energy correlator and compute it in different gravitational theories, including tree-level string theory. The resulting correlators exhibit analyticity and polynomial boundedness, allowing for the formulation of dispersion relations, which we explore. Finally, we discuss additional singularities of the gravitational energy correlators, absent in QCD, that originate from the long-range nature of the gravitational interactions.
}

\maketitle

\newpage

\section{Introduction and summary of the results}

In this paper, we seek the simplest infrared-finite observables in four-dimensional quantum gravity.
Long-range forces---both electromagnetic and gravitational---render standard scattering theory inapplicable \cite{Prabhu:2022zcr}.
In particular, a nontrivial S-matrix between plane-wave states does not exist.\footnote{In massive QED, this statement applies to matrix elements with charged particles in the initial or final state.
By contrast, the photon S-matrix is infrared-finite and nontrivial. 
The simplest example is the four-photon scattering amplitude ${\cal M}_{\gamma\gamma\to\gamma\gamma}$.  So, nonperturbative S-matrix bootstrap methods apply \cite{Haring:2022sdp}.}
This failure can be traced to the asymptotic dynamics of the particles, which are no longer free \cite{Dollard:1964cok,dollard1971quantum,Chung:1965zza,Kibble:1968sfb,Kulish:1970ut}.
Incorporating the correct asymptotic dynamics leads to the dressed S-matrix formalism \cite{Kulish:1970ut,Ware:2013zja,Hannesdottir:2019opa,Lippstreu:2025jit}.\footnote{In the weak-coupling regime, S-matrix bootstrap methods were recently applied to this setting in \cite{Bellazzini:2025bay}.}

A useful way to organize scattering in four dimensions in the presence of long-range forces is via asymptotic symmetries \cite{Strominger:2017zoo}.
The key distinction between four and higher dimensions is that, in four dimensions, an infinite-dimensional symmetry emerges and imposes additional selection rules \cite{Strominger:2013jfa,Strominger:2014pwa}:
any change in the energy flux on the celestial sphere must be accompanied by nontrivial memory.
According to the Coleman-Mandula theorem \cite{Coleman:1967ad}, such a symmetry enhancement is expected to trivialize exclusive amplitudes.
Indeed, one finds that fixed-multiplicity (exclusive) $m \to n$ amplitudes vanish as one removes the infrared regulator.
In perturbation theory, this emergent BMS symmetry manifests itself through infrared divergences.

A complementary approach, which connects more directly to standard collider-physics tools, is to focus on sufficiently inclusive observables, such as Sterman-Weinberg jet cross sections \cite{PhysRevLett.39.1436} and inclusive differential cross sections.
In this setting, infrared divergences cancel, order by order in perturbation theory, between real emissions and virtual corrections \cite{Bloch:1937pw}.
This strategy has been applied to gravity in \cite{Donoghue:1999qh,Gonzo:2020xza,Gonzo:2023cnv,Herrmann:2024yai}, and it is the one we adopt in the present paper.

\begin{figure}[h!]
\centering
\begin{tikzpicture}[scale=1.2]
	\node [] at (0,0) {\includegraphics[scale=0.2]{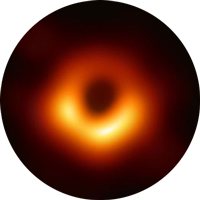}};
	\draw [energy] (-0.1,0.1) -- (-1.3,1.3);
	\draw [energy] (0.1,0.1) -- (1.2,1.2);
	\draw [energy] (-0.025,0.137) -- (-0.25,1.9);
	\draw [energy] (0.06,0.13) -- (0.6,1.2);
	\draw [energy] (-0.06,0.13) -- (-0.5,1);
	\draw [energy] (0.02,0.137) -- (0.2,1.5);
    \draw [initial] (-1.5,-1.5) -- (-0.6,-0.6);
	\draw [initial] (1.5,-1.5) -- (0.6,-0.6);
	\draw [] (-3,0) -- (0,3) -- (3,0) -- (0,-3) -- cycle;
	\draw [] (-3,0) to[out=-45,in=-135,distance=1.5cm] (3,0);
	\draw [dashed] (-3,0) to[out=45,in=135,distance=1.5cm] (3,0);
\begin{scope}[thick,blue,decoration={
    markings,
    mark=at position 0.5 with {\arrow{>}}}]
	\draw[postaction={decorate}] (-1.1,-0.72) -- (0,3);
\end{scope}	
	\node [above] at (1.8,1.65) {$\mathscr{I}^+$};
	\node [below] at (1.8,-1.55) {$\mathscr{I}^-$};
	\node [above] at (0,3) {$i^+$};
	\node [below] at (0,-3) {$i^-$};
	\node [right] at (3,0) {$i^0$};
\end{tikzpicture}
\caption{
We consider an initial state of two gravitons colliding in the center-of-mass frame with total energy $2E$.
We then compute, perturbatively in $(\kappa E)$, where $\kappa^2 = 32 \pi G_N$, the angular distribution of the radiation in the final state, as measured by calorimeters (blue) placed at null infinity and labeled by points on the celestial sphere.
This figure is a slight modification of Figure~4 in \cite{Kologlu:2019mfz}, adapted to the initial state studied here. %
}
\label{eq:eventshapekinematics}
\end{figure}
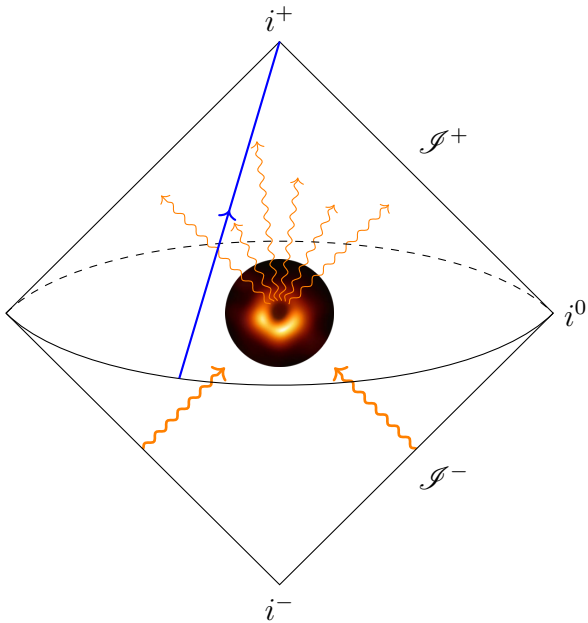

We consider an initial state consisting of two gravitons with fixed momenta, see Figure~\ref{eq:eventshapekinematics}.
We work in the center-of-mass frame, so the total spatial momentum vanishes.
The state is therefore characterized by a total energy $2E$ and a beam axis $\vec n$.\footnote{In ordinary QFT it is natural to consider states created by local operators acting on the vacuum. In a theory with dynamical gravity, this option is not available. For this reason, we instead study the simplest scattering process with two gravitons in the initial state.}
We study the differential cross section $d\sigma_{p_1+p_2\to q_1+q_2+X}$, which defines 
the probability for this state to produce a specified set of massless particles in the final state, after tracing over any unobserved radiation, see Figure~\ref{eq:eventshapekinematics}. We can use this cross section to introduce the energy correlators $\langle{\cal E}(n_1)\rangle$ and $\langle{\cal E}(n_1){\cal E}(n_2)\rangle$. They are defined as differential cross sections, weighted with the energies of the particles detected by calorimeters located on the celestial sphere in the direction indicated by the unit vectors $\vec n_1$ and $\vec n_2$, as shown in Figure~\ref{fig:angles-sphere}. Standard arguments \cite{Weinberg:1965nx} suggest that the energy correlators are infrared finite away from the forward limit, when $(q_i\!\cdot p_j) \neq 0$. Unlike confining gauge theories, we do not normalize the energy correlators by the total cross section $\sigma_{\rm tot}=\int d\sigma_{p_1+p_2\to q_1+q_2+X}$, since it diverges in the presence of long-range forces.

We examine the energy correlators in two gravitational theories: ${\cal N}=8$ supergravity (SG) and pure gravity (i.e. Einstein's gravity without matter). 
We compute the one- and two-point energy correlators at the first nontrivial order (NLO) in the gravitational coupling. They are given by the sum of virtual and real  particle contributions. The former involves the one-loop four-point amplitude, while the latter is given by the tree-level five-point amplitude squared and integrated over the final state phase space. Although virtual and real contributions to the energy correlators are separately infrared divergent, we explicitly show that the infrared divergences cancel in their sum.
As an important consistency check, we verify that the resulting energy correlators satisfy the energy- and momentum-conservation sum rules (Ward identities).
We also view ${\cal N}=8$ SG as the low-energy limit of type~II string theory compactified on $T^6$, and compute the leading stringy corrections to the two-point energy correlator.

The energy correlators $\langle{\cal E}(\vec n_1)\rangle$ and $\langle{\cal E}(\vec n_1){\cal E}(\vec n_2)\rangle$ depend on the angles between the unit vectors $\vec n$ and $\vec n_i$, which define the directions of the beams of the incoming particles and the detected particles, respectively (see Figure~\ref{fig:angles-sphere}).  
We find that the first two terms of the perturbative expansion of the one-point
energy correlator in ${\cal N}=8$ SG are given by
\be
\label{eq:ECintro}
& \langle{\cal E}(\vec n_1)\rangle =
 E \left({\kappa E \over 2} \right)^4
 \bigg[ {1\over 2 \pi^2 y_1^2(1-y_1)^2}+ \left({\kappa E \over 2} \right)^{2} \text{EC}^{(1)}(y_1) +O((\kappa E)^4) \bigg]\,, \nonumber \\
& \text{EC}^{(1)}(y_1) ={1 \over 2 \pi^4} 
\Bigg[
\frac{\log(y_1) \log(1-y_1)}{y_1^{2}(1-y_1)^{2}}
+ \frac{\pi^{2}}{3 y_1(1-y_1)} + \frac{2(1-y_1)^{3}\,\text{Li}_{2}(y_1)}
{y_1^{2}(1-y_1)^{2}(1-2y_1)} + \frac{y_1\log^{2}(1-y_1)}
{y_1(1-y_1)(1-2y_1)} 
\Bigg]   \nonumber \\
& \phantom{\text{EC}^{(1)}(y_1)} + (y_1\to 1-y_1) \ . 
\ee
where $2E$ is the total energy of incoming particles, $y_1=(1-\vec n \cdot\vec n_1)/2$ and $\kappa^2 = 32 \pi G_N$. As mentioned above, this correlator receives contributions from the one-loop $2\to2$ amplitude
and the tree-level $2\to3$ amplitude.
Each contribution is infrared divergent, but the divergences cancel in their sum, yielding an infrared-finite result for $\langle{\cal E}(\vec n_1)\rangle$.
The symmetry of \re{eq:ECintro} under $y_1\to 1-y_1$ reflects the invariance of the observable  under the exchange of the incoming particles. The double poles of \re{eq:ECintro} at $y_1=0$ and $y_1=1$ 
originate from the familiar ${1}/{(tu)^2}$ singularity 
of the squared tree-level gravitational amplitudes in the forward limit $t\to 0$ or $u \to 0$. We also observe that the expression inside the brackets 
in \re{eq:ECintro} is given by a sum of terms, each of the same transcendental weight two. 

The two-point energy correlator $\langle{\cal E}(\vec n_1){\cal E}(\vec n_2)\rangle$  depends on the  three angular variables $y_1$, $z$ and $\beta$ (see Figure~\ref{fig:angles-sphere}). Averaging this correlator over the direction of the incoming beam $\vec n$, we can define a simpler function 
$\widebar{\langle{\cal E}(\vec n_1){\cal E}(\vec n_2)\rangle}$ which  depends only on the angle between the calorimeters $z=(1-\cos\theta)/2$. 
Since the two functions have distinct properties, we discuss them separately.

In this paper, we study the two-point energy correlator
(EEC)
in different kinematic limits, corresponding to two distinct
configurations of the incoming beam and the calorimeters:

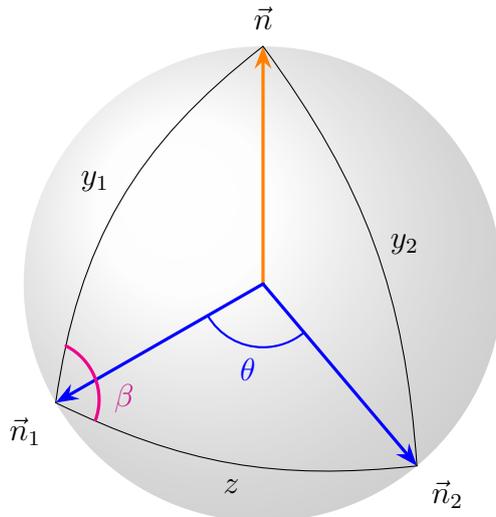
\begin{figure}[t!]
    \centering
\begin{tikzpicture}[
    scale=2.1,
    >=Stealth,
    line cap=round,
    line join=round
]

\def\R{1.5}
\coordinate (O) at (0,0);
\shade[ball color=white!95!gray, opacity=0.30] (O) circle (\R);

\coordinate (N)  at (90:\R);   %
\coordinate (N1) at (210:\R);  %
\coordinate (N2) at (310:\R);  %
\coordinate (M) at ($(N1)!0.45!(N2)+(0,-0.18)$);
\draw[very thick,->,draw={rgb,255:red,255; green,128; blue,0}] (O) -- (N)  node[above=2pt] {$\vec n$};
\draw[very thick,->,draw=blue] (O) -- (N1) node[below left=1pt] {$\vec n_{1}$};
\draw[very thick,->,draw=blue] (O) -- (N2) node[below right=1pt] {$\vec n_{2}$};
\draw[thin,black]
  (N1) to[bend left=22]
  node[midway,above left=0pt] {$y_1$} (N);
\draw[thin,black]
  (N) to[bend left=16]
  node[midway,right=2pt] {$y_2$} (N2);
\draw[thin,black]
  (N1) to[bend right=16]
  node[midway,below=2pt] {$z$} (N2);
\draw[magenta,very thick]
  (-1.05,-0.86)  arc[start angle=-20, end angle=63, radius=0.38];
\node[magenta!80!blue] at ($(-0.87,-0.72)$) {$\beta$};
\draw[blue,thick] (210:0.40) arc[start angle=210, end angle=310, radius=0.40];
\node[blue] at (260:0.55) {$\theta$};
\end{tikzpicture}
    \caption{Geometry of the two-point energy correlator. In orange, we plot the direction of the beam. In blue, the location of the calorimeters at $\vec n_i$ and $\vec n_2$, and the angle  between them, $\vec n_1 \cdot \vec n_2 = \cos \theta$, $0\le \theta \le \pi$. The angle between the  planes spanned by  $(\vec n_1, \vec n_2)$ and $(\vec n, \vec n_1)$ is denoted by $0\le \beta \le \pi$.  We also introduce the variables $y_i = (1 - \vec n \cdot \vec n_i)/ 2$ and $z=(1 - \vec n_1 \cdot \vec n_2)/2$.
    }
    \label{fig:angles-sphere}
\end{figure}

\begin{itemize}
\item In the \emph{collinear limit} $\theta \to 0$, or equivalently $z \to 0$, the EEC decomposes into the sum of a contact term and a 
smooth function that admits a regular expansion in powers of $\sqrt{z}\sim \theta$, see also \cite{Herrmann:2024yai},
\be
\label{eq:OPE}
\big\langle \mathcal E(\vec n_1)\, \mathcal E(\vec n_2) \big\rangle \sim \# \delta(z) + \text{regular} \ .
\ee
This contrasts sharply with gauge theories and is tied to the absence of
collinear divergences in gravity.
The contact term receives contributions from both virtual and real radiation,
each individually infrared divergent.
We show explicitly that these divergences cancel at one loop, and we compute
the finite coefficient of the $\delta(z)$ term.

\item In the \emph{back-to-back limit} $\theta\to\pi$, or equivalently  $z\to 1$, the behavior of the EEC is controlled by soft graviton 
radiation, see \cite{Herrmann:2024yai}.
We derive an all-order expression for the leading behavior,
\begin{align} 
\big\langle \mathcal E(\vec n_1) \mathcal E(\vec n_2)\big\rangle
=\frac{C(y_1,\beta)}{(1-z)^{1-B_{\rm gr}(y_1,E)/2}}\,, 
\label{EE-fin1Intro}
\end{align}
where $B_{\rm gr}(y_1,E)$ is the gravitational Bremsstrahlung function
\cite{Weinberg:1965nx}
\begin{align}
  B_{\rm gr}(y_1,E) = - { (\kappa E)^2\over 2\pi^2} (y_1 \log y_1 + (1-y_1) \log(1-y_1))\,.
\end{align}
The residue $C(y_1,\beta)$ encodes the hard scattering data (including its
dependence on the matter content of the gravitational theory),
 and can be computed
systematically order by order; we determine it explicitly at NLO in the cases
studied here. %
\end{itemize}

Let us next present explicit results for the EEC in ${\cal N}=8$ SG. It is convenient to introduce dimensionless EEC as follows
\be\label{EE}
\big\langle \mathcal E(\vec n_1) \mathcal E(\vec n_2)\big\rangle = E^2 \left({\kappa E \over 2} \right)^4  \bigg[{\rm EEC}^{(0)} + \left({\kappa E \over 2} \right)^{2}  {\rm EEC}^{(1)} +O((\kappa E)^4) \bigg].
\ee
The leading-order result  ${\rm EEC}^{(0)}$ is localized at the endpoints $z=0,1$ (see \eqref{eq:treelevelEEC} below). Here we quote the NLO result
\be
\label{eq:EECSGintro}
{\rm EEC}^{(1)} &= 
{1 \over 8 \pi^5} \delta(z) \biggl[  \frac{\pi^2}{6 y_1^2 (1-y_1)^2} + \frac{2{\rm Li}_2(y_1)}{(1-y_1)^2 (1-2 y_1)} +\frac{\log^2(1-y_1)}{y_1 (1-y_1) (1-2 y_1)} + (y_1\to 1-y_1)\biggr] \nonumber \\ 
&+{1 \over 4\pi^5} \delta(1-z)  \frac{\log(y_1) \log(1-y_1)}{ y_1^2(1-y_1)^2} + {\rm EEC}_{\text{reg}} ,
\ee
where the regular part ${\rm EEC}_{\text{reg}}$ is nonzero away from the endpoints. It is given in \eqref{eecgr3pt} and  has also been reported in \cite{MuratKologluSeminar2024}. In contrast to ${\rm EEC}_{\text{reg}}$, computing the contact terms in 
\eqref{eq:EECSGintro} requires a careful interplay between virtual corrections 
and real emissions, with a delicate cancellation of infrared divergences that 
appear at intermediate stages as we explain in the main text.\footnote{The $\delta(1-z)$ contact term emerges from the finite-coupling formula \eqref{EE-fin1Intro} when expanding it at small $(\kappa E)$.}

As a stringent consistency check of our results, we verify that the EEC~\re{EE}
satisfies the Ward identities associated with energy and momentum conservation.
Importantly, the contact terms play a crucial role in ensuring these identities.
Explicitly, we find
\begin{align}\notag\label{sr}
&2\int_0^1 dz \int_0^{\pi} d\beta\,
{\rm EEC}^{(1)}(y_1,\beta,z)
= {\rm EC}^{(1)}(y_1)\,, \\
&2\int_0^1 dz \int_0^{\pi} d\beta\,
(1-2z)\,{\rm EEC}^{(1)}(y_1,\beta,z)
= 0 \, .
\end{align}
The derivation of these identities is presented in
Section~\ref{sect:obs}.

When analyzing the beam-averaged energy correlator,
$\overline{\rm EEC}=\widebar{\langle{\cal E}(\vec n_1){\cal E}(\vec n_2)\rangle}$,
obtained by performing the angular average
$\int d\Omega_{\vec n} \,{\rm EEC}/(4\pi)$, it is essential to keep the detectors away from the collinear and back-to-back configurations. This amounts to restricting to the kinematic region $0<z<1$.
The reason is that the angle between the detectors $\theta$ acts as a regulator of the otherwise singular forward region, which is inevitably probed once the average over the beam direction is taken.
As a representative example, we quote the NLO result for the beam-averaged EEC in ${\cal N}=8$ SG,
\be
\label{eq:introform}
\overline{\rm EEC}^{(1)} &= \frac{(1+u^2)^2}{2 \pi^5 u^2}\biggl( \frac{\pi^2}{3} - 2 u \log(u) \arctan(u) - (1+i u) {\rm Li}_2(i u) - (1-i u) {\rm Li}_2(-i u) \biggr),
\ee
where $u=\tan(\theta/2)$. %
Using the terminology of \cite{Arkani-Hamed:2008owk}, we can view \eqref{eq:introform} as the simplest observable in the simplest theory.

The result \eqref{eq:introform} exhibits several nontrivial properties, including positivity, analyticity, and polynomial boundedness. We find the same qualitative features for the corresponding correlator in pure gravity, as well as of the leading stringy corrections.
In  addition, in  ${\cal N}=8$ SG the energy correlators exhibit maximal transcendentality.

\begin{figure}[t]
  \centering
 \includegraphics[width=0.75\linewidth]{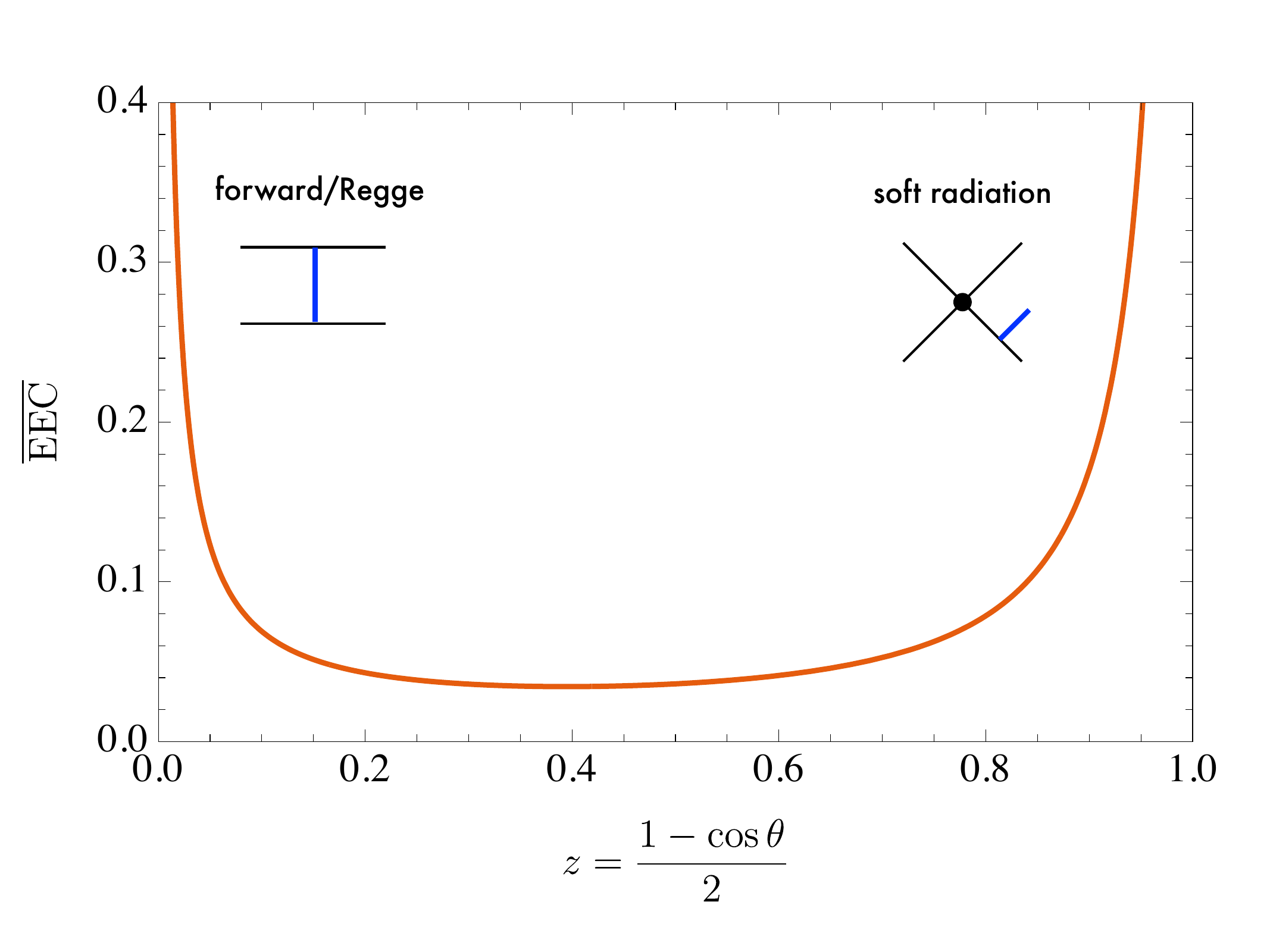}
 \caption{The leading contribution to the
beam-averaged EEC in ${\cal N}=8$ supergravity, see \eqref{eq:introform}.
Its overall shape is reminiscent of the familiar gauge-theory result.
In gravity, however, the singular behavior as $z\to 0$ is tied to the forward-scattering singularity (as opposed to the gauge-theory collinear divergences).
In the back-to-back limit $z\to 1$, the EEC is controlled by soft radiation. 
}
  \label{fig:aveec}
\end{figure}

In the collinear limit $\theta\to 0$, the divergence of the beam-averaged EEC is tied to the singular nature of forward scattering in gravity (rather than the collinear singularities of the gauge-theory type).
In the back-to-back limit $\theta\to\pi$, the behavior of the EEC is instead governed by soft radiation, as discussed above. We plot \eqref{eq:introform} in Figure~\ref{fig:aveec}.

Since ${\cal N}=8$ SG arises as the low-energy limit of type~II string theory compactified on $T^6$, we expect \eqref{eq:introform} to be the leading weak-coupling approximation to a function that is well defined nonperturbatively.
It would be interesting to find an alternative nonperturbative prescription for computing this observable, potentially within flat-space/celestial holography \cite{Pasterski:2017kqt,Pasterski:2017ylz,Pasterski:2021rjz}.

Let us also emphasize that the observables considered in the present paper differ from those typically measured in gravitational-wave experiments \cite{Kosower:2018adc}. In principle, given the waveform, one can calculate the one-point function of the energy flux. More broadly, in the classical limit the multi-point energy correlators factorize \cite{Gonzo:2020xza}. We instead focus on the quantum correlators produced in the collision of two gravitons.

\subsection*{Outline}

In Section~\ref{sect:obs} we define the observables of interest in detail. They are energy-weighted inclusive cross sections that can be written in terms of suitably regularized scattering amplitudes. We begin by describing the kinematics of the incoming beams and the energy detectors. We then discuss the basic properties of the energy correlators, including their symmetries, Ward identities, and the appearance of contact terms.  

In Section~\ref{sec:SG}, we consider the energy correlators 
in ${\cal N}=8$ SG in four dimensions. The amplitudes in this theory have been studied extensively \cite{Bern:1998ug,Bern:2006kd,Bern:2007hh,Bern:2008pv,Bern:2009kd,Bern:2012uf,Bern:2018jmv}. %
We compute the one- and two-point energy correlators at one loop, including the contact terms, and explicitly demonstrate the cancellation of infrared divergences. We verify that the energy correlators satisfy the energy-momentum conservation Ward identities. We also compute the beam-averaged energy correlator $\overline{\rm EEC}$, and extend the analysis to more general detectors by considering correlators weighted by arbitrary powers of the energy.

In Section~\ref{sec:puregravity}, we repeat the analysis for pure gravity. In this case the results are more complicated but exhibit the same qualitative features: IR finiteness, regularity at small angles, universal singular behavior in the back-to-back region.

In Section~\ref{sec:stringyEEC}, we calculate the stringy corrections to the ${\cal N}=8$ SG results. Indeed, ${\cal N}=8$ SG admits a simple UV completion in terms of superstring theory compactified on $T^6$. We study the behavior of the stringy corrections in the collinear and back-to-back limits.

In Section~\ref{sec:disprel}, we study the energy correlators as functions of a complex $z$. We find that they are analytic and polynomially bounded in the complex $z$-plane, and hence admit dispersion relations. We then use these dispersion relations to investigate the positivity properties of (i) the Taylor coefficients of the beam-averaged energy--energy correlator about the midpoint $z=1/2$, and (ii) its multipole expansion.

In Section~\ref{sec:backtoback}, we analyze the behavior of the energy--energy correlator in the back-to-back limit $z \to 1$. This limit is known to be controlled by soft radiation, which exponentiates in gravity \cite{Weinberg:1965nx}. We show that the same exponentiation takes place at the level of the energy--energy correlator for perturbative corrections enhanced by powers of $\log(1-z)$. We use the properties of soft gravitons to derive the explicit all-order result for the gravitational energy--energy correlator in the back-to-back limit.

We conclude with a discussion and a list of open directions. Technical details are collected in several appendices.

\subsection*{Summary of main results}

As mentioned above, the study of energy correlators in gravity was initiated in Refs.~\cite{Gonzo:2020xza,Herrmann:2024yai}.
The present work substantially extends these studies in several directions, which we summarize below.\footnote{We thank the referee for suggesting that we clarify the main results of this work.}
 
\begin{enumerate}[label=(\roman*)]
\item Rather than the scattering of massive scalar particles analyzed in Ref.~\cite{Herrmann:2024yai}, we consider a different physical setup: energy correlators in the scattering of two gravitons in four-dimensional gravity.

\item We formulate, for the first time, sum rules (equivalently, Ward identities) for energy correlators that follow from total energy and momentum conservation (see \re{sr}). We explicitly verify these relations for the  \emph{one-point} and  \emph{two-point} energy correlators at \emph{one-loop order} in two different gravitational theories: maximally supersymmetric $\mathcal N=8$ supergravity and pure Einstein gravity without matter.\footnote{In Ref.~\cite{Herrmann:2024yai} only the leading contribution to the two-point correlator for $0<z<1$ has been calculated, which is not sufficient to check the energy-momentum conservation sum rules.}

\item 
We demonstrate that the validity of the sum rules crucially depends on the inclusion of contact terms in the two-point energy correlators localized at $z=0$ and $z=1$. These terms arise from the interplay between infrared-divergent virtual corrections and real graviton emissions. Infrared safety of the energy correlators requires the corresponding contact terms to be infrared finite. By computing these terms explicitly at one loop, we establish the infrared finiteness of gravitational energy correlators.
 
\item We introduce beam-averaged energy correlators and compute them at one loop (see \re{eq:introform}). We analyze their analytic structure in the complex $z$-plane, establish analyticity and polynomial boundedness, and derive the corresponding dispersion relations and positivity constraints.

\item We compute the leading string-theoretic corrections to gravitational energy correlators and discuss their properties. We also discuss the expected form of gravitational energy correlators at high energies due to creation and evaporation of black holes.

\item We derive the all-order asymptotic behaviour~\re{EE-fin1Intro} of the two-point energy correlator in the back-to-back limit $z\to 1$. This behaviour originates from the resummation of logarithmically enhanced contributions generated by soft graviton radiation.

\end{enumerate}
 
{\it 
{\bf Note added.} We would like to clarify the partial overlap of the material discussed in the present paper with the earlier work of Ref.~\cite{Herrmann:2024yai} (see also  Ref.~\cite{Gonzo:2020xza}) on a similar subject. We thank our referee for requesting this clarification. 

The present work goes substantially beyond Ref.~\cite{Herrmann:2024yai} in several independent directions:
\begin{enumerate}[label=(\roman*)]
\item We study a different physical setup: two incoming gravitons in four-dimensional gravity,
rather than a massive-scalar initial state.
\item Whereas Ref.~\cite{Herrmann:2024yai} computed a leading \emph{two-point} correlator, we compute both the
\emph{one-point} and \emph{two-point} energy correlators at the first nontrivial
\emph{loop} order in two different gravitational theories, namely the maximally supersymmetric $\mathcal N=8$ supergravity and pure Einstein gravity without matter.
\item We compute the contact terms at $z=0$ and $z=1$, which are not a minor detail. They arise from the interplay of virtual corrections and real emission and are essential for the
consistency of the perturbative expansion. Their calculation requires a careful treatment of the cancellation of
infrared divergences, a problem that was not addressed in Ref.~\cite{Herrmann:2024yai}. One cannot claim that the energy correlators are well-defined  IR finite observables without explicitly computing the contact terms and demonstrating that they are themselves IR finite. 
\item We formulate for the first time and explicitly verify the Ward identities associated with energy and momentum conservation. We show
that the contact terms are required for these identities to hold.
\item We introduce and compute the beam-averaged energy-energy correlator, which is a new
observable not analyzed in Ref.~\cite{Herrmann:2024yai}.
\item We study the analytic structure of the beam-averaged correlator in the complex domain for $z$, establish
analyticity and polynomial boundedness, and derive dispersion relations together with positivity
constraints.
\item We compute the leading stringy correction to the gravitational energy correlators.
\item We derive the all-order back-to-back asymptotics governed by soft-graviton dynamics,
including the exponentiation of the logarithmically enhanced terms and the universal function
$\gamma(y_1)$ controlling the singular behavior.
\end{enumerate}

In our view, these results are not contained in Ref.~\cite{Herrmann:2024yai}, and together they constitute the main
novel physics content of our paper. What overlaps with Ref.~\cite{Herrmann:2024yai} is the general direction,
some of the detector language, and the expectation that the back-to-back limit is controlled by
soft physics.

}

\section{Observables}\label{sect:obs}

In this section we introduce the observables of interest---the energy-weighted cross sections---which are expressed in terms of integrated squared scattering amplitudes. Our discussion is perturbative, and we assume that the gravitational theory under consideration has been renormalized to any desired order. Making sense of the theory beyond a perturbative expansion of low-energy observables requires a UV completion, such as string theory. This will not concern us here, since we focus on the IR aspects of our observables.

\subsection{Kinematics}
Let us consider the scattering process $\text{graviton}+\text{graviton}\to\text{anything}$, 
\be\label{proc}
p_1 + p_2 \to q_1 + q_2 + X ,
\ee
in which one or two particles  with momenta $q_{1,2}$ in the final state are detected by calorimeters located at different points on the celestial sphere.  Here $X$ represents an arbitrary number of undetected particles in the final state.\footnote{When only the particle with  momentum $q_1$ is detected, the second particle with  momentum $q_2$ is treated as belonging to $X$.}  In the theories considered in the present paper, all particles are massless.

In the center-of-mass frame, the lightlike momenta of the incoming and detected particles can be parametrized as follows:
\begin{align}\label{kin0}
{}& p_1^\mu=E (1,\vec n)\,, && p_2^\mu=E (1,-\vec n)\,, 
&& q_1^\mu = E_1 ( 1, \vec n_1)\,, && q_2^\mu = E_2 (1, \vec n_2) \,,
\end{align}
where $\vec n$, $\vec n_1$ and $\vec n_2$ are unit vectors and the total center-of-mass energy is $2E$.

The geometry of the incoming beam and the two calorimeters can be conveniently parametrized by the kinematical variables $0 \leq y_1, y_2,z \leq 1$
as shown in Figure~\ref{fig:angles-sphere}.
These variables depend on the relative angles between the unit vectors introduced in \re{kin0},
\begin{align}\label{ys}
    y_1={1-(\vec n\vec n_1)\over 2}\,,\qquad  y_2={1-(\vec n\vec n_2)\over 2}\,,\qquad z={1-(\vec n_1\vec n_2)\over 2} \ .
\end{align}
In addition, it is convenient to introduce the angle $0\le \beta \le\pi$ between the unoriented planes spanned by $(\vec n_1,\vec n_2)$ and $(\vec n,\vec n_1)$, such that
\begin{align}\label{beta}
\cos\beta ={(\vec n \vec n_2) - (\vec n_1 \vec n_2) (\vec n \vec n_1) \over \sqrt{1-(\vec n_1 \vec n_2)^2} \sqrt{1-(\vec n \vec n_1)^2}} \ . 
\end{align}
In what follows, we analyze three distinct kinematic limits, corresponding to different configurations of the beams and the calorimeters:
\begin{itemize}
\item \emph{collinear limit}, $z \to 0$;
\item \emph{back-to-back limit}, $z \to 1$;
\item \emph{forward limit}, $y_1 \to 0$.
\end{itemize}
Introducing the angle $\theta$ between the calorimeters, defined by $\cos\theta = 1 - 2z$, the first two limits correspond to $\theta \to 0$ and $\theta \to \pi$, respectively. In the forward limit, one of the calorimeters becomes aligned with the incoming beam.  

The energy correlators measure the energy flux carried by the particles in the final state independently of their quantum numbers. It is therefore convenient to sum over these quantum numbers and introduce the probability density for producing a final state characterized by a set of particle momenta $(q_1,\ldots,q_L)$. It is given by the squared scattering amplitude, summed over the internal quantum numbers of the particles in the final state 
\be\label{calM}
 \mathbb{M}_{2 \to L} &= \sum_{\text{helicity}} |{\cal M}_{2 \to L}|^2 \ .
\ee
In terms of this  probability density, the one- and two-point energy correlators are given by
\be
\label{eq:defec}
& \text{EC}(\vec n | \vec n_1)=\sum_{L=2}^\infty {1 \over L!} \int d\text{PS}_{L}\, \mathbb{M}_{2 \to L} \Big( \sum_{i=1}^L E_i \,\delta(\Omega_{\vec q_i} - \Omega_{\vec n_1}) \Big) \ , \\
&\text{EEC}(\vec n | \vec n_1,\vec n_2)= \sum_{L=2}^\infty {1 \over L!} \int d\text{PS}_{L}\, \mathbb{M}_{2 \to L} \Big( \sum_{i=1}^L E_i \,\delta(\Omega_{\vec q_i} - \Omega_{\vec n_1}) \Big) \Big( \sum_{j=1}^L E_j \,\delta(\Omega_{\vec q_j} - \Omega_{\vec n_2}) \Big) \ , 
\label{eq:defeec}
\ee
where the Lorentz invariant phase-space integration measure $d \text{PS}_{L}$ includes the overall momentum\hyp{}conserving delta function 
$(2 \pi)^4 \delta^{(4)}(p_1+p_2-\sum_{i=1}^{L} q_i)$, see \p{eq:PS}. 
The symmetry factor 
$1/L!$ in the formulas above is required to avoid overcounting contributions from  detected particles.

The relations \re{eq:defec} and \re{eq:defeec} admit an interpretation in terms of the one- and two-point correlation functions of the energy flow operators \cite{Sveshnikov:1995vi,Korchemsky:1997sy,Korchemsky:1999kt}
\begin{align}\label{eec-E}\notag
  {}&  \text{EC}(\vec n | \vec n_1) = \langle {\cal E}(\vec n_1) \rangle\,,
    \\[2mm]
{}& \text{EEC}(\vec n | \vec n_1,\vec n_2) = \langle {\cal E}(\vec n_1) {\cal E}(\vec n_2) \rangle \,,
\end{align}
where the operator ${\cal E}(\vec n)$ is defined by its action on a multi-particle state,
\be
\label{eq:detdef}
{\cal E}(\vec n) | X \rangle = \sum_{q_i \in X} E_i \,\delta(\Omega_{\vec q_i} - \Omega_{\vec n}) | X \rangle \ . 
\ee
 
Note that, in contrast with the familiar energy correlators defined for the process $e^+ e^- \to \text{hadrons}$ in QCD  \cite{Basham:1977iq,Basham:1978bw,Basham:1978zq,Fox:1978vw}, the gravitational energy correlators \re{eq:defec} and \re{eq:defeec} are not normalized by the total cross section $\sigma_{\text{tot}}$. The reason is that the latter is infinite  for plane-wave scattering in four dimensions. Another key difference  is the presence of initial-state radiation in gravity, a direct consequence of the universal coupling of gravitons to all particles.

\subsection{Energy--momentum conservation and symmetries}
\label{sec:EMWI}

An immediate consequence of the definition \re{eec-E} is that the one- and two-point energy correlators obey the following  energy- and momentum-conservation Ward identities (sum rules), 
\begin{align}
&{} \int d\Omega_{\vec n_2} \text{EEC}(\vec n | \vec n_1,\vec n_2) = 2E \, \text{EC}(\vec n | \vec n_1) \,, \label{eq:energy} \\
&{} \int d\Omega_{\vec n_2} \vec n_2 \, \text{EEC}(\vec n | \vec n_1,\vec n_2) = 0 \,. \label{eq:momentum}
\end{align}
It follows from the definition of the energy flow operator \re{eq:detdef} that the integral over the orientation of the calorimeter in these two relations yields, respectively, the total energy and spatial momentum of all particles in the final state. In the center-of-mass frame, this is  $(2E,\vec 0)$.~\footnote{Note that the relation \re{eq:momentum} only holds for final states containing  massless particles. }

The energy correlators have the following symmetry properties as functions of the unit vectors $\vec n$ and $\vec n_i$,
\begin{align}\notag
 {}&    \text{EC}(\vec n | \vec n_1) = \text{EC}(-\vec n | \vec n_1)\, , 
 \\[2mm]
{}& \text{EEC}(\vec n | \vec n_1,\vec n_2) = \text{EEC}(-\vec n | \vec n_1,\vec n_2)= \text{EEC}(\vec n | \vec n_2,\vec n_1) \ . \label{eq:EECdiscrsym}
\end{align}
They follow from the symmetry of the observables under the exchange of the particles in the initial state, $p_1\leftrightarrow p_2$, and the detected particles in the final state, $q_1\leftrightarrow q_2$. In terms of the angular variables introduced in \eqref{ys}, this is the symmetry under $y_1 \leftrightarrow y_2$ 
and $y_i \to 1-y_i$.

\subsection{Perturbative expansion}

We compute the energy correlators \re{eq:defec} and \re{eq:defeec} in perturbation theory by expanding the squared matrix element \re{calM} in powers 
of the gravitational coupling constant
 $\kappa^2 = 32\pi G_N$, 
\begin{align}\label{M-weak}
\mathbb{M}_{2 \to L} = \left(\frac{\kappa}{2} \right)^{L+2} \mathbb{M}^{(0)}_{2 \to L} + \left(\frac{\kappa}{2} \right)^{L+4} \mathbb{M}^{(1)}_{2 \to L} + \ldots \ . 
\end{align}
The corresponding expressions for the energy correlators are series in
the dimensionless parameter $(\kappa E)^2$,
\be
& \text{EC} = E \left({\kappa E \over 2} \right)^4  \sum_{\ell=0}^\infty \left({\kappa E \over 2} \right)^{2 \ell} \text{EC}^{(\ell)}  , \cr
& \text{EEC} =E^2 \left({\kappa E \over 2} \right)^4 \sum_{\ell=0}^\infty \left({\kappa E \over 2} \right)^{2 \ell} \text{EEC}^{(\ell)} , \label{eq:ECpertexp}
\ee
where $\text{EC}^{(\ell)}$ and $\text{EEC}^{(\ell)}$ are dimensionless functions of the angles. 

These functions are obtained from \re{eq:defec} and \re{eq:defeec} by replacing the integration densities with their perturbative expansion. %
The first two terms in \re{M-weak} are given by the tree-level and one-loop 
$(L+2)$-particle amplitudes squared, summed over the helicities of the particles in the final state,
\begin{align}\notag
& \mathbb{M}_{2 \to L}^{(0)} = \sum\limits_{\rm helicity}|{\cal M}^{(0)}_{2 \to L}|^2 \,, \\
& \mathbb{M}_{2 \to  L}^{(1)} = \sum\limits_{\rm helicity} 2{\rm Re} \left( {\cal M}_{2 \to L}^{(1)} \left( {\cal M}^{(0)}_{2 \to L} \right)^* \right) . \ \label{eq:MM1loop}
\end{align}

The calculation of the energy correlators \re{eq:ECpertexp} requires an intermediate  infrared (IR) regulator. The reason is that,  
at any fixed order in perturbation theory, each term in the sums \eqref{eq:defec} and \eqref{eq:defeec}  is individually IR divergent, but these divergences cancel in the total sum. This is the familiar cancellation mechanism  of IR divergences between virtual corrections and real emissions. In what follows we employ dimensional regularization with $d=4-2\epsilon$ and $\epsilon <0$.

\subsection{Contact terms}

\label{sec:contact}

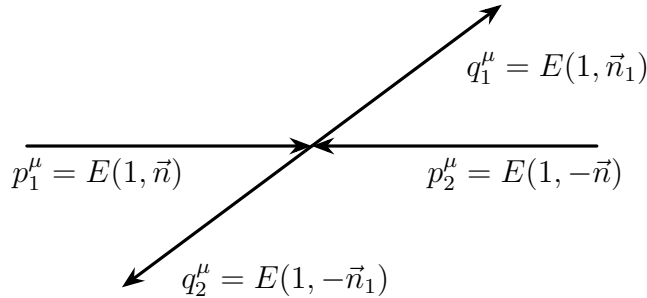
\begin{figure}
\begin{center}
\begin{tikzpicture}[
    scale=1.25,
    >=Stealth,
    line cap=round,
    line join=round
]

\coordinate (O) at (0,0);

\draw[very thick,->]
  (-3,0) -- (O)
  node[near start,below] {$p_1^\mu = E(1,\vec n)$};

\draw[very thick,->]
  (3,0) -- (O)
  node[near start,below] {$p_2^\mu = E(1,-\vec n)$};

\draw[very thick,->]
  (O) -- (2,1.5)
  node[near end,below right] {$q_1^\mu=E(1,\vec n_1)$};

\draw[very thick,->]
  (O) -- (-2,-1.5)
  node[near end,below right] {$q_2^\mu=E(1,-\vec n_1)$};

\end{tikzpicture}
\end{center}
\caption{Kinematics of the two-to-two scattering in the center-of-mass frame. In a two-particle final state, the only way to obtain a nontrivial correlator between the particles is to place the detectors in the collinear or back-to-back configurations, see  \eqref{conterm}.}
\label{fig:2to2}
\end{figure}

In order to gain further insight into the structure of the energy correlators,  let us consider the leading-order contribution $\text{EEC}^{(0)}$ to \re{eq:ECpertexp}.
It arises from the tree-level two-to-two scattering $p_1+p_2\to q_1+q_2$ and is given by the sum of two contact terms;
 see Figure~\ref{fig:2to2},
\begin{align} 
\text{EEC}^{(0)} = {\mathbb{M}_{2 \to 2}^{(0)}  \over 32 \pi^2 E^4} \Big( \delta(\Omega_{\vec{n}_{1}}-\Omega_{\vec{n}_{2}}) + \delta(\Omega_{\vec{n}_{1}}+\Omega_{\vec{n}_{2}}) \Big) . \label{conterm}
\end{align}
They are
localized at the collinear $(\vec n_1=\vec n_2)$ and back-to-back $(\vec n_1=-\vec n_2)$ configurations of  calorimeters.

The first (collinear) contact term in \re{conterm} is already manifest in the definition \eqref{eq:defeec}. It arises from a single particle being detected by both calorimeters.
We expect such terms to persist at any loop order in $(\kappa E)^2$ and at finite coupling. This is different from energy correlators in CFTs (or perturbative QCD). In these cases, we expect that the contact term is a feature of the perturbative expansion and at finite coupling turns into an integrable power, ${\gamma / z^{1-\gamma}}$ with $\gamma>0$, at finite coupling. In gravity, due to the absence of collinear divergences \cite{Weinberg:1965nx,Akhoury:2011kq}  we expect the collinear delta function to survive at finite coupling (the same is true in gapped theories, such as QCD in the IR), see also the discussion in \cite{Herrmann:2024yai}. Below we check this statement explicitly at NLO in ${\cal N}=8$ SG (plus stringy corrections) and pure gravity.

The second (back-to-back/anti-collinear) contact term in \eqref{conterm} is specific to the perturbative expansion. Like the collinear contact term, it persists to any order in perturbation theory. We will show, however, that \emph{at finite coupling} it turns into an integrable power-like function ${1 / (1-z)^{1-B_{\rm gr}/2}}$, where $B_{\rm gr}$ is the gravitational Bremsstrahlung function \cite{Weinberg:1965nx}.  

The importance of the contact terms stems from the fact that they contribute to the energy--momentum conservation sum rules \eqref{eq:energy} and \eqref{eq:momentum}. %
These sum rules provide a powerful consistency check of the energy-correlator calculation. For this reason, we compute the contact terms explicitly and demonstrate their infrared finiteness.

\subsection{Generalized energy fluxes}

According to their definitions \re{eq:defec} and \re{eq:defeec}, the energy correlators are linear in the energy of the detected particles.  
We observe, however, that the contact term in the two-point energy correlator \eqref{eq:defeec} involves the second power of the energy. Thus, when considering multi-point energy correlators it is also natural to introduce detectors that measure higher powers of the energy.

Let us therefore generalize the definitions \re{eq:defec} and \re{eq:defeec} as follows:
\be
\label{eq:defecJ} 
&\text{EC}_{J_1}(\vec n | \vec n_1)=\sum_{L=2}^\infty {1 \over L!} \int d\text{PS}_{L}\, \mathbb{M}_{2 \to L} \Big( \sum_{i=1}^L E_i^{J_1} \,\delta(\Omega_{\vec q_i} - \Omega_{\vec n_1}) \Big) \ , \\\notag
&\text{EEC}_{J_1, J_2}(\vec n | \vec n_1,\vec n_2)= \sum_{L=2}^\infty {1 \over L!} \int d\text{PS}_{L}\, \mathbb{M}_{2 \to L} \Big( \sum_{i=1}^L E_i^{J_1} \,\delta(\Omega_{\vec q_i} - \Omega_{\vec n_1}) \Big) \Big( \sum_{j=1}^L E_j^{J_2} \,\delta(\Omega_{\vec q_j} - \Omega_{\vec n_2}) \Big) \ , 
\ee
where $J_i$ can be complex. Note that our convention for labeling the detectors differs from the standard one in the QCD/CFT literature. Indeed, if we start with a twist-2 local operator of spin $J$ in the free theory and place it at null infinity, it  produces \eqref{eq:defecJ} with $J_1 = J-1$. In gauge theory, the free-field definition of the detectors introduced above undergoes nontrivial renormalization \cite{Caron-Huot:2022eqs}, which can be traced to the presence of collinear divergences in the theory, or to the anomalous dimension of twist-two operators. We do not expect such effects in gravity \cite{Herrmann:2024yai}, and our explicit one-loop calculations below confirm this expectation. Let us also point out that the generalized energy fluxes do not obey simple sum rules like \eqref{eq:energy} and \eqref{eq:momentum}. %

We do expect generalized energy flow correlators to develop divergences for certain values of $J_i$. One notable example is $J_i=0$, which is related to measuring the fluxes of the total number of particles. It has been pointed out recently that energy flux correlators obey analogs of soft theorems \cite{Chang:2025zib}. In gravity, the corresponding detector soft theorem has been discussed recently in \cite{Gonzalez:2025ene,Moult:2025njc}.

\section{Energy correlators in ${\mathcal N}=8$ supergravity}

\label{sec:SG}

As a first example, we study energy correlators in ${\cal N}=8$ supergravity (SG). This theory has numerous appealing analytic features, providing a unique testing ground for perturbative quantum-gravity observables: its amplitudes are tightly constrained by symmetry and exhibit uniform transcendentality, and potential UV divergences are postponed to very high loop order \cite{Bern:2018jmv}. In this sense, we find that energy correlators in SG take a simpler analytic form than their pure-gravity counterparts, while their infrared behavior is completely analogous.

The tree-level contribution to the energy correlators is finite. We refer to it as leading order (LO).  It requires the tree-level four-point amplitude, see \re{eq:MM1loop}. Our goal is to compute the first correction (NLO) to the energy correlators in the perturbative expansion \re{eq:ECpertexp}. To this end, using \re{eq:MM1loop}, it is enough to know the tree-level four- and five-point amplitudes together with  the one-loop four-point amplitude. These amplitudes are well known in the literature, see, for example, \cite{Berends:1988zp,Green:1982sw,Dunbar:1994bn}. For completeness, we briefly recall them below.

The NLO correction receives contributions from both virtual and real processes. Each of these contributions is individually infrared divergent and therefore requires regularization. The divergences in the virtual contribution originate from the one-loop amplitude and arise from the integration over soft loop momenta. In contrast, the divergences in the real contribution emerge upon integrating over the one-particle phase space of undetected soft radiation.
We compute the virtual and real contributions explicitly below and verify that their sum is infrared finite. Further details on the phase-space parametrization, as well as explicit phase-space integrals for the single- and double-detector energy correlators, are provided in Appendix~\ref{app:PSint}.

\subsection{Amplitudes}

The ${\cal N}=8$ supergravity multiplet contains $2^{8}$ on-shell states. They are conveniently packed in a single superstate $\Phi(\eta)$ with the help of eight  Grassmann variables $\eta^A$ (with $A=1,\ldots,8$) carrying helicity $+1/2$. The gravitons of helicity $+2$ and $-2$ are the bottom and the top components of the multiplet, respectively,  and their superpartners of helicities $\pm 3/2, \pm 1, \pm 1/2,0$ lie in between, 
\begin{align}
\Phi(\eta) = (\eta)^0 \ket{\text{grav},+2} +\ldots + (\eta)^8 \ket{\text{grav},-2} \,.
\end{align}
The scattering amplitudes of the various  states are combined in a superamplitude. We need the four- and five-point MHV superamplitudes $\mathcal{M}_n$ with $n=4,5$. They have the following perturbative expansion: 
\begin{align}
\mathcal{M}_n = \delta^{16}(Q) \left( \frac{\kappa}{2}\right)^{n-2} \left( {M}^{\rm (0)}_n+ \left( \frac{\kappa}{2} \right)^2 {M}_{n}^{(1)} + \ldots \right) . \label{eq:amplSG}
 \end{align}
Here $Q^{\alpha A} = \sum_{i=1}^{n} \la^{\alpha}_i \eta_i^{A}$ is the ${\cal N}=8$ supercharge built from the Grassmann variables at each point, as well as the spinor-helicity variables $\lambda^\alpha_i$  carrying helicity weight $-1/2$. The latter are  defined via the representation $p^{\alpha\dot\alpha}_i = \lambda^{\alpha}_i \tilde\lambda^{\dot\alpha}_i$ of the on-shell particle momenta.\footnote{In this subsection we temporarily treat all the momenta  as incoming and label them $p_i^\mu$, $i=1,\ldots,n$.} The superamplitude $\mathcal{M}_n$  carries  helicity weight  $+2$ at each point.  

The relevant tree-level MHV functions  are
\begin{align}\notag
{}&  {M}^{(0)}(1234) = i \frac{[12]}{\vev{12}\vev{13}\vev{14}\vev{23}\vev{24}\vev{34}^2} \,, 
\\[2mm] {}& 
 {M}^{(0)}(12345) = i \frac{\varepsilon(1234)}{\prod\limits_{1 \leq i < j \leq 5} \vev{ij}}\,,   \label{eq:Mtree5}
\end{align}
where we employ the spinor-helicity bracket notation
\begin{align}
\vev{ij} = \la^{\alpha}_i \la_{j \alpha}\,, \qquad  [ij] = \tilde\la_{i \dot\alpha}  \tilde\la^{\dot\alpha}_{j}\,, \qquad s_{ij} =[ij] \vev{ji} = (p_i+p_j)^2\,, 
\end{align}
and $\varepsilon(1234) \equiv 4 i \ep_{\mu\nu\rho\sigma} p_1^\mu p_2^\nu p_3^\rho p_4^\sigma$. 
The amplitudes have Bose symmetry under permutations of the particles. 

Squaring the tree-level amplitudes and summing over all final states of the supermultiplet, see \p{eq:MM1loop}, we obtain 
\begin{align}
 \mathbb{M}_{2 \to 2}^{(0)} =\frac{s_{12}^6}{s_{13}^2 s_{23}^2}  \,, \qquad 
 \mathbb{M}_{2 \to 3}^{(0)} = - \frac{2s_{12}^8 }{\prod\limits_{1 \leq i \leq j \leq 5} s_{ij}} \times 16\, {\rm Gram}(p_1,p_2,p_3,p_4) \, . \label{eq:MMtree2pt}
\end{align}
In the five-point case, the numerator is given by the Gram determinant, which is a degree-4 polynomial in the Mandelstam variables $s_{ij}$. %
The summation over the supermultiplet of the final states results in overall factors of $s_{12}^8$ and $2s_{12}^8$ for the four- and five-point amplitudes, respectively. \footnote{If we remove this factor, the expressions in \re{eq:MMtree2pt} become symmetric under the permutations of all particles. \label{foot8}} This is obtained by a Grassmann Fourier transform of $\delta^{16}(Q)\delta^{16}(\bar{Q})$ to all-chiral Grassmann variables and by a subsequent Grassmann integration over the latter. The factor of $2$ in the five-point case comes from the anti-MHV helicity superamplitude, which is related to the MHV amplitude by charge conjugation. We explain this in detail in Appendix~\ref{app:SU_N}.
 
We would like to emphasize that, as a consequence of supersymmetry, the squared amplitude summed over final states, $\mathbb{M}_{2\to L}$, is independent of the helicity configuration of the incoming particles and is identical for all two-particle initial states. Thus, we do not have to specify the initial state in the energy correlators in supergravity. 

We will also need the one-loop four-point amplitude ${M}^{(1)}(1234)$. It is given by the crossing-invariant sum of  one-loop Feynman integrals 
\begin{align} \label{eq:M1loop}
& {M}^{(1)}(1234) = 
i \frac{[12]^2 [34]^2}{\vev{12}^2 \vev{34}^2} \, \left( I(s_{12},s_{13}) + I(s_{12},s_{23}) + I(s_{13},s_{23}) \right) \,,
\end{align}
where $I(s,t)$ is the zero-mass-box integral  in the dimensional regularization with $d = 4 -2 \ep$ \cite{Bern:1992em}
\begin{align}
& I(s,t) =  c_{\Gamma}  \frac{1}{st}\left[ \frac{2}{\ep^2}\left( \left(\frac{-s}{\mu^2} \right)^{-\ep} +  \left(\frac{-t}{\mu^2}\right)^{-\ep} \right) - \log^2\left(\frac{s}{t} \right) - \pi^2 \right] \,,\notag
\\ 
& c_{\Gamma} \equiv  \frac{1}{(4\pi)^{2-\ep}}\frac{\Gamma(1+\ep)\Gamma^2(1-\ep)}{\Gamma(1-2\ep)} \label{eq:zmbox} \,.
\end{align}
Here the Mandelstam variables come with the prescription $s \to s+i0$ and $t \to t + i0$. It  specifies the analytic continuation of $I(s,t)$ from the Euclidean region $s<0$ and $t<0$ to the physical region of interest. 

Substituting \re{eq:M1loop} into \eqref{eq:MM1loop} we find that
the one-loop contribution to the four-point squared matrix element $\mathbb{M}_{2 \to 2}^{(1)}$ takes the form
\begin{align}
& \mathbb{M}_{2 \to 2}^{(1)} = \frac{2 s_{12}^8}{s_{12}s_{13}s_{23}} {\rm Re}\, \left( I(s_{12},s_{13}) + I(s_{12},s_{23}) + I(s_{13},s_{23}) \right) \, . \label{6.11}
\end{align} 
The double poles in $\epsilon$ of the loop integral \re{eq:zmbox}, originating from the soft-collinear divergences, cancel out in the sum \eqref{6.11}, so that the function $\mathbb{M}_{2 \to 2}^{(1)}$ only has a simple pole in $\epsilon$.

In what follows, we denote the final state momenta $ q_i = -p_{2+i}$ for $i=1,2,3$ as in Section~\ref{sect:obs}, choosing them outgoing, and rewrite the squared amplitudes in the Mandelstam variables $2(p_i q_j)$ and $2(q_i q_j)$. 

\subsection{Tree level}

For the two-particle contribution
$p_1 + p_2 \to q_1 + q_2$, with final-state momenta $q_1^\mu =E(1,\vec n_1)$ and $q_2^\mu = E(1,-\vec n_1)$,  the Mandelstam invariants take the following form:
\begin{align}
& 2(p_1 p_2) = (2E)^2 \,,\qquad 2(p_1 q_1) =  (2E)^2 y \,,\qquad 2(p_2 q_1) =  (2E)^2 (1-y) \,,
\end{align}
where $y \equiv y_1$ is the angular variable defined in \p{ys}. The tree-level (LO) expression for the averaged squared matrix element  \eqref{eq:MMtree2pt} becomes
\begin{align}
\mathbb{M}_{2 \to 2}^{(0)}= {16 E^4 \over y^2 (1-y)^2} \,. \label{eq:M0}
\end{align}
This expression exhibits double poles in the forward ($y=0$) and backward ($y=1$) directions, which arise from the massless exchange in the $t-$ and $u-$channels, respectively.

\subsection{Virtual correction}

At NLO, the two-particle contribution corresponds to the virtual correction. The one-loop squared amplitude \p{6.11} contains an IR pole $1/\ep$ and takes the following form: 
\begin{align}
&\mathbb{M}^{(1)}_{2 \to 2} = \mathbb{M}_{2 \to 2}^{(0)} \frac{E^{2-2\ep} e^{-\ep \gamma_E}}{\pi^{2-\ep}} 
\left[- \frac{1}{\ep} \left( y\log(y) + (1-y)\log(1-y)\right) + \log(y) \log(1-y)+O(\epsilon) \right].  \label{M41loop}
\end{align}
Up to a normalization factor, it represents the virtual contribution to the energy correlators, see \p{eq:EJC2virt} (up to $O(\epsilon)$ terms),
\begin{align}
\text{EC}_{J}^{\text{virt}} = 
\frac{1}{2\pi^4}\left(\frac{2\pi}{E} \right)^{4\ep} {e^{-\ep \gamma_E} (4\pi)^{-\ep}  \over y^2 (1-y)^2} 
\left[- \frac{1}{\ep} \left( y\log(y) + (1-y)\log(1-y)\right) + \log(y) \log(1-y) \right].\label{eq:virtpole}
\end{align}

\subsection{Real correction}\label{s:3particle}

For the three-particle contribution $p_1 + p_2 \to q_1 + q_2+q_3$, the final-state momenta are $q_i^\mu =E x_i (1,\vec n_i)$ and 
the Mandelstam invariants reduce to (for $i=1,2,3$)
\begin{align}
& (p_1 p_2) = 2E^2 \,, \qquad
(p_1 q_i) = E^2 x_i (1-\vec n \vec n_i)  \,, \qquad (p_2 q_i) = E^2 x_i (1+\vec n \vec n_i)  \,.
\end{align}
The energy fractions satisfy $x_3 = 2 -x_1 - x_2 > 0$, $x_1 >0$, $x_2>0$, and  energy--momentum conservation requires
\begin{align}
-1+x_1+x_2-x_1 x_2\, z =0\, .
\end{align}
We can then rewrite the three-particle matrix element squared \eqref{eq:MMtree2pt} in terms of the energy fraction $x\equiv x_1$ and the angular variables $z, y_1 , y_2$ \p{ys},
\begin{align}
\mathbb{M}_{2 \to 3}^{(0)} & = \frac{8E^4 \Delta(z,y_1,y_2)}{z(1-z)y_1 y_2 (1-y_1)(1-y_2)} \frac{(1-z x)^4}{x^2 (1-x)^2  P(x) Q(x)} \,, \label{M5simpl}
\end{align}
where $\Delta(z,y_1,y_2)$ is the tetrahedron volume formed by the unit vectors $\vec n, \vec n_1 ,\vec n_2$,
\begin{align}\label{A}
\Delta \equiv  {\rm Vol}(\vec n,\vec n_1,\vec n_2) = 1- (\vec n \vec n_1)^2 - (\vec n \vec n_2)^2 - (\vec n_1 \vec n_2)^2 + 2 (\vec n_1 \vec n_2) (\vec n \vec n_1)(\vec n \vec n_2) \,,
\end{align}
and $P(x;z,y_1,y_2)$ and $Q(x;z,y_1,y_2)$ are quadratic polynomials in $x$,
\begin{align}\notag
& P(x;z,y_1,y_2) \equiv (1-z)(1-y_1) + (z+y_1-2 z y_1-y_2)(1-x) + z y_1 (1-x)^2 \,,\\
&Q(x;z,y_1,y_2) \equiv (1-z)y_1 + (y_2-z-y_1+2z y_1)(1-x) + z (1-y_1) (1-x)^2\,.
\end{align}

The squared matrix element $\mathbb{M}_{2 \to 3}^{(0)}$ is needed for the calculation of the real corrections in the single- and double-calorimeter energy correlators, see \p{eq:ECreal} and \p{eq:EECreal}.  In Appendix~\ref{app:PSint} we provide an explicit parametrization of these phase space integrals. The real correction contains an IR pole $1/\ep$ coming from the emission of a soft graviton, which cancels  the IR pole \p{eq:virtpole} of the virtual correction. 

\subsection{Infrared-finite differential cross section}

To elucidate the cancellation of infrared divergences, we consider the real contribution to the differential cross section for two-to-two graviton scattering,
\begin{align}
d \sigma_{\rm real}(q_1) = \frac{1}{2!}\int\limits_{q_2,q_3} d\text{PS}_{3}(q_1,q_2,q_3) \ \mathbb{M}_{2 \to 3}^{(0)}(q_1,q_2,q_3)\, ,
\end{align}
where $y\equiv y_1$, $x$ is the energy fraction from $q_1 = x E(1,\vec n_1)$, and $\frac{1}{2!}$ is a symmetry factor.  We find (see the details in Appendix~\ref{app:PSint})
\begin{align}\label{321}
& d \sigma_{\rm real}(q_1) = -   (1-x)^{-\ep} R(x,y) \,d\Pi(q_1) \,, \\[2mm]
& R(x,y) \equiv \frac{x \log (x) + (1-x) \log (1-x)+x \left( y \log (y)+(1-y) \log (1-y) \right)}{x^2 y (1-x) (1-y) (1-x(1-y)) (1-x y)} + O(\ep) \,, \notag
\end{align}
where we employ the shorthand notation  
\be
d\Pi(q_1) \equiv 2^{2\ep} e^{-\ep \gamma_E} \pi^{-4+3\ep} E^{-2-2\ep} \delta_+(q_1^2) d^{4-2\epsilon} q_1 .
\ee
The singularity of $R(x,y)$ for $x \sim 0$ corresponds to hard collinear radiation. The soft radiation corresponds to the pole at $x \sim 1$,
\begin{align}
R(x,y) \sim {1 \over 1-x} {y \log y + (1-y) \log (1-y) \over y^2 (1-y)^2}\, .
\end{align}
Together with the prefactor in \re{321},  this pole gives rise to a singular distribution, 
\begin{align}\label{trad}
{1 \over (1-x)^{1+\ep}} = -{1 \over \epsilon} \delta(1-x)+{1 \over (1-x)_+}+O(\epsilon)\,.
\end{align}
The divergent part cancels against the IR pole of the virtual correction,
\begin{align}
d \sigma_{\rm virt}(q_1) = \int_{q_2} d\text{PS}_{2}(q_1,q_2) \ \mathbb{M}_{2 \to 2}^{(1)}(q_1,q_2)\,,
\end{align}
which we have calculated previously, see \p{M41loop}, 
\begin{align}\notag
& d \sigma_{\rm virt}(q_1) = \delta(1-x) V(x,y) \,d\Pi(q_1) \, , \\[2mm]
& V(x,y) \equiv - \frac{1}{\ep}\frac{y\log(y) + (1-y)\log(1-y)}{ y^2 (1-y)^2} + \frac{\log(y)\log(1-y)}{y^2 (1-y)^2} + O(\ep) \,.
\end{align}

\subsection{One-point energy correlator}

Let us first consider the one-point energy correlator. At the leading order, only the two-particle final state contributes,
so that, using \p{eq:M0} and \p{eq:EJC2}, we have
\begin{align}
\label{eq:onepointEC}
&\text{EC}^{(0)}(y) ={1 \over 2 \pi^2}{1 \over y^2(1-y)^2} \ .
\end{align}

At the next-to-leading order (NLO), the virtual correction is given by \p{eq:virtpole}. The real correction \p{eq:ECreal} requires integration over the phase space of the undetected particles. It is done using the van Neerven integrals \cite{vanNeerven:1985xr}, as explained in Appendix~\ref{app:PSint}. As expected, the IR divergences cancel out in the sum of real and virtual corrections and we get 
\begin{align}
\text{EC}^{(1)}(y){}&={1 \over 2 \pi^4} 
\Bigg[
\frac{2 \log(y) \log(1-y)}{y^{2}(1-y)^{2}}
+ \frac{2\pi^{2}}{3 y(1-y)}   \label{first corr} \\
{}&+ \frac{2(1-y)^{3}\,\text{Li}_{2}(y) - 2 y^{3}\,\text{Li}_{2}(1-y)}
{y^{2}(1-y)^{2}(1-2y)} + \frac{y\log^{2}(1-y) - (1-y)\log^{2}(y)}
{y(1-y)(1-2y)} \nonumber 
\Bigg] ,
\end{align}
where $y \equiv y_1$ is the angular variable between the beam axis and the calorimeter, see \p{ys}.

The following comments are in order about the properties of the energy correlator \re{first corr}. The expression inside the brackets in \re{first corr} is given by a linear combination of functions of uniform transcendental weight two, with rational coefficient functions.
In addition, it is symmetric under the exchange of the incoming particles, 
\begin{align}
\text{EC}(y) = \text{EC}(1-y)\,,
\end{align}
and is analytic in the complex $y$-plane, with the exception of the branch points  $y=0$ and $y=1$.  

It is interesting to consider the behavior of \re{first corr} in the limit $y\to0$ when the calorimeter is aligned with the beam of incoming particles. In this limit, the tree-level amplitude of $p_1+p_2\to q_1+q_2$ develops a $t-$channel pole  $1/(p_1q_1)=O(1/y)$ due to graviton exchange and, as a consequence, the energy correlator \eqref{eq:onepointEC} exhibits a double pole $\text{EC}^{(0)}(y)\sim 1/y^2$. The first correction to this behavior follows from   (\ref{first corr}) 
\begin{align}
\text{EC}^{(1)}(y) & = \frac{1}{2 \pi^4 y} \left( 2 + \frac{2\pi^{2}}{3} - 2\log y - \log^{2} y \right)
+ O(\log^2 y) \ . \label{eq:ECy0}
\end{align}
It is suppressed by the factor of $y$ as compared with $\text{EC}^{(0)}(y)$ and exhibits a double logarithmic behavior  
$\text{EC}^{(1)}(y)/\text{EC}^{(0)}(y)\sim y\log^2 y$. This behavior comes from both the real and virtual corrections to $\text{EC}^{(1)}(y)$. The real correction originates from integration over soft graviton momentum $q_3$ in the process $p_1+p_2\to q_1+q_2+q_3$ where $q_1=E_1(1,\vec n_1)$ and $q_2=E_2(1,-\vec n_1)$ are the momenta of energetic particles with $E_1\sim E_2\sim E$. As we discuss in Section~\ref{sec:backtoback}, the contribution of soft gravitons can be analyzed using the eikonal approximation and is universal. 
By contrast, the virtual correction is sensitive to the details of the gravitational theory.

\subsection{Two-point energy correlator}

Next we analyze the two-point energy correlator \re{eq:defeec}.
According to the discussion in Section~\ref{sec:contact}, to lowest order in the coupling, the energy correlator ${\rm EEC}^{(0)}$ is given by the sum of two contact terms \re{conterm} localized at the collinear  ($\vec n_1 = \vec n_2$) and back-to-back ($\vec n_1 = - \vec n_2$) configurations of the calorimeters. 

In terms of the angular variable $z$ defined in \p{ys}, these contact terms are given by
\begin{align}\label{omega}
\delta(\Omega_{\vec{n}_{1}}-\Omega_{\vec{n}_{2}}) = \frac{1}{4\pi} \delta(z) \,,\qquad
\delta(\Omega_{\vec{n}_{1}}+\Omega_{\vec{n}_{2}}) = \frac{1}{4\pi} \delta(1-z) \,,
\end{align}
and relation \re{conterm} takes the form
\begin{align}
\label{eq:treelevelEEC}
&{\rm EEC}^{(0)} ={1 \over 8 \pi^3  y_1^2(1-y_1)^2} \left( \delta(z)+\delta(1-z) \right) \,.
\end{align}
Here the angular variable $y_2$ takes the value $y_2=y_1$ or $y_2=1-y_1$ for the first and second delta function,  respectively.

At NLO, the energy correlator $\text{EEC}^{(1)}$ 
receives contributions from the virtual-particle exchange and from the real particle emission. As a result, it takes nonzero values for $0<z<1$ and exhibits  contact terms in the collinear and back-to-back configurations. %
It is convenient to split the resulting expression for $\text{EEC}^{(1)}$ into a sum of two contact terms localized at $z=0$ and $z=1$, as well as a regular term,
\begin{align}
{\rm EEC}^{(1)} = {\rm EEC}_{\text{coll}}(y_1,z) + {\rm EEC}_{\text{b-to-b}}(y_1,z)+ {\rm EEC}_{\text{reg}}(y_1,\beta,z)  \,. \label{eq:EEC1}
\end{align}
The crossing symmetries \p{eq:EECdiscrsym} of the two-point energy correlator take the following form in the angular variables, 
\begin{align}
{\rm EEC}^{(1)}(y_1,y_2,z)={\rm EEC}^{(1)}(y_2,y_1,z) = {\rm EEC}^{(1)}(1-y_1,1-y_2,z) \,. 
\end{align}
In what follows, we describe each term on the right-hand side of \re{eq:EEC1} separately. 

We recall that the contact term at $z=0$ describes the possibility for the particle to go through the two detectors aligned along the same vector $\vec n_2=\vec n_1$. The corresponding contribution to the energy correlator is given by the phase space integral \re{eq:defeec} weighted with the square of the energy of the detected particle. It is easy to see that this integral coincides with the one-point correlator $\text{EC}^{(1)}_{J=2}$ defined in \re{eq:defecJ}. This correlator can be obtained by repeating the calculation in the previous subsection. 
As in the previous case, the IR divergences  cancel between the real and virtual contributions to $\text{EC}^{(1)}_{J=2}$ and the result is  
\begin{align}
\label{eq:collEEC}
&{\rm EEC}_{\text{coll}} = {1 \over 4 \pi}\delta(z)\,  \text{EC}_{J=2}^{(1)}(y_1) \ , \\
&\text{EC}_{J=2}^{(1)}(y_1) ={1 \over 2 \pi^4}  \biggl[  \frac{\pi^2}{3 y_1^2 (1-y_1)^2}   \label{eq:E2C1L} \nonumber\\ 
& \qqqquad + \frac{2{\rm Li}_2(y_1)}{(1-y_1)^2 (1-2 y_1)} - \frac{2{\rm Li}_2(1-y_1)}{y_1^2 (1-2 y_1)} +\frac{\log^2(1-y_1)-\log^2(y_1)}{y_1 (1-y_1) (1-2 y_1)} \biggr] \ .  
\end{align}
Notice that  $\text{EC}_{J=2}^{(1)}(y_1)$ is very similar to the function  $\text{EC}^{(1)}(y_1)=\text{EC}_{J=1}^{(1)}(y_1)$ defined in \eqref{first corr}. Both functions have transcendental weight two, are symmetric under $y_1 \to 1-y_1$, and exhibit double-logarithmic $\log^2 y_1$ behavior, in the limit $y_1 \to 0$.

Let us now discuss the back-to-back limit $z \to 1$. In general, the three angles $z,y_1,y_2$ are independent. However, when the calorimeters are in the back-to-back configuration $z=1$, only one angular variable is required to specify their orientation with respect to the beam axis. To describe the vicinity of this degenerate configuration, we employ the coordinates $(z,y_1,\beta)$ where $\beta\in[0,\pi]$ is the angle formed by the unoriented planes $( \vec n_1, \vec n_2 )$ and $(\vec n , \vec n_1 )$, see \eqref{beta} for the precise definition. Expressed in terms of $(z,y_1,y_2)$, the angle $\beta$ becomes
\be
\label{eq:rely1y2beta}
\cos \beta =
\frac{y_1 - y_2 + z - 2 y_1 z}
{2 \sqrt{y_1(1 - y_1)} \sqrt{z(1 - z)}} \ .
\ee

In the limit  $z\to 1$ the set of coordinates $(z,\beta)$ is degenerate, in the sense that $\beta$ becomes arbitrary. This is similar to the polar coordinates $(r,\beta)$ on a plane: the polar angle $\beta$ becomes arbitrary when the radius $r\to0$. The   squared amplitude $F(z,\beta)$ is treated as  a distribution on the $(2-2\ep)-$dimensional unit sphere, acting on smooth test functions $\vp(z,\beta)$. If $F(z,\beta)$ has a pole $\sim f(\beta)/(1-z)$, the interplay with  the measure yields a contact term $\sim \delta(1-z)$. On its support the test functions $\vp(z=1,\beta) \equiv \vp(1)$ become constants, due to the degeneracy. Then the integration over $\beta$ results in the average of the function $f(\beta)$  (see Appendix~\ref{Gelfand} for a detailed explanation).

We now apply this treatment to the $\text{EEC}$ in the back-to-back limit $z \to 1$. To reveal the presence of the contact term, we need the leading asymptotics of the  real contribution. Analyzing  the  squared  amplitude, integrated over the three-particle phase space in $d=4-2\epsilon$ dimensions, see \eqref{eq:EECintx}, and using the method of regions \cite{Jantzen:2012mw}, we find 
\begin{align}
{\rm EEC}_{\text{real}} = \frac{1}{4\pi^5} \left( \frac{2\pi}{E}\right)^{4\ep}  \frac{f(y_1,\beta)}{1-z}  + O(1/\sqrt{1-z})\, .\label{EEC3ptGRz1f} 
\end{align}
Here we introduced the short-hand notation for the residue at $z=1$,
\begin{align}
f(y_1,\beta) \equiv \frac{(1-2y_1) \log\left(\frac{1-y_1}{y_1}\right)\sin^2\beta +\sin(2\beta) ({\pi \over 2}-\beta) }{y_1(1-y_1)(1- y_1 (1-y_1) 4 \sin^2 \beta)} \, , \label{fdef}
\end{align}
and  $0\le \beta \le \pi$. The pole at $z=1$ is treated in the  distributional sense described above. This results in  the following distribution on the sphere $S^{2-2\epsilon}$ (neglecting the ${\cal O}(\ep)$ remainders), 
\begin{align}\label{contact}
& {f(y_1,\beta) \over 1-z} = \delta(1-z) \biggl[  -{1\over \epsilon} (4\pi e^{\gamma_E})^{-\ep}  \int_0^{\pi} {d\beta\over\pi} f(y_1,\beta) +2 \int_0^{\pi} {d\beta\over\pi} \log(2\sin\beta) f(y_1,\beta) \biggr]+ {f(y_1,\beta) \over (1-z)_+},
\end{align}
where $\gamma_{E}$ is the Euler constant.
Notice the presence of an IR {divergence} in the contact term $\delta(1-z)$. 
Averaging $f(y_1,\beta)$ over $\beta$, we find
\begin{align}
& \int_0^{\pi} {d \beta\over\pi} \, f(y_1,\beta) = -\frac{y_1 \log(y_1) + (1-y_1) \log(1-y_1)}{2 y_1^2 (1-y_1)^2} \,, 
\label{averfbeta0} \\
& \int_0^{\pi} {d \beta\over\pi} \, \log(2\sin \beta ) f(y_1,\beta) = \frac{\log(y_1) \log(1-y_1)}{4y_1^2(1-y_1)^2} \,.
\label{averfbeta}
\end{align}

We now demonstrate that the infrared divergence in \re{EEC3ptGRz1f} cancels against the corresponding divergence in the virtual contribution. Recall that the $2\to2$ scattering amplitude contributes to both the collinear and the back-to-back contact terms in the EEC. The back-to-back contact term from the virtual correction to the $2\to2$ scattering amplitude  takes the form
\begin{align}
\text{EEC}_{\text{virt}} =  {1 \over 4\pi} \delta(1-z)  \,   \text{EC}_{J=2}^{\text{virt}} \, , \label{eq:virtEEC1}
\end{align}
where $\text{EC}_{J=2}^{\text{virt}}$ is given in \eqref{eq:virtpole}. 
It follows that the IR-divergent part of \eqref{EEC3ptGRz1f}, when combined with \eqref{contact} and \eqref{averfbeta0}, cancels precisely against the divergence in \eqref{eq:virtEEC1}.

{The finite contribution to the contact term $\delta(1-z)$ from the real emission} has the same functional form as the virtual contribution. In this way, we find the finite back-to-back contact term at one loop
\begin{align}
\label{eq:anticollEEC}
{\rm EEC}_{\text{b-to-b}} = {1 \over 4\pi} \delta(1-z)  \frac{\log(y_1) \log(1-y_1)}{\pi^{4} y_1^2(1-y_1)^2} \,.
\end{align}

In order to find the regular contribution to the EEC, it is sufficient to calculate the real correction at $d=4$, see also the discussion around \p{eq:EECreal}. The corresponding phase space integral \p{eq:EECreal} is reduced to the univariate integral \p{eq:EECintx}, and the relevant tree-level squared matrix element $\mathbb{M}_{2 \to 3}^{(0)}$ is given in \p{M5simpl}. The calculation is very similar to the one in \cite{Herrmann:2024yai}.\footnote{The same result was also reported in \cite{MuratKologluSeminar2024}. See also \cite{Chen:2025rjc} for an analogous QCD calculation.} 
Performing the phase space integration we obtain an infrared finite result for $0<z,y_1,y_2<1$. In the back-to-back region, the real contribution has a pole $1/(1-z)$. Replacing the pole by the $1/(1-z)_{+}$ distribution according to \p{fdef}, we find the
regular contribution in the EEC,
\begin{align}
{\rm EEC}_{\text{reg}} &(y_1,\beta,z) = \frac{1}{32\pi^5} \frac{1}{(1-z)_{+}}\frac{\sqrt{\Delta}}{z  y_1 (1-y_1) y_2 (1-y_2)(\Delta-4z(1-z))}    \label{eecgr3pt} \\[2mm]
& \times\biggl[  (4z(1-z)-\Delta) \arctan\left( \frac{1-2z-(1-2y_1)(1-2y_2)}{\sqrt{\Delta}}\right) \notag\\[2mm]
& -\sqrt{\Delta} (y_1-y_2) \log \left( \frac{y_1 (1-y_2)}{y_2(1-y_1)}\right) +\frac{\pi}{2} \left( \Delta-4z(1-z)\right) -2\pi z |1-y_1-y_2|  \notag\\
& - 4z(1-y_1-y_2)\arctan\left( \frac{2(1-y_1-y_2)^2 - (1-z)(1+(1-2y_1)(1-2y_2))}{\sqrt{\Delta}(1-y_1-y_2)} \right)   \Biggr],   \notag
\end{align}
where $\Delta \equiv \Delta(z,y_1,y_2) > 0$,
\begin{align}
\Delta = 16 z(1-z) y_1 (1-y_1)\sin^2 \beta     
\end{align}
is the tetrahedron volume function  defined in \eqref{A} and the function $y_2=y_2(y_1,\beta,z)$ satisfies \eqref{eq:rely1y2beta}.

As already observed in \cite{Herrmann:2024yai} in a similar context, the result \eqref{eecgr3pt} is regular in the collinear limit $z \to 0$,
\begin{align} \label{eq:EECregz0}
\text{EEC}_{\text{reg}} = \frac{1}{4\pi^5} \frac{\sin^2 \beta}{y_1^2(1-y_1)^2} + O(\sqrt{z}) \,.
\end{align}
In the back-to-back limit $z\to 1$ we have instead
\begin{align}\label{b2b-limit}    \text{EEC}_{\text{reg}} = \frac{1}{4\pi^5}   \frac{f(y_1,\beta)}{1-z}\lr{1 + O(\sqrt{1-z})}\,.
\end{align}

 Summarizing, the two-point energy correlator at NLO is given by the sum \p{eq:EEC1} of three infrared-finite contributions: the regular term \p{eecgr3pt}, the collinear contact term \p{eq:E2C1L}, and the back-to-back contact term \p{eq:anticollEEC}. Both contact terms receive contributions from real emissions and virtual corrections. While each individual contribution is infrared divergent, these divergences cancel precisely in the total sum, yielding an infrared-finite result. 

 We would like to emphasize that the NLO result \p{eq:EEC1} is incomplete without the contact terms. These terms are essential to ensure that ${\rm EEC}^{(1)}$ satisfies the Ward identities associated with energy and momentum conservation, a property that we verify explicitly in the next subsection.

\subsection{Sum rules}

As a powerful check of our results, we can verify the sum rules (Ward identities) \p{eq:energy} and \p{eq:momentum}. Integrating over the sphere, we prefer to parametrize it in terms of $z$ and the angle  $\beta$ defined in  \p{beta}, so the measure is  $d\Omega_{\vec{n}_2} = 4d\beta dz$. One can immediately see that the leading-order energy correlators \p{eq:onepointEC} and \p{eq:treelevelEEC} satisfy these identities,
\begin{align}\notag
{}& 2\pi\int^1_0 dz \, \text{EEC}^{(0)}(y_1,z) = \text{EC}^{(0)}(y_1) \,, 
\\
{}& \int^1_0 dz \,(1-2z)\text{EEC}^{(0)}(y_1,z) = 0 \, ,
\end{align}
where the integration over $\beta$ is trivial.

At NLO, the Ward identities \p{eq:energy} and \p{eq:momentum} lead to non-trivial relations between the one-loop functions \p{first corr}, \p{eq:E2C1L}, \p{eq:anticollEEC}, and \p{eecgr3pt}. Let us introduce the coefficient functions of the contact terms,
\begin{align}\label{}
{\rm EEC}_{\text{coll}}(y_1,z) =  \frac{1}{\pi} a_{\text{coll}}(y_1)\, \delta(z)  \,, \qquad {\rm EEC}_{\text{b-to-b}}(y_1,z) = \frac{1}{\pi} a_{\text{b-to-b}}(y_1)\, \delta(1-z)  \,.
\end{align}
The two Ward identities can be recast in the equivalent form
\begin{align}
& \int\limits_0^1 dz  \, (1-z) \int\limits_0^{\pi} d \beta\, {\rm EEC}_{\text{reg}}(y_1,y_2(\beta),z)=\frac1{4}{\rm EC}^
{(1)}(y_1) - a_{\text{coll}}(y_1)  \,,           \label{wi3}\\
&  \int\limits_0^1 dz  \, z \int\limits_0^{\pi} d \beta\, {\rm EEC}_{\text{reg}}(y_1,y_2(\beta),z)       =\frac1{4}{\rm EC}^
{(1)}(y_1) -  a_{\text{b-to-b}}(y_1) \,. \label{wi4}
\end{align}
Note that the weight $(1-z)$ in \re{wi3} cancels the distribution $1/(1-z)_+$ in ${\rm EEC}_{\text{reg}}$ from \p{eecgr3pt}. This is natural because \re{wi3} probes the regime $z\sim0$, i.e. away from $z\sim 1$. 

We conclude that, in principle, the Ward identities  determine unambiguously the two contact-term coefficient functions  in terms of EC and two moments of EEC. This statement holds to any order in the coupling. In particular, the Ward identities predict the existence of the Born-level contact terms \p{eq:treelevelEEC} without any further analysis of the singular distributions. 
In practice however the double integrals on the left-hand sides of \p{wi3} and \p{wi4} may be difficult to work out. At NLO, we have checked both relations numerically.

\subsection{Averaging over the beam}

In order to simplify the expression for the EEC above, it is instructive to average over the beam direction $\vec n$. In this way we get a function which only depends on the relative angle between the two calorimeters on the sphere,
\begin{align}
\label{eq:avEECdef}
\overline{\rm EEC}^{(1)}(z) \equiv \frac{1}{4\pi}\int d\Omega_{\vec n} \, {\rm EEC}_{\rm reg} \,.
\end{align}
This averaging can be performed for $0<z<1$. 
In terms of the calorimeter variables we have $d\Omega_{\vec n} = 4 d\beta d y_1$. 

The leading-order EEC is given by the sum of contact terms \p{eq:treelevelEEC}. 
As a result, both the ${\rm EEC}^{(0)}$ and its angular average vanish for $0<z<1$. Integrating \eqref{eecgr3pt} we obtain the following NLO result,
\begin{align}
\overline{\rm EEC}^{(1)}_{\text{SG}}(z) 
= &  \frac{1}{2 \pi^5} \frac{(1+u^2)^2}{u^2}\biggl( \frac{\pi^2}{3} - 2 u \log(u) \arctan(u) - (1+i u) {\rm Li}_2(i u) - (1-i u) {\rm Li}_2(-i u) \biggr) ,\label{EECav}
\end{align}
where we find it convenient to introduce the variable, cf. \p{ys},
\begin{align} \label{eq:defu}
u \equiv \sqrt{\frac{z}{1-z}} = \tan \frac{\theta}{2}>0 \,,
\end{align} 
which parametrizes the angle between the two calorimeters.

The averaged EEC has the following behavior for $z \to 0$, 
\begin{align} 
\overline{\rm EEC}_{\rm SG}^{(1)}  = & \frac{1}{2 \pi^5} \left[ \frac{\pi^2}{3z} + \left( -\log(z) +\frac{5}{2} + \frac{\pi^2}{3}\right) + O(z)  \right] \,, \label{EECbarSG0} 
\end{align}
and $z \to 1$,
\begin{align}
\overline{\rm EEC}_{\rm SG}^{(1)} = & \frac{1}{2 \pi^5} \left[ \frac{1}{1-z} \left( \frac14 \log^2(1-z) - \log(1-z) + 2 +\frac{5\pi^2}{12} \right) \right. \notag\\
& \left. + \frac14 \log^2(1-z) -\frac16 \log(1-z) + \frac{5}{18} + \frac{5\pi^2}{12} + O(1-z)  \right]  , \label{EECbar}
\end{align}
where all the $\sqrt{z}$ and $\sqrt{1-z}$ terms disappear in the expansions. 
Notice that the averaged EEC is non-integrable as $z \to 0$. We expect that this behavior is not a perturbative artifact, but rather a consequence of the fact that the total cross section for scattering of plane waves is infinite in $4d$ gravitational theory.

\subsection{Generalized energy correlators}

The calculations of the energy correlators presented above can be straightforwardly generalized to arbitrary powers of the energy weight, see \re{eq:defecJ}. We begin by considering the one-point correlator $\text{EC}_{J}(y)$.

At the leading order, it follows from \p{eq:ECJpert} and  \p{eq:EJC2} that the dimensionless function $\text{EC}_{J}^{(0)}(y)$ is independent of $J$ and coincides with \eqref{eq:onepointEC},
\begin{align}\label{EC-B1}
&\text{EC}_{J}^{(0)}(y) ={ 1 \over 2 \pi^2}{1 \over y^2(1-y)^2} \,.
\end{align} 
At NLO, the function $\text{EC}_{J=2}^{(1)}(y)$ already appeared in the computation of the collinear contact term of the EEC, see \p{eq:collEEC}.  
Upon computing the NLO correction $\text{EC}_{J}^{(1)}(y)$, we observe the cancellation of IR divergences between the real-emission contribution \p{eq:ECreal} and the virtual contribution \p{eq:virtpole} (see also \p{eq:EJC2virt}). As a result, we obtain an analytic expression for $\text{EC}_{J}^{(1)}(y)$ that is valid for arbitrary integer $J \geq 1$.

For convenience, we collect our results for $\text{EC}_{J}^{(1)}(y)$ into a compact generating function, presented in Appendix~\ref{app:EJC}, see \eqref{eq:GenFunEJC}. 
The functions $\text{EC}_{J}^{(1)}(y)$ are crossing symmetric, and their asymptotic behavior as $y \to 0$ is analogous to that of \p{eq:ECy0}, see \p{eq:EJCasypmy0}. 
Moreover, these functions have uniform transcendental weight two for $J \leq 3$ (see \p{first corr} and \p{eq:E2C1L} for the explicit results for $J=1$ and $J=2$). 
By contrast, for $J \geq 4$ the NLO corrections $\text{EC}_{J}^{(1)}(y)$ contain contributions of lower transcendental weight.

In Figure~\ref{fig:EJC} we display the NLO energy correlators $\text{EC}_{J}^{(1)}(y)$ for several values of $J$. 
For large $J$, they take a simple asymptotic form,
\begin{align}
& \text{EC}_{J}^{(1)} (y) \underset{J\to \infty}{\sim}  \frac{1}{\pi^4} \log(J) \, \frac{y\log(y)+(1-y)\log(1-y)}{y^2 (1-y)^2} \,. \label{eq:EJClargeJ}
\end{align}
This asymptotic behavior is governed by the emission of an arbitrary number of soft gravitons. Their resummation to all loops is performed in Section~\ref{sect:large-J}, resulting in the finite-coupling counterpart \p{eq:large-J-all-loops} of the one-loop asymptotics \p{eq:EJClargeJ}.

Analogous to the one-point energy correlator, the leading-order two-point energy correlator $\text{EEC}^{(0)}_{J_1,J_2}$  (see \p{eq:EECJ1J2pert} and \p{eq:EEC_LO}) is independent of the spins $J_i$ and coincides with \p{eq:treelevelEEC}. 
At NLO, the function $\text{EEC}^{(1)}_{J_1,J_2}$ is given by the sum of three terms, in close analogy with \re{eq:EEC1}
\begin{align}
\text{EEC}^{(1)}_{J_1,J_2} = \text{EEC}^{\text{reg}}_{J_1,J_2} +  {\rm EEC}_{J_1,J_2}^{\text{coll}} + {\rm EEC}_{J_1,J_2}^{\text{b-to-b}}\,.
\end{align}
As before, we observe the cancellation of the infrared divergences in the collinear and back-to-back contact terms.

The collinear contact term is proportional to the one-point energy correlator, cf. \p{eq:collEEC},
\begin{align}
{\rm EEC}_{J_1,J_2}^{\text{coll}} & = {1 \over 4 \pi}\delta(z)\,  \text{EC}_{J_1+J_2}^{(1)}(y_1)  \,.  
\end{align}
The back-to-back contact term receives contributions from the virtual correction \p{eq:virtEEC1}, \p{eq:virtpole} and the real emission  \p{contact}. Both expressions are the same as in the case $J_1 = J_2=1$, so their sum is identical to \p{eq:anticollEEC},
\begin{align}
{\rm EEC}_{J_1,J_2}^{\text{b-to-b}} & =  {1 \over 4\pi} \delta(1-z) \frac{\log(y_1) \log(1-y_1)}{\pi^{4} y_1^2(1-y_1)^2} \,.    
\end{align}
Finally, the regular part $\text{EEC}^{\text{reg}}_{J_1,J_2}$ is given by the univariate integral of the real-emission contribution \p{eq:EECintx}, with the back-to-back pole $1/(1-z)$ replaced by the distribution $1/(1-z)_{+}$.

\section{Energy correlators in pure gravity}
\label{sec:puregravity}

The calculations of energy correlators in the previous section are readily extended to pure gravity. The one-loop counterterms in this theory do not contribute to the on-shell matrix elements \cite{tHooft:1974toh}, and the NLO calculations are free of UV divergences. The pure gravity results exhibit similar analytic behavior and belong to the same function space as their supergravity counterparts. However, the analytic expressions in pure gravity are more cumbersome and involve lengthy polynomials.  
In particular, they are not of  homogeneous transcendentality. 

In addition, the helicities of the initial states do play a role in the pure gravity case. Indeed, the squared amplitudes averaged over the final states depend on the initial-state helicity configuration. In what follows, we consider polarized event shapes and explicitly specify the initial-state helicities, e.g.
\begin{align}
\text{EC}_{J}^{++} \,,\qquad \text{EEC}_{J_1,J_2}^{+-}\,,
\end{align} 
where the superscripts $+$ and $-$ refer, respectively, to the helicities $+2$ and $-2$ of the gravitons in the initial state. Due to parity conjugation, it suffices to consider the initial states $++$ and $+-$, consisting of two gravitons of the same or opposite helicity, respectively, 
\begin{align}
\text{EC}_J^{--} = \text{EC}_J^{++} \,, \qquad
\text{EC}_J^{-+} = \text{EC}_J^{+-} \,. 
\end{align}

In what follows, we briefly recall the expressions for the relevant amplitudes. We then report all pure gravity counterparts of the supergravity observables considered in the previous section: the NLO corrections to the EC for arbitrary weight $J \geq 1$, the EEC including the contact terms, and their beam-direction average. The explicit expressions are collected in the ancillary file.

\subsection{Amplitudes and squared matrix elements}

The tree-level amplitudes in pure gravity are the same as the graviton helicity components of the supergravity amplitudes in \eqref{eq:amplSG}. The squared matrix elements, averaged over the helicities of the final-state gravitons and evaluated for different helicity configurations of the initial-state gravitons, are
\begin{align}
& \mathbb{M}^{(0)}_{++\boldsymbol{\to} 2} = \frac{s_{12}^6}{s_{13}^2 s_{23}^2}  \,, \qquad
\mathbb{M}^{(0)}_{+- \boldsymbol{\to}\, 2} = \frac{s_{13}^8 + s_{23}^8}{s_{12}^2 s_{13}^2 s_{23}^2}  \,,\notag \\
& \mathbb{M}^{(0)}_{++\boldsymbol{\to}\, 3} =(s_{12}^8 + s_{34}^8 + s_{35}^8 + s_{45}^8) \, |{M}_5^{(0)}|^2 \,,\notag \\[2mm]
&  \mathbb{M}^{(0)}_{+- \boldsymbol{\to} 3} =(s_{13}^8 + s_{14}^8 + s_{15}^8 + s_{23}^8 + s_{24}^8+ s_{25}^8) \,|{M}_5^{(0)}|^2 \label{eq:MM++and+-}\,,
\end{align}
where the five-particle MHV function ${M}_5^{(0)}$ is given in \eqref{eq:Mtree5}. The remaining initial-state helicity configurations are obtained by parity conjugation.

In addition to the MHV case, the one-loop two-to-two pure gravity amplitudes are also nontrivial for the all-plus and the single-minus helicity configurations. Nevertheless, since these helicity amplitudes vanish at tree level, they do not contribute to the NLO corrections to the event shapes, see \p{eq:MM1loop}. Thus, only the MHV helicity configurations are required \cite{Dunbar:1994bn}, 
\begin{align}
&M^{(1)}(1^+ 2^-3^- 4^+) = 
\frac{i}{(4\pi)^2} 
\left(\vev{23}[14]\right)^4 \, \biggl\{ (4\pi)^2 \left( I(s_{12},s_{13}) + I(s_{12},s_{23}) + I(s_{13},s_{23}) \right) \notag\\
& + \frac{1}{s_{23}^8} \left( 4 s_{12}^6 + 14 s_{12}^5 s_{13} + 28 s_{12}^4 s_{13}^2 + 35 s_{12}^3 s_{13}^3 + 28 s_{12}^2 s_{13}^4 + 14 s_{12} s_{13}^5 + 4 s_{12}^6 \right) \left( \log^2\left(\frac{s_{12}}{s_{13}}\right) + \pi^2\right) \notag \\ 
& + \frac{(s_{12}-s_{13})}{30 s_{23}^7} \left( 261 s_{12}^4 + 809 s_{12}^3 s_{13} + 1126 s_{12}^2 s_{13}^2 + 809 s_{12} s_{13}^3 + 261 s_{13}^4 \right) \log\left(\frac{s_{12}}{s_{13}}\right) \notag\\
&  +  \frac{1}{180 s_{23}^6} \left( 1682 s_{12}^4 + 5303 s_{12}^3 s_{13} + 7422 s_{12}^2 s_{13}^2 + 5303 s_{12} s_{13}^3 + 1682 s_{13}^4 \right)
\biggr\}  + O(\ep) \label{eq:M1loopGrav} \,,
\end{align}
where $I(s,t)$ is the one-loop zero-mass box integral \eqref{eq:zmbox}. The previous expression has to be analytically continued from the Euclidean region (where we consider $s_{12},s_{13},s_{23}<0$ as independent variables) to the physical channel; this is accomplished by $\log(-s) \to \log(s)-i\pi$. Other helicity configurations are obtained by permuting the labels of the gravitons in Eq.~\eqref{eq:M1loopGrav}. The pure gravity amplitude differs from the supergravity expression \eqref{eq:M1loop} by the last three lines in \eqref{eq:M1loopGrav}, which involve terms of subleading transcendentality.
Using these expressions, we calculate, according to \p{eq:MM1loop}, the NLO squared matrix elements for the two-particle contribution $\mathbb{M}_{+\pm \boldsymbol{\to} 2}^{(1)}$.

\subsection{One-point energy correlators}

Using the tree-level two-to-two squared amplitudes and rewriting them in terms of the calorimeter variable  $y \equiv y_1$, we immediately find the LO energy correlators in pure gravity,    
\begin{align}\label{EC-B2}\notag
{}& \text{EC}^{(0)}_{++}(y) ={1 \over 2 \pi^2}{1 \over y^2(1-y)^2} \,, 
\\
{}& \text{EC}^{(0)}_{+-}(y) ={1\over 2 \pi^2}\frac{(y^8 + (1-y)^8)}{y^2(1-y)^2} \,.
\end{align}

\begin{figure}[t]
\begin{center}
\includegraphics[height=6.4cm]{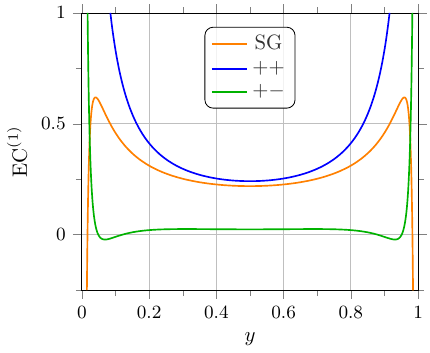}\quad
\includegraphics[height=6.4cm]{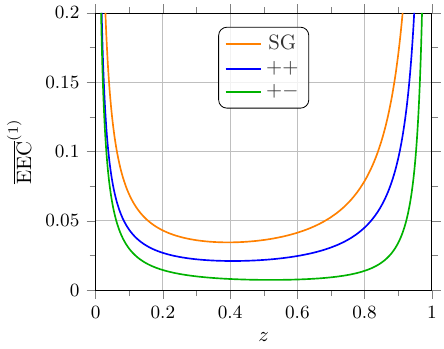}
\end{center}
\caption{Comparison of energy correlators in maximal supergravity and in pure gravity, for two-graviton pure initial states of polarizations $(+,+)$ and $(+,-)$. The angle variables are defined in \p{ys} and $y \equiv y_1$. Left: the NLO correction to the EC. The functions are symmetric under $y \to 1 - y$ and exhibit different asymptotic behavior near the endpoints. Note that the presence of negative values does not contradict the positivity of the EC, see \re{eq:ECpertexp}.  
Right: the NLO correction  to the beam-averaged EEC, see \p{eq:avEECdef}. All three functions 
take positive values for $0<z<1$,  grow as $O(1/z)$ at small $z$ and display a universal behavior for $z\to 1$, see \re{EECbar} and \re{b2b-gr}.}
\label{fig:EC}
\end{figure}

\begin{figure}

\begin{center}   
\includegraphics[width=0.7\textwidth]{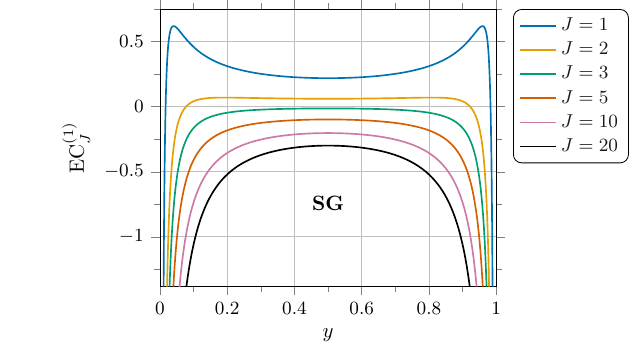}
\end{center}

\begin{center}   
\includegraphics[width=0.7\textwidth]{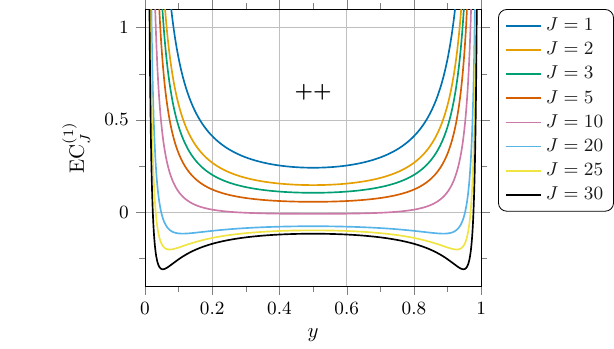}
\end{center}

\begin{center}   
\includegraphics[width=0.7\textwidth]{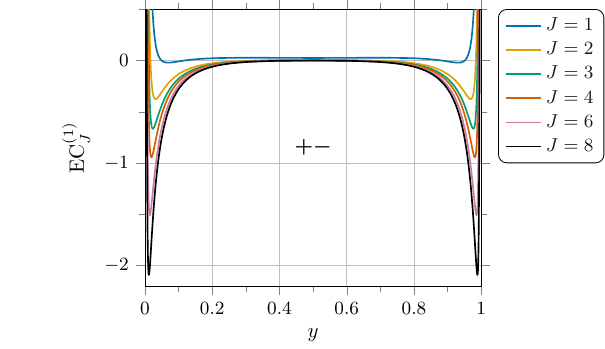}
\end{center}

\caption{NLO energy correlators $\text{EC}_{J}^{(1)}$ of weight $J$ in supergravity \p{eq:GenFunEJC} and in pure gravity, for polarized initial states, as functions of $y \equiv y_1$ \p{ys}. They approach the asymptotics \p{eq:EJClargeJ}, \p{eq:EC+-largeJ} at large $J$.}
\label{fig:EJC}
\end{figure}

Integrating $\mathbb{M}^{(1)}_{2 \to 2}$ and $\mathbb{M}^{(0)}_{2 \to 3}$ over the two- and three-particle phase spaces, we find the NLO virtual and real contributions to the EC. We confirm the cancellation of the IR poles. In Figure~\ref{fig:EC}, we plot the EC in pure gravity and supergravity. The $\text{EC}_{J}^{+\pm}$ with $J \geq 1$ are combined in generating functions, which are the analogs of the supergravity expression \eqref{eq:GenFunEJC}. The explicit results for the EC are provided in the ancillary file. We plot several higher-weight NLO EC in Figure~\ref{fig:EJC}.

For $J=1$, the EC have the following collinear calorimeter-beam limit, cf. the supergravity asymptotics \eqref{eq:ECy0},
\begin{align}
& \text{EC}_{++}^{(1)}(y) = \frac{1}{2 \pi^4} \frac{1}{y} \left( \frac{6821}{450} + \frac{47}{20}\log(y) + \log^{2}(y) \right)
+ O(\log^2(y)) \,,\notag\\
& \text{EC}_{+-}^{(1)}(y) = \frac{1}{2 \pi^4} \frac{1}{y} \left( \frac{6821}{450} -2\pi^2 + \frac{47}{20}\log(y) + \log^{2}(y) \right)
+ O(\log^2(y)) \,.
\end{align}
For large $J$, they behave like  their supergravity counterparts \eqref{eq:EJClargeJ},
\begin{align}
& \text{EC}^{(1)}_{J,++} \underset{J\to \infty}{\sim}  \frac{1}{\pi^4} \log(J) \, \frac{1}{y^2 (1-y)^2} (y\log(y)+(1-y)\log(1-y))\,,\notag\\
& \text{EC}^{(1)}_{J,+-} \underset{J\to \infty}{\sim}  \frac{1}{\pi^4}  \log(J) \,  \frac{(y^8 + (1-y)^8)}{y^2 (1-y)^2} (y\log(y)+(1-y)\log(1-y))\,. \label{eq:EC+-largeJ}
\end{align}
Note that the coefficients of $\log J$ in both relations are proportional to the product of the tree-level amplitudes \re{EC-B2} and the function $y\log y+(1-y)\log(1-y)$. The same structure appears in \re{eq:EJClargeJ}. The underlying reason for this is explained in Section~\ref{sect:large-J}.

\subsection{Two-point energy correlators}

Here we calculate $\text{EEC}^{(1)}_{+\pm}$ at NLO, including the regular part and the collinear and back-to-back contact terms, and obtain the analogs of Eqs.~\eqref{eecgr3pt}, \eqref{eq:E2C1L}, \eqref{eq:anticollEEC} in pure gravity. The back-to-back contact term $\delta(1-z)$ receives a virtual correction determined by the one-loop graviton amplitude \eqref{eq:M1loopGrav}, and a real correction specified by the $z \to 1$ asymptotics of the squared amplitudes $\mathbb{M}^{(0)}_{+\pm \to 3}$ \eqref{eq:MM++and+-} integrated over the three-graviton phase space. We find that the latter is analogous to the supergravity asymptotics \eqref{EEC3ptGRz1f}, since the corresponding squared amplitudes differ only by a simple prefactor,
\begin{align}
{\rm EEC}^{\rm real}_{+\pm} = \frac{1}{4\pi^5} \left( \frac{2\pi}{E}\right)^{4\ep}  \frac{f(y_1,\beta)}{1-z} \left\{ \begin{array}{cc}
  1   &  \text{,\quad for $++$}\\
  y_1^8 + (1-y_1)^8   & \text{,\quad for $+-$} 
\end{array} \right. + O(1/\sqrt{1-z}) \,,
\end{align}
where $f(y_1,\beta)$ is defined in \eqref{fdef}. Converting $\frac{1}{1-z}$ into a  distribution according to \eqref{contact}, we find that the IR poles cancel out in the sum of the real and virtual corrections. The finite contribution to the back-to-back contact term from the real correction is
\begin{align}
{1 \over 4\pi} \delta(1-z)  \frac{\log(y_1) \log(1-y_1)}{2 \pi^{4} y_1^2(1-y_1)^2} \left\{ \begin{array}{cc}
  1   &  \text{,\quad for $++$}\\
  y_1^8 + (1-y_1)^8   & \text{,\quad for $+-$} 
\end{array} \right.  \ ,
\end{align}
which is supplemented by the finite contribution from the virtual correction. We also check that the  pure-gravity $\text{EEC}^{(1)}_{+\pm}$ satisfy the Ward identities for energy and momentum conservation, see \eqref{wi3}, \eqref{wi4}. As in the supergravity case, the real contribution is regular in the collinear limit $z \to 0$, cf. \p{eq:EECregz0},
\begin{align}
{\rm EEC}^{\rm reg}_{+ \pm} = \frac{11}{72\pi^5}  \frac{\sin^2\beta}{y_1^2 (1-y_1)^2} \left\{ \begin{array}{cc}
  1   &  \text{,\quad for $++$}\\
  y_1^8 + (1-y_1)^8   & \text{,\quad for $+-$} 
\end{array} \right. 
+ O(\sqrt{z}) 
\,.
\end{align}
In the back-to-back limit we get
\begin{align}
{\rm EEC}^{\rm reg}_{+\pm} = \frac{1}{4\pi^5} \frac{f(y_1,\beta)}{(1-z)_+} \left\{ \begin{array}{cc}
  1   &  \text{,\quad for $++$}\\
  y_1^8 + (1-y_1)^8   & \text{,\quad for $+-$} 
\end{array} \right. + O(1/\sqrt{1-z}) \, .
\end{align}
Note that the limiting behavior is very similar to the one we obtained in supergravity \eqref{b2b-limit}. We will elucidate the reason for this similarity in Section \ref{sec:backtoback}.

Finally, we consider the energy correlators \p{eq:avEECdef} averaged over the beam direction, $\overline{\text{EEC}}^{(1)}_{+\pm}$, which are the analogs of \eqref{EECav}. They have the following collinear ($z\to 0$) asymptotics 
\begin{align}\label{EECbar-z}
& \overline{\rm EEC}^{(1)}_{+\pm}(z) = \frac{1}{2\pi^5} \frac{1}{z}\left( -\frac{419017}{352800} + \frac{\pi^2}{3}\right) + O\left(\log(z)\right)\,, 
\end{align}
and back-to-back ($z \to 1$) asymptotics,
\begin{align}
& \overline{\rm EEC}^{(1)}_{++}(z) = \frac{1}{2\pi^5}  \frac{1}{1-z} \left( \frac14 \log^2(1-z) - \log(1-z) + \frac{11237}{1800} -\frac{11\pi^2}{12} \right) + O\left(\log^2(1-z)\right) \,, \notag\\
& \overline{\rm EEC}^{(1)}_{+-}(z) = \frac{1}{2\pi^5}  \frac{1}{1-z} \left( \frac14 \log^2(1-z) - \log(1-z) + \frac{10037}{1200} -\frac{19\pi^2}{12} \right) + O\left(1/\sqrt{1-z}\right) \,. \label{b2b-gr}
\end{align}
Let us note that, compared to the supergravity case \re{EECbar}, the square roots $\sqrt{1-z}$ do not cancel out in \re{b2b-gr}. 

\begin{figure}[t]
\begin{center}   
\includegraphics[width=0.49\textwidth]{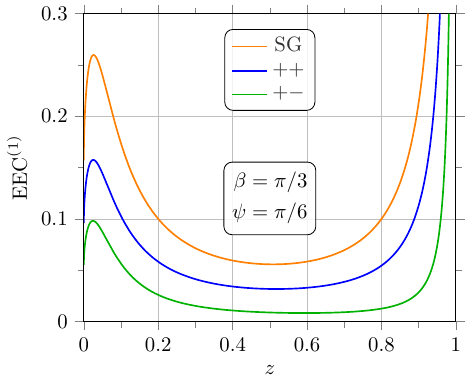}\quad
\includegraphics[width=0.48\textwidth]{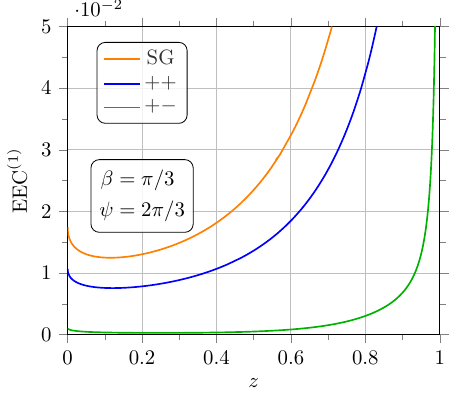}
\end{center}
\caption{EEC with $J_1=J_2=1$ in supergravity \p{eecgr3pt} and in pure gravity. The angle variables $z,\beta$ are defined in \p{ys}, \p{beta}, and $\cos\psi = 1-2y_1$. The functions display similar behavior: they remain positive throughout the interval $0<z<1$, tend to a finite value at $z=0$, and grow like $1/(1-z)$ as $z\to 1$. }
\label{fig:EEC}
\end{figure}

We plot the functions $\overline{\rm EEC}^{(1)}_{+\pm}(z)$ and $\overline{\rm EEC}^{(1)}_{\rm SG}(z)$ in Figure~\ref{fig:EC}, and the regular part of the energy correlators, ${\rm EEC}^{\rm reg}_{+\pm}(z)$ and ${\rm EEC}^{\rm reg}_{\rm SG }(z)$, in Figure~\ref{fig:EEC}. 

In contrast to the NLO expressions in supergravity, which exhibit uniform transcendentality, the energy correlators in pure gravity contain contributions of lower transcendental weight and depend on the helicity configuration of the incoming particles. Their explicit expressions are provided in the ancillary file.

\section{Stringy EEC}
\label{sec:stringyEEC}

We have argued that energy correlators define well-posed, infrared finite observables in four-dimensional gravity. Extending this construction to higher perturbative orders, and more ambitiously to the fully non-perturbative regime, requires a UV completion of gravity. Four-dimensional Minkowski string theory vacua furnish a rich class of such UV completions.

Let us consider the simplest case of a string theory on $R^{1,3} \times M_6$. Tree-level gravitational scattering in such theories has a large degree of universality \cite{Chowdhury:2019kaq}, with the conjectural existence of only two consistent stringy gravitational S-matrices, type II and heterotic. This universality, however, is not sufficient for our purposes because, when computing energy correlators, we sum over all possible final states. These include states beyond the universal sector considered in \cite{Chowdhury:2019kaq}.\footnote{The universal sector consists of states whose worldsheet operator is the identity on $M_6$ and which are even under $(-1)^{F_L}$, $(-1)^{F_R}$, $\Omega$, where $F_{L/R}$ are the left/right-moving worldsheet fermion number operators, and $\Omega$ is the worldsheet orientation reversal operator.} Therefore, already at tree level, we do not expect the same level of universality for the gravitational energy correlators in string theory, as the one enjoyed by the gravitational amplitudes.

In order to connect to the previous sections, we therefore consider type II string theory compactified on $T^6$. At low energies, it becomes ${\cal N}=8$ SG considered in Section~\ref{sec:SG}. We would like to calculate the stringy correction to the EEC away from the end points,  thus ignoring possible contact terms. We introduce a new dimensionless parameter, 
\be
a= (2E)^2\app ,
\ee
where $\alpha'$ is the Regge slope  controlling the string states  mass. If we restrict our consideration to $a<1$, then the only states that appear in the final states at tree level are those of the graviton supermultiplet. In this regime the only new ingredient needed, compared with the previous sections, is the tree-level five-point amplitude. For simplicity we focus on the energy--energy correlators averaged over the beam direction, $\overline{\rm EEC}^{(1)}(z)$ for $0<z<1$, see \eqref{eq:avEECdef}.

Let us notice that, apart from stringy modes, due to the presence of compact extra dimensions, the theory also contains Kaluza-Klein modes, which should be taken into account \cite{Russo:1997mk}. Their mass is controlled by the size of the six-torus $T^6$, so that $m_{\text{KK}} \sim {1 \over R}$. In particular, for $s \geq 4 m_{\text{KK}}^2$ there are tree-level amplitudes ${\cal M}_{g,g\to \text{KK},\text{KK},g}$ at the same order in $G_N$ as the ones considered in this section. Therefore, we restrict our analysis to low energies $s < 4 m_{\text{KK}}^2$ and to the leading order in $G_N$. In this regime, the contribution of the KK modes to the energy correlator for $0<z<1$ can be ignored.\footnote{For $z=0,1$ the KK modes will contribute through the one-loop diagram.} It would be very interesting to analyze the effect of the Kaluza-Klein modes on the energy correlators in gravity.

\subsection{Five-point amplitude}

Let us review the relevant tree-level superstring amplitudes. %
The KLT relations \cite{Kawai:1985xq} allow us to express tree-level closed-string amplitudes on the sphere in terms of open-string disk amplitudes. In the five-point case, the MHV  graviton amplitude is given by 
\begin{align}
M^{\text{string}} (12345) =  &   g_1 A_L(12345)A_R(21435)  + g_2 A_L(13245)A_R(31425) \,, \label{stringAmpl5}
\end{align}
where we employ the shorthand notation
\begin{align} \label{eq:defg1g2}
g_1 \equiv (\app \pi)^{-2} \sin(\app \pi s_{12})\sin( \app \pi s_{34})   \,, \qquad
g_2 \equiv (\app \pi)^{-2} \sin( \app \pi s_{13})\sin( \app \pi s_{24}) \,,
\end{align}
and the open-string amplitudes $A_L$ and $A_R$ correspond to the left- and right-moving modes of the closed string. For the sake of presentation all momenta are inflowing, and we omit the coupling $\left( \frac{\kappa}{2}\right)^3$. 
Both have the following form in four-dimensional spinor-helicity notation \cite{Medina:2002nk,Barreiro:2005hv,Stieberger:2006te} for MHV  scattering of gluons,\footnote{For $A_L$,  the spinor-helicity factors coming from the expansion of $\delta^{8}\left( \sum_{i=1}^5 \la_i \eta_i \right)$ (with  $\eta^A$ carrying an $SU(4)$ index $A=1,\ldots,4$), which specify the helicity configuration, have to be taken into account as well. Similarly, $A_R$ is accompanied by $ \delta^{8}\left( \sum_{i=1}^5 \la_i \hat{\eta}_i \right)$,  where  $\hat\eta^{A'}$ corresponds to another $SU(4)$ with $A'=5,\ldots,8$, see Appendix~\ref{app:SU4xSU4}.}
\begin{align}
A(12345) = \frac{i \app^2}{\vev{12}\vev{34}\vev{45}} \Bigl( [15][32] f_1(12345) + [12][35] f_2(12345) \Bigr)  \,. \label{AYM5}
\end{align}
We have omitted the gauge coupling.
The disk world-sheet integrals, 
\begin{align}
& f_1(12345) = \int\limits^1_0 dx \int\limits^1_0 dy \, x^{-1-\app s_{23}}y^{-1-\app s_{15}} (1-x)^{-\app s_{34}}  (1-y)^{-\app s_{45}}  (1-x y)^{-\app s_{35}}\,,   \label{eq:wsDiskInt} \\
& f_2(12345) = \int\limits^1_0 dx \int\limits^1_0 dy \, x^{-\app s_{23}}y^{-\app s_{15}} (1-x)^{-\app s_{34}}  (1-y)^{-\app s_{45}}  (1-x y)^{-1-\app s_{35}}\,,\label{eq:wsDiskInt2}
\end{align}
are expressible in terms of the hypergeometric function ${}_{3}F_{2}$ accompanied by a ratio of products of gamma functions, which can be found in \cite{Stieberger:2006te}. 
The low-energy expansion of the open string amplitudes \p{AYM5} has the following cyclic symmetric form~\cite{Stieberger:2006te},
\begin{align}
& A(12345) = \frac{i}{\vev{12}\ldots\vev{51}} \left[ 1 + \app^2 \frac{ \zeta_2}{2} \left( - \sum_i s_{i,i+1} s_{i+1,i+2}- \varepsilon(1234)\right) \right. \label{eq:Aalphexpand}\\
& \left. + \app^3 \frac{\zeta_3}{2} \left(-\frac12 \sum_{i<j<k} s_{ij} s_{jk} s_{ik} + 3\sum_i s_{i,i+1} s_{i+1,i+2} s_{i+3,i+4}  + \varepsilon(1234) \sum_i s_{i,i+1}\right) + O(\app^4) \right] ,\notag
\end{align}
where the one-fold sums contain five cyclic terms, and the parity-odd factor $ \varepsilon(1234)$ is defined after \p{eq:Mtree5}. Substituting the permutations of \p{eq:Aalphexpand} in \p{stringAmpl5}, and expanding the coefficient functions \p{eq:defg1g2} at small $\app$, one obtains the low-energy expansion of the closed-string amplitude. The even zeta values cancel out from the latter~\cite{Stieberger:2009rr,Schlotterer:2012ny}. Also, the graviton amplitude has Bose symmetry, which is not manifest in the KLT representation \eqref{stringAmpl5}.

Next, we calculate the squared five-point amplitude $\mathbb{M}^{(0)}_{2 \to 3}$, summing over the final states of all helicities from the ${\cal N}=8$ supermultiplet. We employ the supersymmetric extension ${\cal M}^{\text{string}}$ (see \p{eq:M5super}) of the graviton MHV scattering amplitude \p{stringAmpl5}, which is given by the KLT relations in their supersymmetric form \cite{Elvang:2010kc}. The R-symmetry $SU(8)$ of supergravity is broken down to $SU(4)\times SU(4)$ by stringy corrections. This results in a more cumbersome summation formula, compared with the supergravity case, which we discuss in Appendix~\ref{app:SU4xSU4}:
\begin{align}
\mathbb{M}_{2\to 3}^{(0)} = \sum\limits_{\rm helicity}|{\cal M}^{\rm string}|^2 ={}&  2 s_{12}^8 \Bigl( 2g_1^2 |A_{12345}|^2 |A_{21435}|^2 + 2g_2^2 |A_{13245}|^2 |A_{31425}|^2 \\
& + g_1 g_2 \left( A_{13245} A_{12345}^* + A_{12345} A_{13245}^* \right) \left( A_{31425} A_{21435}^* + A_{21435} A_{31425}^* \right) \Bigr) ,
\notag
\end{align}
where $A_{ijklm} \equiv A(ijklm)$.

The square of \eqref{stringAmpl5} has a complicated functional dependence on the kinematic variables. In order to obtain analytic results for the EEC at $0<z<1$, we series-expand the world-sheet integrals \eqref{eq:wsDiskInt} in powers of the string tension $\alpha'$  with the help of \texttt{HypExp} \cite{Huber:2005yg}, and obtain the following expansion for the squared amplitude, up to order $\alpha{'}^{12}$,
\begin{align}
\mathbb{M}_{2 \to 3}^{(0)} =  & \app^0 I_0 +\app^3 \zeta_3 I_3 +\app^5 \zeta_5 I_5  + \app^6 (\zeta_3)^2 I_6 +  \app^7 \zeta_7 I_7 + \app^8 \zeta_3\zeta_5 I_8 + \app^9 (\zeta_9 I_{9,1} + (\zeta_3)^3 I_{9,2}) 
\notag\\ 
& + \app^{10} (\zeta_3 \zeta_7 I_{10,1} + (\zeta_5)^2 I_{10,2}) + \app^{11} (\zeta_{11} I_{11,1} + \zeta_5 (\zeta_3)^2 I_{11,2}) \notag\\
& + \app^{12} (\zeta_{5}\zeta_{7} I_{12,1} + \zeta_3 \zeta_9 I_{12,2} +  (\zeta_3)^4 I_{12,3}) + O(\app^{13})\,. \label{eq:MMstring}
\end{align}
The coefficients $I_{n,m}({\bf s})$ are Bose-symmetric rational functions of the Mandelstam variables,
\begin{align}
I_{n,m}({\bf s}) = \frac{2s_{12}^8 N_{n,m}({\bf s})}{\prod\limits_{1 \leq i \leq j \leq 5} s_{ij}} \,,
\end{align}
with polynomial numerators $N_{n,m}({\bf s})$ of degree $n+4$. In particular, we recover the supergravity squared amplitude, see Eq.~\eqref{eq:MMtree2pt}, $N_{0}({\bf s}) = -16{\rm Gram}(p_1,p_2,p_3,p_4)$, as the leading term of the low-energy expansion. The numerator for the leading stringy correction,
\begin{align}
N_{3}({\bf s}) =  -16\, {\rm Gram}(p_1,p_2,p_3,p_4) \sum_{i<j<k} s_{ij} s_{jk} s_{ik} \,.
\end{align}
The label $n$ counts the transcendental weight, which coincides with the order in the $\alpha'$-expansion. The label $m$ distinguishes terms of the same transcendental weight that are proportional to different products of zeta values.

We also observe that the transcendental numbers in this expansion are products of the odd zeta values, $\zeta_{2n+1}$, i.e. the even zeta values $\zeta_{2n}$ and MZV are absent. This is in contrast with the amplitude itself, where MZVs do appear, starting from weight 11, see (6.24) in \cite{Schlotterer:2012ny}. It would be interesting to understand better the origin of this simplification at the level of EEC.

We do not have to rely on the series expansion \p{eq:MMstring} when calculating the EEC in the collinear and back-to-back regimes.

\begin{figure}[t]

\begin{center}   
\includegraphics[width=0.75\textwidth]{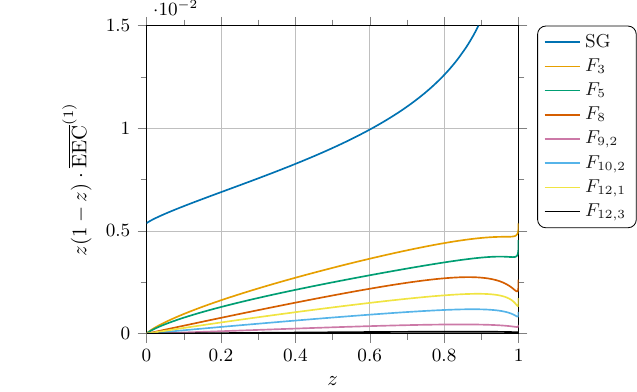}
\end{center}

\caption{Beam-direction averaged EEC in supergravity and their stringy corrections $F_n$, \p{eq:EECavString}. We multiply by $z(1-z)$ to emphasize the dominance of the supergravity contribution, see the asymptotics \p{eq:stringz0} at $z\to 0$ and \p{eq:EECstringz1} at $z\to 1$. For clarity, we display only a few of the stringy corrections from \p{eq:EECavString}; the remaining terms exhibit a similar behavior.}
\label{fig:EECavstring}
\end{figure}

\subsection{Beam-averaged EEC}

Next, we integrate the squared amplitude \eqref{eq:MMstring} term-by-term over the phase space $d{\rm PS}_3$, with the EEC weight according to \p{eq:EECintx}, and average over the beam directions \eqref{eq:avEECdef}. In this way, we find the low-energy expansion of the EEC for $0<z<1$ up to order $a^{12}$,
\begin{align}
\label{eq:EECavString}
\overline{\rm EEC}^{(1)}(z)  = &  a^0 \overline{\rm EEC}^{(1)}_{\text{SG}} +a^3 \zeta_3 F_3 +a^5 \zeta_5 F_5  + a^6 (\zeta_3)^2 F_6 +  a^7 \zeta_7 F_7 + a^8 \zeta_3\zeta_5 F_8  \\ 
& + a^9 (\zeta_9 F_{9,1} + (\zeta_3)^3 F_{9,2}) + a^{10} (\zeta_3 \zeta_7 F_{10,1} + (\zeta_5)^2 F_{10,2}) \notag\\
&  + a^{11} (\zeta_{11} F_{11,1} + \zeta_5 (\zeta_3)^2 F_{11,2}) + a^{12} (\zeta_{5}\zeta_{7} F_{12,1} + \zeta_3 \zeta_9 F_{12,2} +  (\zeta_3)^4 F_{12,3}) + O(a^{13}) \,. \notag
\end{align}
The leading term is the supergravity approximation \eqref{EECav}. The leading stringy correction 
\begin{align}
F_3(u) = & \frac{1}{\pi^5}
\frac{(1+u^2)^3}{u^6}\biggr(   (4+u^2) \log(1+u^2) - 2 u^2 \frac{(3+u^2)}{(1+u^2)} \log(u) -2u\arctan(u) \notag\\
&  + 
(1+u^2) \arctan^2(u) + (6-u^2) \log(u)\log(1+u^2) - \frac14 (1-3 u^2) \log^2(1+u^2) 
\notag\\ 
&
+ (6+u^2) \left( {\rm Li}_2(i u)
+ {\rm Li}_2(-i u) \right) \biggr) \, ,
\end{align}
where $u = \sqrt{\frac{z}{1-z}}$, see \p{eq:defu}.
We omit the energy factor $E^8$ and the coupling $\left( \frac{\kappa}{2}\right)^6$, according to our conventions in \eqref{eq:ECpertexp}.
The higher terms in the expansion ($n \geq 3$) have a similar form when written in terms of the variable $u$,
\begin{align}
F_n (u) = & \frac{1}{\pi^5}   \frac{(1+u^2)^3}{u^{2n}} \biggl(  i u^{n+\frac{1+(-1)^n}{2}} R^{(a)}_{n-5-\frac{1+(-1)^n}{2}}(u^2) \left( {\rm Li}_2(i u) - {\rm Li}_2(-i u)-2i \log(u)\arctan(u) \right)  \notag\\
& +  R^{(b)}_{2n-5-(-1)^n}(u^2) \left( {\rm Li}_2(i u) + {\rm Li}_2(-i u) \right) + (1+u^2) R^{(c)}_{2n-7-(-1)^n}(u^2)\arctan^2(u)\notag\\
& + R^{(d)}_{2n-5-(-1)^n}(u^2)  \log^2(1+u^2)  + R^{(e)}_{2n-5-(-1)^n}(u^2) \log(u)\log(1+u^2) \notag\\
& +  u R^{(f)}_{2n-6}(u^2) \arctan(u)  + \frac{u^2 }{1+u^2} R^{(g)}_{2n-4}(u^2) \log(u) +  R^{(h)}_{2n-4}(u^2) \log(1+u^2) \notag\\
& +\frac{u^2}{(1+u^2)^{n-3-\frac{1+(-1)^n}{2}}} R^{(k)}_{4n-13-(-1)^n}(u^2) \biggr)\,,
\end{align}
where $R^{(a)},\ldots,R^{(k)}$ are even polynomials in $u$ of specified degree. Let us note the presence of low-transcendentality contributions to the stringy corrections,  compared to the supergravity approximation. It would be interesting to obtain a closed-form expression for them at any order of the low-energy expansion. We attach the first few corrections to the submission. The plots are shown in Figure~\ref{fig:EECavstring}.

\subsection{Collinear and back-to-back limits}

The closed-string amplitude \eqref{stringAmpl5} significantly simplifies in the collinear and back-to-back regimes, so that we are able to obtain the closed-form asymptotics for the  $\overline{\rm EEC}^{(1)}(z)$, without relying on the low-energy expansion \p{eq:MMstring}. In both regimes, the leading asymptotics comes from supergravity, and the string corrections are softer.

In the collinear region,
\begin{align}
\overline{\rm EEC}^{(1)}(z)  = & \frac{ 1}{2 \pi^5} \biggl( \frac{\pi^2}{3z} -\log(z)  - \log(z) \sum_{n \geq 1} \frac{n+2}{n+1} \zeta_{2n+1} a^{2n+1} \biggr) + O(z^0)\, , \label{eq:stringz0}
\end{align}
where the series can easily be summed in terms of  polygamma functions. The latter fact is in agreement with the supergravity asymptotics \p{EECbarSG0}. We also notice that all stringy terms in \p{eq:EECavString}, accompanying products of several $\zeta$-values, are finite at $z = 0$.

Let us discuss in more detail the calculation of the back-to-back asymptotics. To simplify the calculations, we permute the points $(31245)$ in the KLT relation \eqref{stringAmpl5}. The leading contribution in the back-to-back regime comes from the region $x=1$ of the phase-space integration in \p{eq:EECintx}. We reveal this contribution by the change of integration variable $x \to \frac{x}{x+\sqrt{1-z}(1-x)}$. Then, we find that the world-sheet integral $f_2$ \p{eq:wsDiskInt2} does not contribute to the leading term of the $z\to 1$ asymptotics. Moreover, in this limit, one of the integrations in $f_1$ \eqref{eq:wsDiskInt} is localized, and the remaining integration in $f_1$ is expressed in terms of gamma functions. Then we rewrite the ratio of products of gamma functions, using the well-known formula for $\log \Gamma(1+ax)$ as a Taylor series expansion with zeta-valued coefficients. The remaining phase space integration is a univariate integral over the energy fraction $x$. Thus, we find that the stringy corrections in the back-to-back asymptotics diverge as $\frac{1}{1-z}$,
\be
\overline{\rm EEC}^{(1)}(z) =& \frac{1}{2 \pi^5} \biggl\{ \frac{1}{1-z} \biggl( \frac14 \log^2(1-z) \label{eq:EECstringz1} \\ 
& - \log(1-z) + 2 +\frac{5\pi^2}{12} \biggr) + \frac{G(a) }{1-z} \biggr\} + O\left((1-z)^{-\frac{1}{2}}\right), \notag
\ee
and they are controlled by the function $G(a)$ with the following integral representation,
\begin{align}
G(a)\equiv -\frac{1}{4}\int\limits^1_0  dx & \frac{ x \log(x) + (1-x) \log(1-x)}{x^2 (1-x)^2}  \notag\\
& \times \left( \exp\left( 4 \sum_{n \geq 1} \frac{\zeta_{2n+1}}{2n+1} a^{2n+1} \left(1-x^{2n+1}-(1-x)^{2n+1} \right)\right) -1 \right) . \label{eq:Gint}
\end{align}
In order to reveal the transcendental numbers appearing in $G(a)$, we expand it at small $a$ and perform the univariate integrations,
\begin{align}
& G(a) = \frac{\pi^2}{3}\sum_{n\geq 1} \zeta_{2n+1} \, a^{2n+1} + \sum_{k \geq 5}  a^{k} \sum_{m \geq 1} \sum\limits_{\substack{w_1,\ldots,w_m \; \text{odd}\\w_1 + \ldots +w_m =k}}c_{k}^{(\vec{w})} \zeta_{w_1} \ldots \zeta_{w_m} \,, \label{eq:Gaexpand}
\end{align}
where the second term contains products
of odd zeta values, and the $c_k$'s are rational numbers. Namely, the linear $\zeta$-terms in the expansion of \p{eq:Gint} integrate to $\pi^2$, and their products integrate to rational numbers. Let us notice that the leading ${\log^2(1-z) \over 1-z}$ term does not acquire any stringy corrections, see Figure~\ref{fig:EECavstring}. This is related to the fact that it arises from soft radiation, which is completely universal as we explain in Section~\ref{sec:backtoback}.

\section{Dispersion relations and positivity}
\label{sec:disprel}

In this section we explore the positivity and dispersive properties of the energy correlators computed in the previous sections. For simplicity, we focus on the beam-averaged EEC for $0<z<1$. We find it convenient to introduce the following regularized EEC,
\be
\text{eec}(z) = (1-z) \overline{\rm EEC}(z) \ . 
\ee
As we will see in a moment, the advantage of introducing $\text{eec}(z)$ is that its discontinuity is integrable close to the endpoints $z=0,1$ inside the dispersion relations.

\subsection{Analyticity and polynomial boundedness}

To the best of our knowledge, all the known results for the EEC exhibit what we can call \emph{maximal collider analyticity}: as a function of the complexified angle $z$, the EEC is an analytic function with a pair of branch points located at $z=0,1$. This is also true for the results obtained in the present paper. 

In perturbation theory, this property is directly connected to the maximal analyticity of the scattering amplitudes, which is known to hold perturbatively for the scattering of the lightest particles in both QFT and gravity. At finite coupling, this is less clear, but the bootstrap results of \cite{Dempsey:2025yiv} also appear to be consistent with this hypothesis.

Let us discuss the behavior of the EEC at large complex $z$. Using the formulas in the present paper, we find the following asymptotic behavior  in the limit $|z| \to \infty$:
\be
\text{eec}_{\text{SG}}(z), \text{eec}_{++}(z)  &\sim {\log z \over z} \ , ~~~\text{eec}_{+-}(z) \sim z^4 \ . 
\ee
For the stringy corrections we find the following pattern:
\begin{equation}
\begin{aligned}
&F_3 \sim z^{-2} \ , ~~~F_5, F_6 \sim z^{-1} \ ,~~~
F_7, F_8, F_{9,2} \sim z  \ , \\
&F_{9,1}, F_{10,1}, F_{10,2}, F_{11,2}, F_{12,3} \sim z^3 \ , ~~~
F_{11,1}, F_{12,1}, F_{12,2} \sim z^5 \ \ . 
\end{aligned}
\end{equation}
It can be traced back to the worsening Regge scaling of higher  powers of $\alpha'$ in the low-energy behavior of the stringy amplitude.

At finite coupling, the leading large-$z$ asymptotics is not known. In ${\cal N}=4$ super Yang-Mills (SYM), however, the existence of bootstrap bounds \cite{Dempsey:2025yiv}  allows one to probe the EEC indirectly in the planar limit, and the results are consistent with $\lim_{|z| \to \infty} {\text{eec}_{{\cal N}=4}(z) \over z} = 0$. 

\subsection{Positivity and dispersion relations}

Given the analyticity and polynomial boundedness of the EEC, it is interesting to explore the dispersive representation of the various corrections. Here we focus on the cases in which we can write dispersion relations without subtractions, namely $\text{eec}_{\text{SG}}, \text{eec}_{++}$, and $\text{eec}_{3} \equiv (1-z) F_3(z)$.

Using the formulas of the previous section, we see that the large-$|z|$ limit of the eec in this case satisfies
\be
\lim_{|z| \to \infty} \text{eec}(z) = 0 ,
\ee
which implies that it admits a zero-subtracted dispersion relation,
\be
\label{eq:dispEEC}
\text{eec}(z) = \oint {d z' \over 2 \pi i} {\text{eec}(z') \over z' - z} = \int_0^\infty {dw \over \pi} \left( { \sigma_0(w) \over w+z} + {\sigma_1(w) \over w+(1-z)} \right) .
\ee
Here we have defined the discontinuities as follows:
\be
\sigma_0(w) &\equiv {\text{eec}(-w-i 0) - \text{eec}(-w+i 0) \over 2 i}  , \\
\sigma_1(w) &\equiv {\text{eec}(1+w + i 0) - \text{eec}(1+ w -i 0) \over 2 i} \ . 
\ee
Notice that the presence of the ${1 \over z}$ term in $\text{eec}(z)$ at small $z$ implies that $\sigma_0(w)$ contains a $\delta(w)$ piece.
What makes the dispersive representation \eqref{eq:dispEEC} useful are its corollaries, 
\be
\sigma_{0,\text{SG}},\sigma_{0,++},\sigma_{0,3} > 0 \ , 
~~~ \sigma_{1,\text{SG}},\sigma_{1,++} > 0 , 
\ee
while $\sigma_{1,3}(w)$ develops a small region of negativity close to $w=0$.

The positivity of the discontinuities immediately implies several interesting properties of the eec. First, since the integrand in \eqref{eq:dispEEC} is nonnegative, we get the familiar positivity of the EEC,
\be
\text{eec}(z) \geq 0, ~~~ 0<z<1 \ .
\ee
Second, considering the expansion around the orthogonal configuration $z=1/2$, we notice that
\be
\label{eq:cndef}
\text{eec}(z) = \sum_{n=0}^\infty c_n \left({1 \over 2}-z \right)^n  \ ,
\ee
where ${1 \over 2}-z = {\cos \theta \over 2}$. The coefficients $c_n$ can be written as moments of the discontinuity,
\be
c_n = \int_{0}^{\infty} {d w \over \pi} { \sigma_0(w)+(-1)^n \sigma_1(w) \over (w+1/2)^n} \ .
\ee
It is convenient to consider odd and even $n$ separately. In this way we get
\begin{equation}
\begin{aligned}
c_n^+ &= \int_{0}^{\infty} {d w \over \pi} {\sigma_0(w) + \sigma_1(w) \over (w+1/2)^n} \ , \\
c_n^- &= \int_{0}^{\infty} {d w \over \pi} {\sigma_0(w) - \sigma_1(w) \over (w+1/2)^n} \ .
\end{aligned}
\label{eq:twomoments}
\end{equation}
The representation \eqref{eq:twomoments} with positive discontinuities implies that we get a pair of moment problems associated with $\text{eec}(z)$, see e.g. \cite{Bellazzini:2020cot}. %

We also find that $\sigma_{0,\text{SG}}(w) \pm \sigma_{1,\text{SG}}(w) \geq 0$, which immediately yields $c_n > 0$. Despite the negativities present in the analogous combinations for the $++$ scattering in pure gravity and in the leading stringy corrections, we find that in all three cases $c_n >0$ for $n \geq 0$ and integer. The positivity of $c_n$ is closely related to the complete monotonicity observed in the perturbative studies of scattering amplitudes \cite{Henn:2024qwe}.

Let us also notice that in ${\cal N}=4$ SYM, the LO result is $\text{eec}(z) = - {\log(1-z) \over z^2}$ and it has all the properties observed above as well. Using the findings of \cite{Dempsey:2025yiv}, we have observed that the finite-coupling results in ${\cal N}=4$ SYM are consistent with $c_n > 0$.

\subsection{Energy multipoles}

It is natural to consider a multipole expansion of the eec. Due to the presence of a pole at $z=0$, it is only well defined for $d>4$.\footnote{Recall that the orthogonality of the Gegenbauer polynomials in $d$ dimensions implies\\  $c_J^{(d)} \sim \int_0^{1} dz (z(1-z))^{{d-4 \over 2}} P_J^{(d)}(1-2z)\text{eec}(z)$, which is finite in $d>4$.} We can therefore write
\be
\label{eq:multipole}
\text{eec}(z) = \sum_{J=0}^\infty c_J^{(d)} P_J^{(d)}(1-2z), ~~~ c_J^{(d)} \geq 0 .
\ee
In QFT$_d$ or gravity in $d>4$, the positivity of $c_J^{(d)}$ follows from unitarity \cite{Fox:1978vu}. In $4d$ gravity the divergence of the total cross section requires to consider the multipole expansion \eqref{eq:multipole} $d>4$.

Curiously, there is a direct relationship between the positivity of the coefficients $c_n$ defined in \eqref{eq:cndef} and the energy multipoles $c_{J}^{(d)}$, see \cite{Gneiting2013StrictlyNonStrictlyPD}: non-negativity of $c_n$ implies that $c_{J}^{(d)}$ are also non-negative in any $d$ for which the multipoles are well defined.\footnote{Theorem 1 in \cite{Gneiting2013StrictlyNonStrictlyPD} requires regularity which is violated for gravitational energy correlators, which however can be relaxed, see \cite{bilyk2024positive}.}
We can therefore physically think of the positivity of $c_n$ as a strong, i.e. $d \to \infty$, version of unitarity.

Of course, we do not have first-principle arguments for the various positivity properties discussed in this section. It would be very interesting to understand if the positivity properties observed here persist at higher loops, or maybe even at finite coupling. It would also be very interesting to explore the dispersion relations for the energy correlators and the positivity properties of $c_n$ in other theories.

\section{Back-to-back asymptotics of the EEC}\label{sec:backtoback}

In this section, we analyze the behavior of the energy--energy correlator (EEC) in the limit $z =(1 - \vec n_1 \cdot \vec n_2)/2 \to 1$, or equivalently $\vec n_2 \simeq -\vec n_1$, corresponding to the configuration where the two detected particles
move back-to-back. 

From the explicit one-loop computation, we have observed that the EEC simplifies considerably in this kinematic regime, taking the same functional form in both gravity and ${\mathcal N}=8$ supergravity, see \re{b2b-limit}, \re{EECbar} and \re{b2b-gr}. Below we show that the EEC behavior in the back-to-back limit is governed by the emission of soft gravitons and is insensitive to contributions from lower-spin particles. By exploiting the universal properties of soft graviton radiation \cite{Weinberg:1965nx}, we demonstrate the cancellation of infrared divergences between virtual and real corrections. Building on these results, we apply techniques originally developed in QCD \cite{COLLINS1981381,KODAIRA198266,Sterman:1986aj,Korchemsky:1993uz}  (and later applied to gravity,  \cite{Brandhuber:2008tf,Akhoury:2011kq,White:2011yy,Naculich:2011ry}) to derive an all-order expression for the EEC in the limit $z \to 1$.\footnote{In gauge theories analogous formulas for the EEC have been derived in \cite{deFlorian:2004mp,Moult:2018jzp,Korchemsky:2019nzm}.}

\subsection{Eikonal approximation}

To make the presentation self-contained, we begin by reviewing the well-known properties of scattering amplitudes involving soft gravitons.

Let us consider the scattering process \re{proc},
where the momenta of incoming and outgoing particles are given by \re{kin0}
and \( X \) represents an arbitrary number of undetected particles in the final state. In the absence of such radiation (\( X = \varnothing \)), the detected particles move exactly back-to-back and carry equal energies, \( E_1 = E_2 = E \). The corresponding contribution to the EEC is then localized at $z=0$ and \( z = 1 \). For \( z < 1 \), the final state necessarily includes undetected particles whose emission induces a recoil, causing the detected particles to deviate from the strictly back-to-back configuration.

In the limit \( z \to 1 \), the recoil momentum vanishes, so the undetected radiation \( X \) consists of an arbitrary number of soft particles. The energies of these particles are much smaller than those of the incoming and detected particles, so their contribution can be treated in the eikonal approximation. In this regime, the dominant contribution to the scattering amplitude arises from the particles with maximal spin, namely the gravitons, while the contribution from soft fermions and scalars is suppressed by additional powers of their energy.

In general, energy-energy correlators receive contributions from both soft particles (real and virtual) and hard particles (virtual), whose momenta scale with the total center-of-mass energy \(2E \). To disentangle these contributions in the scattering amplitude, it is convenient to treat the soft graviton as an external long-range field \( h_{\mu\nu} (x)\). The calculation can then be organized in two steps. First, we factorize the soft-graviton contribution to the scattering amplitude \( {\cal M}_{2\to 2} \) of the process \re{proc}. Next, we average the squared amplitude \( |{\cal M}_{2\to 2}|^2 \) over the fluctuations of the field \( h_{\mu\nu} \). This averaging procedure effectively incorporates both the virtual and real corrections to the energy correlators arising from soft-particle emission.
 
In the eikonal approximation, the scattering amplitude \({\cal M}_{2\to 2} \) factorizes into the product of a hard function describing the short-distance \( 2 \to 2 \) scattering and an eikonal phase that depends on the background field of the soft gravitons, %
\begin{align}\label{eik}
 {\cal M}_{2\to 2} = H(E)\, \textrm{Texp}\left[{i\kappa\over 2} \int \frac{d^4 k}{(2\pi)^4}\, 
\tilde h_{\mu\nu}(k)\, j_{\rm eik}^{\mu\nu}(k)\right] + \dots\,,
\end{align}
where \( \tilde h_{\mu\nu}(k) = \int d^4 x\, e^{ikx} h_{\mu\nu}(x) \), $\kappa^2=32 \pi G_N$ and $\textrm{T}$ stands for the time ordering of the graviton fields. 
This approximation is valid up to corrections suppressed by powers of the soft-graviton energy (indicated by the dots), and it holds independently of the specific matter content of the gravitational theory. 
The dependence on the matter enters only through the hard function $H(E)$. The factorization in~\eqref{eik} applies provided the underlying \(2 \to 2\) process is hard, meaning that the momentum transfers \( 2(p_i q_j) = O(E^2) \) are much larger than the characteristic energy of the soft radiation. This condition further implies that the angular variables \( y_1 \) and \( y_2 \) defined in~\eqref{ys} must not vanish in the limit \( z \to 1 \).

The soft factor in \eqref{eik} involves the coupling between the soft graviton field \( \tilde h_{\mu\nu}(k) \) and the eikonal current,  
\begin{align}\label{soft}   
j_{\text{eik}}^{\mu\nu}(k)
= \frac{i\, p_1^\mu p_1^\nu}{(p_1 k) - i0}
+ \frac{i\, p_2^\mu p_2^\nu}{(p_2 k) - i0}
- \frac{i\, q_1^\mu q_1^\nu}{(q_1 k) + i0}
- \frac{i\, q_2^\mu q_2^\nu}{(q_2 k) + i0}\,.
\end{align}
This current is conserved, $k_\mu j_{\rm eik}^{\mu\nu}(k)=0$, for $p_1+p_2=q_1+q_2$.   
The physical meaning of the `\(\pm i0\)` prescription in the eikonal propagators in \eqref{soft} becomes clear upon Fourier transforming the eikonal phase to configuration space:  
\begin{align}\label{Js}\notag
J_{\rm soft}(x){}& =\int \frac{d^4 k}{(2\pi)^4}\, e^{-ikx}\, \tilde h_{\mu\nu}(k)\, j_{\text{eik}}^{\mu\nu}(k)
\\
{}&= \int_{-\infty}^0 \! ds\, \big[\, h_{\mu\nu}(x + p_1 s)\, p_1^\mu p_1^\nu
+ h_{\mu\nu}(x + p_2 s)\, p_2^\mu p_2^\nu \big]
\notag\\
{}&\quad + \int_0^{\infty} \! ds\, \big[\, h_{\mu\nu}(x + q_1 s)\, q_1^\mu q_1^\nu
+ h_{\mu\nu}(x + q_2 s)\, q_2^\mu q_2^\nu \big] \,.
\end{align}
This representation shows that the `\(\pm i0\)` prescription selects the direction of time flow along the classical trajectories of the incoming (\(s < 0\)) and outgoing (\(s > 0\)) particles.

The relation \re{Js} admits a simple classical interpretation in terms of the particle trajectories involved in the scattering process \( p_1 + p_2 \to q_1 + q_2 \). For a classical particle moving along a worldline \( x_\mu(s) \), the eikonal phase is given by  \cite{Brandhuber:2008tf,White:2011yy,Naculich:2011ry}
\begin{align}\label{WL}
\exp\!\left[{i \kappa\over 2} \int_C ds \, \dot x^\mu(s)\, \dot x^\nu(s)\, h_{\mu\nu}(x(s))\right].
\end{align}
The eikonal phase appearing in \re{eik} takes precisely this form, with the contour \( C \) corresponding to the concatenation of the worldlines of the incoming and outgoing particles. %

\subsection{Energy--energy correlator}

Let us apply \re{eik} to compute the energy--energy correlators $\vev{\mathcal E(\vec n_1) \mathcal E(\vec n_2)}$.
These correlators are given by the weighted squared scattering amplitude \eqref{eik}, integrated over the energies of the detected particles and averaged over the soft graviton field $h^{\mu\nu}$, 
\begin{align}\label{ds}\notag
{}& \vev{\mathcal E(\vec n_1) \mathcal E(\vec n_2)}
= \int_0^\infty {dE_1\,E_1\over 2(2\pi)^3} \int_0^\infty {dE_2\,E_2\over 2(2\pi)^3}
\\
{}&\qquad \times E_1E_2\sum_X  \big|\bra{0}  {\cal M}_{2\to 2}\ket{X}_h\big|^2 \, (2\pi)^4 
\delta^{(4)}(p_1+p_2-q_1-q_2-k_X)\,.
\end{align}
Here the sum runs over the final states $X$ containing an arbitrary number of soft gravitons carrying the total momentum $k_X$.
The particle momenta $p_i$ (incoming) and  $q_i$ (outgoing) 
are given by \re{kin0}.
The integral over the energies in the first line of \re{ds} comes from the integration over the phase space of the detected particles. The factor of $E_1E_2$ in the second line in \eqref{ds} comes from the definition of the energy--energy correlators.

In general, the evaluation of the expectation value in \eqref{ds} is complicated by the graviton self-interaction. A major simplification arises, however, if the gravitons are soft. The strength of the gravitational interaction scales with the energy, hence the soft-graviton self-interaction can be neglected. As a result, in evaluating \eqref{ds} we may treat the gravitons as free particles. 

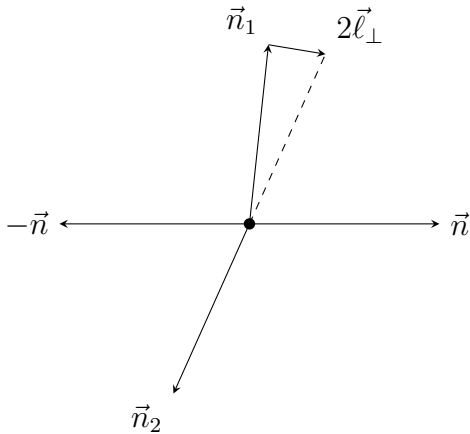
\begin{figure}
\begin{center}
\begin{tikzpicture}[>=stealth, scale=2.5]

\fill (0,0) circle (0.03);

\draw[->] (0,0) -- (1,0) node[anchor=west] {$\vec n$};
\draw[->] (0,0) -- (-1,0) node[anchor=east] {$-\vec n$};

\draw[->] (0,0) -- (0.1,0.95) node[anchor=south east] {$\vec n_1$};
\draw[-,dashed] (0,0) -- (0.4,0.9) node[anchor=south west] {$ $};
\draw[->] (0,0) -- (-0.4,-0.9) node[anchor=north east] {$\vec n_2$};

\draw[->] (0.1,0.95) -- (0.4,0.9) node[anchor=south west] {$2\vec \ell_\perp$};

\end{tikzpicture}
\end{center}
\caption{Kinematical configuration of unit vectors in the back-to-back limit. The incoming beams propagate in the directions $\vec n$ and $-\vec n$. The two calorimeters are oriented along $\vec n_1$ and $\vec n_2$. The recoil vector $2\vec \ell_\perp=-(\vec n_1+\vec n_2)$ defines the transverse momentum of soft gravitons.}\label{fig:vectors}
\end{figure}

A further simplification occurs in the back-to-back regime  $\vec n_2 \sim -\vec n_1$ or equivalently $z\to 1$. 
In this limit, the momentum conservation delta function in \eqref{ds} can be simplified as
\begin{align}\notag\label{delta}
{}& \delta (2E-E_1-E_2-k_X^0)\delta^{(3)}(E_1 \vec n_1+E_2\vec n_2+\vec k_X)
\\[2mm]\notag
{}& \qquad\sim 
\delta (2E-E_1-E_2)\delta (E_1 -E_2)\delta^{(2)}(2E_2\vec \ell_\perp -\vec k_{X,\perp})
\\[1.2mm]
{}& \qquad \sim \frac12\delta ( E_1-E)\delta (E_2 - E)\delta^{(2)}(\vec k_{X,\perp}-2E\vec \ell_\perp)\,,
\end{align}
where %
we have introduced the auxiliary recoil vector (see Figure~\ref{fig:vectors})
\begin{align}\label{recoil}
\vec \ell_\perp=-\frac12\lr{\vec n_1+\vec n_2}\,, \qqquad \vec \ell{\,}^2_\perp= 1-z\,, \qqquad (\vec n_1 \vec \ell_\perp) = -(1-z)\,.
\end{align}
In the second relation of \re{delta}, we decomposed the three-dimensional delta function into a product of two delta functions: one corresponding to the projection along the direction $\vec n_1 \sim -\vec n_2$, and the other associated with the orthogonal component $\vec k_{X,\perp}$ satisfying $(\vec n_1, \vec k_{X,\perp}) = 0$. At this step, we also neglected terms subleading in the limit $z \to 1$.

Substituting \re{eik} and \re{delta} into \eqref{ds}, we integrate over the energies $E_i$ of the detected particles and express $\vev{\mathcal E(\vec n_1)\mathcal E(\vec n_2)}$ as a sum over the final states $X$ containing an arbitrary number of soft gravitons with total transverse momentum $\vec k_{X,\perp}=2E\vec \ell_\perp$.
Fourier transforming the two-dimensional delta function in \eqref{delta}, this sum can be evaluated explicitly, leading to the representation 
\begin{align}\label{ds1}
\vev{\mathcal E(\vec n_1)\mathcal E(\vec n_2)}
= H^2(E) \int d^2 x_\perp \, e^{-2iE (\vec x_\perp   \vec \ell_\perp)} 
W(x_\perp)\,,
\end{align}
where $\vec x_\perp$ lies in the plane orthogonal to the vector $\vec n_1$.
The hard function $H(E)$ encodes the virtual corrections to the $2\to2$ scattering process and remains regular as $z\to 1$.~\footnote{Strictly speaking, the hard function in \re{ds1} differs from the analogous function in \re{eik} by a normalization factor, which we drop in order to  simplify the formulae.}

The function $W(x_\perp)$ 
accounts for the soft-graviton contribution.
It is expressed as the product of two eikonal phases~\eqref{Js}, separated in the transverse direction by the two-dimensional vector~$x_\perp$ and averaged over the fluctuations of the soft graviton field,  
\begin{align}\label{M}
     W(x_\perp) = \big\langle 0 \big| \big( T e^{{i \kappa\over 2} J_{\rm soft}(x_\perp)} \big) \big( \bar T e^{-{i \kappa\over 2} J_{\rm soft}(0)} \big) \big| 0 \big\rangle_h \,.
\end{align}
Eq.~\eqref{ds} is obtained from~\eqref{ds1} by inserting the completeness relation  
$\sum |X\rangle\langle X| = 1$
between the two operators in~\eqref{M} and integrating over~$x_\perp$.  
The operators~$(T e^{i {\kappa\over 2} J_{\rm soft}(x_\perp)})$ and~$(\bar T e^{-i {\kappa\over 2} J_{\rm soft}(0)})$ in~\eqref{M} originate from the amplitude and its conjugate in~\eqref{ds}, respectively. They depend explicitly on the graviton field and are time~$(T)$ or anti-time~$(\bar T)$ ordered.

The relations \eqref{ds1} and \eqref{M} describe the leading behavior of $\vev{\mathcal E(\vec n_1) \mathcal E(\vec n_2)}$ for $z\to 1$ to any order in the gravitational coupling. As noted earlier, the dependence of \eqref{ds1} on the specific matter content of the gravitational theory is contained entirely in the hard function, while the soft function in \eqref{ds1} is universal.

\subsection{Leading order}

Let us show that the relation \eqref{ds1} correctly reproduces the asymptotic behavior of the EEC for $z\to 1$ obtained previously in \eqref{EEC3ptGRz1f} and \eqref{fdef}. 

To lowest order in the coupling, we use \eqref{Js} and \eqref{soft} to obtain from \eqref{M}
\begin{align}\notag\label{M-lo}
  W(x_\perp) {}&=1+{\kappa^2\over 4} \vev{J_{\rm soft}(x_\perp) J_{\rm soft}(0)} + O(\kappa^4)  
  \\[2mm]
  {}&=1+{\kappa^2\over 4} \int {d^4 k\over (2\pi)^4} e^{-ikx_\perp} \vev{\tilde h_{\mu_1\nu_1}(k)\tilde h_{\mu_2\nu_2}(-k)} 
   j_{\rm eik}^{\mu_1\nu_1}(k)
  j_{\rm eik}^{\mu_2\nu_2}(-k) + O(\kappa^4)  \,.
\end{align}
After the integration in \eqref{ds1}, the Born ($O(\kappa^0)$) contribution to the EEC  yields a contact term  $\sim\delta(1-z)$.

Substituting \eqref{M-lo} into \eqref{ds1} we find 
\begin{align}\label{EE-lo}
   \vev{\mathcal E(\vec n_1) \mathcal E(\vec n_2)}={\kappa^2\over 4} H^2(E)
 \int {d^4k\over (2\pi)^4} (2\pi)^2 \delta^{(2)}(\vec k_{\perp}-2E\vec \ell_\perp)
 j_{\rm eik}^{\mu_1\nu_1}(k)j_{\rm eik}^{\mu_2\nu_2}(-k)D^+_{\mu_1,\nu_1;\mu_2,\nu_2}(k)  \,,
\end{align}
where $D^+_{\mu_1,\nu_1;\mu_2,\nu_2}(k)=\vev{\tilde h_{\mu_1\nu_1}(k)
    \tilde h_{\mu_2\nu_2}(-k)} $ is the propagator of the real  gravitons (Wightman correlation function). In the de Donder gauge it has the form (see, e.g. \cite{Weinberg:1965nx}) %
\begin{align}\notag\label{prop}
    {}& D^+_{\mu_1,\nu_1;\mu_2,\nu_2}(k) = 2\pi\delta_+(k^2)d_{\mu_1,\nu_1;\mu_2,\nu_2}\,,
    \\[2mm]
    {}& d_{\mu_1,\nu_1;\mu_2,\nu_2}= \frac12\left(g_{\mu_1\mu_2}g_{\nu_1\nu_2}+g_{\mu_1\nu_2}g_{\nu_1\mu_2}- g_{\mu_1\nu_1}g_{\mu_2\nu_2} \right)\,,
\end{align}
where $g_{\mu\nu}$ is the Minkowski metric tensor.
Due to the eikonal current conservation, the relation \eqref{EE-lo} is independent of the gauge choice.  

In order to evaluate the integral in \eqref{EE-lo}, it is convenient to introduce the light-cone variables
\begin{align}\label{lc}
    k^\pm = {1\over \sqrt 2}(k^0\pm (\vec k \vec n_1))\,,
    \qquad \vec k_\perp=\vec k - (\vec n_1 \vec k)\,\vec n_1 \,,
\end{align}
where $\vec n_1$ is a unit vector defining the position of $\mathcal E(\vec n_1)$ on the celestial sphere. 
The  delta functions in \eqref{EE-lo} and \eqref{prop} localize   the integrand  at $\vec k_\perp=2E\vec \ell_\perp$ and  $2k^+k^-=k_\perp^2=(2E)^2(1-z)$. Changing the integration variable as $k^+=E\omega \sqrt{2(1-z)}$, we find
\begin{align}\label{z-as}
{}&\vev{\mathcal E(\vec n_1) \mathcal E(\vec n_2)}={\kappa^2\over 2\pi}H^2(E)((1-y_1) y_1)^2 {f(y_1,\beta)\over 1-z}\left[ 1+ O(\sqrt{1-z})\right]\,, 
\end{align}
where the angular variables $y_1$ and $\beta$ are defined in \eqref{ys} and \eqref{beta}. The function $f(y_1,\beta)$ is given by
\begin{align}\label{f-def}\notag
f(y_1, \beta)={\sin ^2(\beta )\over y_1(1-y_1) }\int_0^\infty{}&  \frac{d \omega \,\omega} {\left(1-y_1\right) \omega^2+y_1-2
   \sqrt{\left(1- y_1\right) y_1}\,\omega\cos \beta}
   \\
  \times {}& {1\over y_1\omega^2+1-y_1+2  \sqrt{\left(1- y_1\right) y_1}\,\omega\cos\beta}\,.
\end{align}
This integral reproduces the expression for the one-loop EEC  obtained previously, see \re{EEC3ptGRz1f} and \eqref{fdef}. Matching the two expressions we can identify the hard function to the lowest order in the coupling
\begin{align}\label{H}
H^2(E) = {(\kappa E)^4 E^4\over 128 \pi^4}{1\over  ((1-y_1) y_1)^2}\lr{1+O(\kappa^2)}\,,
\end{align}
where $O(\kappa^2)$ denotes higher-order corrections. As mentioned above, these corrections depend on the matter content of the gravitational theory.

\subsection{Resummation of soft gravitons}

The main advantage of the representation \eqref{M} is that it can be used to compute the higher-order corrections to \eqref{z-as}. 

As explained above, the soft gravitons behave as free fields. Therefore, the expectation value in \eqref{M} amounts to 
Gaussian integration over the  $h-$fields. This leads to
\begin{align}\label{M1}\notag
  {}&  W(x_\perp) =e^{{1\over 4} \kappa^2F(x_\perp)}\,,
    \\[2mm]
 {}& F(x_\perp) =G_{\rm r}(x_\perp)-\frac12 (G_{\rm v}(0)+\bar G_{\rm v}(0))  \,.
\end{align}
The function $G_{\rm v}(0)$ and its complex conjugate $\bar G_{\rm v}(0)$ describe the virtual graviton contribution. They are obtained by contracting the $h-$fields within each of the operators in \eqref{M} using an (anti-)time-ordered Feynman propagator.  The function  $G_{\rm r}(x_\perp)$ describes the real graviton emission. It is built by cross-contracting fields from the two operators  in \eqref{M} with the help of a cut (Wightman) propagator. Explicitly,
\begin{align}\notag\label{GG}
   {}& G_{\rm v}(0)=\vev{T J_{\rm soft}(0)J_{\rm soft}(0)}_h = \int{d^4k\over (2\pi)^4}j_{\rm eik}^{\mu\nu}(k)
    j_{\rm eik}^{\mu'\nu'}(-k)D_{\mu\nu;\mu'\nu'}(k)\,,
    \\[2mm]
   {}& G_{\rm r}(x_\perp)=\vev{J_{\rm soft}(x_\perp)J_{\rm soft}(0)}_h = \int{d^4k\over (2\pi)^4} e^{-ikx_\perp}   j_{\rm eik}^{\mu\nu}(k)
    j_{\rm eik}^{\mu'\nu'}(-k)D^+_{\mu\nu;\mu'\nu'}(k)\,.
\end{align}
The relation \eqref{M1} reflects the well-known fact that the contribution of real and virtual soft gravitons  exponentiates.

The sum $G_{\rm v}(0)+\bar G_{\rm v}(0)$  in \eqref{M1} simplifies due to the identity (dropping the tensor structure from the Feynman propagators)
\begin{align}
 D(k)+\bar D(k) =  {i\over k^2+i0}- {i\over k^2-i0}  = 2\pi (\theta(k^0)+\theta(-k^0))\delta(k^2) = D^+(k) + D^+(-k)\,. 
\end{align}
Substituting the above into \eqref{M1} and using the  symmetry  under the exchange $k\to -k$, we obtain
\begin{align}\label{M2} 
F(x_\perp)=\int{d^4k\over (2\pi)^4} (e^{-ikx_\perp}-1)   j_{\rm eik}^{\mu\nu}(k)
   D^+_{\mu\nu;\mu'\nu'}(k) j_{\rm eik}^{\mu'\nu'}(-k)\,,
\end{align}
where the real graviton propagator is given by \re{prop}.
Replacing the eikonal current $j_{\rm eik}^{\mu\nu}(k)$ with its expression \eqref{soft} we get
\begin{align}\label{sum-I}\notag
{}&     F(x_\perp)  = I(p_1,p_2|x_\perp)-I(p_1,q_1|x_\perp) - I(p_1,q_2|x_\perp) 
    \\[2mm]\notag
{}& \phantom{F(x_\perp) }   -I(p_2,q_1|x_\perp) - I(p_2,q_2|x_\perp)+I(q_1,q_2|x_\perp)\,,
\\[2mm]%
 {}&    I(p,p'|x_\perp) = \int{d^4 k\over (2\pi)^4} (e^{-ikx_\perp}-1) 2\pi \delta_+(k^2) {2(pp')^2\over (pk)(p'k)}\,.
\end{align}
Here 
$x_\perp^\mu=(0,\vec x_\perp)$ is a two-dimensional space-time vector satisfying $(\vec n_1 \vec x_\perp)=0$ and  $p$ and $p'$ are lightlike vectors. 

The two terms in the factor $(e^{-ikx_\perp}-1)$ in $I(p,p'|x_\perp)$ describe the real and virtual soft graviton contributions, respectively. The integral in \eqref{sum-I} develops both infrared and ultraviolet logarithmic divergences. However the IR divergences cancel in the sum of six integrals in $F(x_\perp)$. Indeed, the infrared divergence of $I(p,p'|x_\perp)$ is proportional to $2(pp')$.  As a consequence, the infrared divergence of $F(x_\perp)$ is proportional to the total energy $(p_1+p_2-q_1-q_2)^2$ and it vanishes in the back-to-back limit $z\to 1$. This ensures that the function $F(x_\perp)$, and hence  the energy correlators \eqref{ds1}, are infrared finite,  by a mechanism similar to that  described by Weinberg \cite{Weinberg:1965nx}.

The ultraviolet divergences in \eqref{sum-I} can be treated in dimensional regularization with $d=4-2\epsilon$ and $\epsilon>0$. They are an artifact of the eikonal approximation. We recall that this approximation correctly captures the contribution of soft gravitons, whose momenta lie below some 
factorization scale $\mu$ acting as the UV cutoff in \eqref{sum-I}. The contribution of particles with larger momenta to \eqref{ds1} is described by the hard function $H(E)$. The two factors on the right-hand side of \eqref{ds1} depend on the factorization scale $\mu$ but this dependence cancels in their product. 
 
Computing the functions $F(x_\perp)$, we combine the six integrals in \eqref{sum-I} together and introduce the light-cone variables \eqref{lc}.
Separating the integrals over $k_\perp$ and $k_\pm$ and changing the integration variable $k_+=\omega |k_\perp|/\sqrt{2}$, we find from \eqref{sum-I}
\begin{align}\label{F-Fourier}
   \frac{\kappa^2}4 F(x_\perp)= \int {d^2k_\perp\over (2\pi)^2}\lr{e^{i\vec k_\perp \vec x_\perp}-1} {2(\kappa E)^2\over k_\perp^2}\widehat F(e_\perp)\,.
\end{align}
Here $\widehat F(e_\perp)$ is a function  of the unit two-dimensional vector $\vec e_\perp=\vec k_\perp/|k_\perp|$, given by the integral
\begin{align}\label{tilde-F}
    \widehat F(e_\perp)={1\over 4\pi}\int_0^\infty  \frac{d\omega\,\omega\,(4{(1-y_1) y_1-(\vec n \vec e_\perp)^2})}{((1-y_1)\omega ^2+ y_1 -(\vec n \vec e_\perp)\,\omega)(y_1 \omega^2 + (1-y_1)+(\vec n \vec e_\perp)\,\omega)}\,.
\end{align}
This function also depends on the angle $y_1=(1-\vec n\vec n_1)/2$ between the momenta of the incoming and outgoing particles, as well as on the angle between the transverse momentum of the soft graviton $\vec k_\perp$ and the incoming particles. %

If the transverse momentum is aligned with the recoil vector \eqref{recoil},  $k_\perp \sim \ell_\perp$,
the function \eqref{tilde-F} becomes closely related to the function $f(y_1,\beta)$ defined in \eqref{f-def}. In this case, for $\vec e_\perp=\vec \ell_\perp/|\vec \ell_\perp|$ and $\vec \ell_\perp^2=1-z$, %
\begin{align}
    (\vec n \vec e_\perp) = {(\vec n \vec \ell_\perp)\over |\vec \ell_\perp|} ={1-y_1-y_2\over \sqrt{1-z}} = 2\sqrt{(1-y_1)y_1}\cos\beta + O(\sqrt{1-z})\,.
\end{align}
We find, up to terms vanishing in the limit $z\to1$, 
\begin{align}\label{F-spec}
    \widehat F(\ell_\perp/|\vec \ell_\perp|) = {1\over \pi} f(y_1, \beta)  (y_1(1-y_1))^2\,.
\end{align}

We can apply the relations \eqref{ds1}, \eqref{M1} and \eqref{F-Fourier} to obtain the all-order resummed expression for the energy correlators in the back-to-back limit $z\to 1$ %
\begin{align}\label{EE-resum}
   \vev{\mathcal E(\vec n_1) \mathcal E(\vec n_2)} 
    = H^2(E) \int d^2 x_\perp \, e^{-2iE(\vec x_\perp \vec \ell_\perp) }\,
    \exp\lr{\int {d^2k_\perp\over (2\pi)^2}\lr{e^{i\vec k_\perp \vec x_\perp}-1} {2(\kappa E)^2\over k_\perp^2}\widehat F(e_\perp)}\,,
\end{align}
 where $\vec \ell_\perp$ is the recoil vector \eqref{recoil} and $\vec e_\perp =\vec k_\perp/|k_\perp|$ is a unit vector. In this relation, the hard function takes into account the  hard-particles contribution
and it is given by \re{H}. 
 
Let us show that, to the lowest order in the coupling, the relation \re{EE-resum} is in agreement with  \eqref{z-as}.
Expanding the integrand of \re{EE-resum} in  powers of  $\kappa^2$, we find that the integral over $k_\perp$ is localized at $\vec k_\perp=2E\vec\ell_\perp$, leading to
\begin{align} \label{1loop}
   \vev{\mathcal E(\vec n_1) \mathcal E(\vec n_2)} = \kappa^2  {H^2(E)\over 1-z}\widehat F(\ell_\perp/|\vec \ell_\perp|)+O(\kappa^4)\,.
\end{align}
Taking into account \eqref{F-spec}, we correctly reproduce \eqref{z-as}.

 The integral in the exponent in  \eqref{EE-resum} converges at small $k_\perp$ and is infrared finite. At the same time, it develops a logarithmic divergence at large $k_\perp$. As explained above, the latter is an artifact of the eikonal approximation and can be treated by dimensional regularization.  
 Setting $\vec k_\perp=\rho \,\vec e_\perp$ (with $\vec e_\perp^{\,2}=1$) 
 and $d^{2-2\epsilon}k_\perp = \rho^{1-2\epsilon}d\rho\,d \vec e_\perp$, we get
\begin{align}  \label{arte} \notag
 {\kappa^2\over 4} F(x_\perp) {}&= \mu^{2\epsilon}\int {d^{2-2\epsilon} k_\perp\over (2\pi)^{2-2\epsilon}}\lr{e^{i\vec k_\perp \vec x_\perp}-1} {2(\kappa E)^2\over k_\perp^2}\widehat F(e_\perp)  
\\
{}& = 2(\kappa E)^2 \int {d \vec e_\perp \over (2\pi)^{2-2\epsilon}}\widehat F(e_\perp) \Gamma(-2\epsilon) (-i(\vec e_\perp \vec x_\perp)\mu )^{2\epsilon} \,.
\end{align}
Expanding this expression as $\epsilon \to 0$ we find that, as a function of the UV cutoff, it satisfies the evolution equation 
\begin{align}\label{F-eq}
\mu {\partial\over\partial \mu}\lr{{\kappa^2\over 4}F(x_\perp)} = -2(\kappa E)^2\gamma(y_1) +O(\epsilon) \,,
\end{align}
where the function $\gamma(y_1)$ is given by the integral of the function \re{tilde-F} over the unit vector $\vec e_\perp$, located in the two-dimensional plane orthogonal to $\vec n_1$,
\begin{align}\label{gamma1}
\gamma(y_1)=\int {d \vec e_\perp \over (2\pi)^{2}}\widehat F(e_\perp)\,. 
\end{align}

Replacing the function $\widehat F(e_\perp)$ with its expression \re{tilde-F} and taking into account the identities 
\begin{align}\label{aux}
    (\vec n \vec e_\perp)=(\vec n_\perp \vec e_\perp) = |\vec n_\perp|\cos\chi = 2\sqrt{y_1(1-y_1)}\cos\chi\,,
    \qquad d\vec e_\perp=d\chi \,,
\end{align}
where $\vec n_\perp=\vec n-\vec n_1 (\vec n\vec n_1)$ and $0\le \chi \le 2\pi$ is the angle between the vectors $\vec n_\perp$ and $\vec e_\perp$, we find after some algebra
\begin{align}\label{gamma}  
\gamma(y_1)= -{1 \over (2\pi)^2} \left( y_1 \log y_1 + (1-y_1) \log (1-y_1) \right).
\end{align} 
This function takes positive values within the physical region $0< y_1<1$ and vanishes at the end points. 
In Appendix~\ref{App:cusp} we show  that the function \re{gamma} is closely related to the gravitational cusp anomalous dimension.

Note that the ultraviolet divergent part of \re{arte} is independent of $x_\perp$ and can be absorbed into the hard function $H(E)$. The resulting renormalized hard function acquires a dependence on the renormalization scale $\mu$.
Keeping only the  finite, $\vec x_\perp$ dependent part of \eqref{arte}, we get from \eqref{EE-resum}
\begin{align}\label{rem}
    \vev{\mathcal E(\vec n_1) \mathcal E(\vec n_2)} 
    = H^2(E,\mu) \int d^2 x_\perp \, e^{-2iE(\vec x_\perp \vec \ell_\perp) }\,
    \exp\lr{- 2(\kappa E)^2   \int {d \vec e_\perp \over (2\pi)^{2}}\widehat F(e_\perp)\log((\vec e_\perp\vec x_\perp)\mu) }\,.
\end{align} 
To understand the behavior for $z\to 1$, we replace $x_\perp \to x_\perp/\sqrt{(2E)^2\ell_\perp^2}$ where $\ell_\perp^2=1-z$. We obtain
\begin{align}\label{EE-fin}
    \vev{\mathcal E(\vec n_1) \mathcal E(\vec n_2)} 
    ={1\over 1-z}\exp\lr{ (\kappa E)^2\gamma(y_1)\log(1-z)} C(y_1,\beta)\,,
\end{align}
where  $\gamma(y_1)$ is defined in \re{gamma1}. %
The function $C(y_1,\beta)$ is independent of $z$ and the renormalization scale $\mu$. It is given by the product of the hard function $H^2(E,\mu)$ and the remaining  $z-$independent part of the integral \eqref{rem}.

\subsection{Energy correlators in the back-to-back limit}

The relation~\eqref{EE-fin} describes the asymptotic behavior of the energy correlator in the back-to-back region and is valid up to corrections that vanish as $z \to 1$. It is convenient to rewrite it in the equivalent form
\begin{align}\label{EE-fin1}
    \big\langle \mathcal E(\vec n_1)\, \mathcal E(\vec n_2) \big\rangle
    = \frac{C(y_1,\beta)}{(1-z)^{\,1- (\kappa E)^2 \gamma(y_1)}} \, .
\end{align}
Away from the forward limit, for $0<y_1<1$, the positivity of the function $\gamma(y_1)$ ensures that the integral 
$\int_{1-\delta}^1 dz\, \langle \mathcal E(\vec n_1)\mathcal E(\vec n_2)\rangle$
over the end-point region $0\le 1-z\le \delta$ is convergent, or equivalently the total energy deposited in the back-to-back region remains finite. 

We emphasize that the relation~\eqref{EE-fin1} holds in any gravitational theory. The function $\gamma(y_1)$, given in~\re{gamma}, captures the contribution of soft graviton radiation in the back-to-back limit $z\to 1$. Expanding~\eqref{EE-fin1} in powers of $\gamma(y_1)$ produces corrections enhanced by powers of $(\kappa E)^2\log(1-z)$.

The coefficient function $C(y_1,\beta)$ receives contributions from virtual particles and depends on the matter content of the theory. It admits an expansion in powers of $(\kappa E)^2$, with coefficients that depend on the angular variables. By comparing~\eqref{EE-fin1} with the one-loop result~\re{EEC3ptGRz1f}, we can determine $C(y_1,\beta)$ in $\mathcal N=8$ SG  to the lowest order in the coupling,  
\begin{align}\label{C-1loop}
C(y_1,\beta)
    = \frac{(\kappa E)^6}{256 \pi^5}\, E^2 f(y_1,\beta)
    + O(\kappa^8)\,,
\end{align}
where the function $f(y_1,\beta)$ is defined in~\re{fdef}.

Determining the $O(\kappa^8)$ correction to this coefficient requires computing the two-loop contribution to the energy correlator in the back-to-back region and matching the result to~\eqref{EE-fin1}. 
Even without performing this calculation, relation~\eqref{EE-fin1} allows us to predict all higher-loop contributions of the form $(\kappa E)^{6+2n}\log^{n}(1-z)$. In~\eqref{EE-fin1} these logarithmically enhanced terms arise solely from expanding the denominator in powers of $\gamma(y_1)$ and are therefore insensitive to higher-order corrections to the coefficient function $C(y_1,\beta)$.

Note that the denominator in~\eqref{EE-fin1} is independent of the angular variable $\beta$ defined in \eqref{beta}. Therefore, averaging both sides of~\eqref{EE-fin1} over~$\beta$ does not modify the leading $z\to 1$ behavior of the EEC,
\begin{align}\label{EE-beta0}
\big\langle \mathcal E(\vec n_1)\,\mathcal E(\vec n_2) \big\rangle_{\!\beta}
\equiv \frac{1}{\pi}\int_0^\pi d\beta \,
\big\langle \mathcal E(\vec n_1)\,\mathcal E(\vec n_2) \big\rangle \, .
\end{align}
This procedure corresponds to averaging over the direction of the outgoing particle $\vec n_2$ while keeping fixed the angle between $\vec n_1$ and $\vec n_2$.

Substituting~\eqref{EE-fin1} and~\eqref{C-1loop} into~\eqref{EE-beta0}, and using~\eqref{averfbeta0}, we obtain
\begin{align}\label{EE-beta}  
{}& 
\big\langle \mathcal E(\vec n_1)\,\mathcal E(\vec n_2) \big\rangle_{\!\beta}
= 
\frac{C(y_1)}{(1-z)^{ 1-(\kappa E)^2 \gamma(y_1)}} \, , \end{align}
where the notation was introduced for the coefficient function
\begin{align}\label{C1}
C(y_1)
= \frac{1}{\pi} \int_0^\pi d\beta\, C(y_1,\beta)
= \frac{E^2 (\kappa E)^6\,\gamma(y_1)}{128\pi^3 \,(y_1(1-y_1))^2}
\lr{1+C^{(1)}(y_1)(\kappa E)^2+ O(\kappa^4)}\,,
\end{align}
and the function $C^{(1)}(y_1)$ parameterizes the subleading correction.

Note that the resummed expression for the energy--energy correlators \re{EE-beta} is integrable at $z=1$ and, therefore, in distinction with the fixed order corrections \re{EEC3ptGRz1f} it does not require any contact terms to be well-defined. In particular, integrating \re{EE-beta} over the end-point region $1-\delta<z\le 1$ we expect to obtain a finite expression which should match the fixed order result.
\begin{align}\label{EE-int}
    \int_{1-\delta}^1 dz\, \big\langle \mathcal E(\vec n_1)\, \mathcal E(\vec n_2) \big\rangle_\beta = {E^2 (\kappa E)^4 e^{(\kappa E)^2 \gamma(y_1)\log\delta}\over 128 \pi^3(y_1(1-y_1))^2}\lr{1+C^{(1)}(y_1)(\kappa E)^2+ O(\kappa^4)}
\end{align}
We observe that the leading term on the right-hand side correctly reproduces the Born-level contribution \eqref{eq:treelevelEEC}.  At one loop, the integral in \re{EE-int} receives the contribution from the contact term \eqref{eq:anticollEEC}. This leads to the prediction for the coefficient $C^{(1)}(y_1)$ in \re{C1} in $\mathcal N=8$ SG
\begin{align}\label{C1-res}
    C^{(1)}(y_1) = {1 \over 2 \pi^2} \log y_1 \log (1-y_1) .
\end{align}

\subsection{Averaging over beam direction}

We can further average the energy correlators over the beam direction $\vec n$, which corresponds to integrating over $0 \le y_1 \le 1$. As discussed at the beginning of this section, the relations~\re{EE-fin1} and~\eqref{EE-beta} were derived under the assumption that the Mandelstam invariants $s_{ij} = 2 (p_i  q_j)$ are much larger than the invariant mass of the soft radiation, $(2E)^2(1-z)$. This condition restricts the allowed range of $y_1$. In particular, for $y_1 \to 0$ or $y_1 \to 1$, the outgoing particle momenta $q_1$ and $q_2$ become aligned with the incoming momenta $p_1$ and $p_2$, causing $s_{ij}$ to vanish and invalidating the above assumption.

Imposing $s_{ij} \gg (2E)^2(1-z)$ effectively amounts to restricting the integration to $1-z \ll y_1 \ll z$. Integrating \eqref{EE-beta}, we obtain %
\begin{align}\label{int-simp}
    \int_{1-z}^{\,z} dy_1 \, \big\langle \mathcal E(\vec n_1)\, \mathcal E(\vec n_2) \big\rangle_{\!\beta}
    \sim \int_{1-z}^{\,z} \frac{dy_1}{(y_1(1-y_1))^2} \, 
    \frac{E^2 (\kappa E)^6 \gamma(y_1)}{(1-z)^{1-(\kappa E)^2 \gamma(y_1)}} \lr{1+C^{(1)}(y_1)(\kappa E)^2+ O(\kappa^4)}\,.
\end{align}
In the limit $z \to 1$, the dominant contribution arises from the regions near the endpoints, $y_1 \sim z$ and $y_1 \sim 1-z$. Since $\gamma(y_1)$ vanishes at the endpoints, it can be safely neglected in the exponent of the $z$-dependent factor in the denominator of~\re{int-simp}. Consequently, the integral simplifies to
\begin{align}\notag\label{C-irr}
    \int_{1-z}^{\,z} dy_1 \, \big\langle \mathcal E(\vec n_1)\, \mathcal E(\vec n_2) \big\rangle_{\!\beta}
   {}& \sim \frac{E^2 (\kappa E)^6}{1-z} \int_{1-z}^{\,z} \frac{dy_1 \, \gamma(y_1)}{(y_1(1-y_1))^2}\lr{1+C^{(1)}(y_1)(\kappa E)^2+ O(\kappa^4)}
    \\
   {}& \sim \frac{E^2 (\kappa E)^6}{4\pi} \frac{\log^2(1-z)}{1-z} \lr{1+ O(\kappa^4)}\,.
\end{align}
This result is in agreement with the one-loop computation~\re{EECbar}.  Note that the coefficient function \re{C1-res} vanishes for $y_1\to 0$ and $y_1\to 1$, hence its contribution to \re{C-irr} is suppressed by a factor of $(1-z)$. 

\subsection{Large-$J$ limit}\label{sect:large-J}
 
We have previously observed that the generalized energy correlators ${\rm EC}_{J}(y)$ take a remarkably simple form at large $J$, where $J$ denotes the power of energy measured by the detector (see \re{eq:defecJ}).
Namely, the one-loop corrections in $\mathcal N=8$ SG and gravity, given by \re{eq:EJClargeJ} and \re{eq:EC+-largeJ}, respectively, have a factorized form
\begin{align}
    {\rm EC}^{(1)}_{J}(y) \sim -8 \gamma(y)\log (J)\times {\rm EC}^{(0)}_{J}(y)
\end{align}
where the cusp anomalous dimension $\gamma(y)$ is given by \re{gamma}. In this subsection, we elucidate the origin of this relation and generalize it to all loops.

The following analysis is very similar to the one for the structure functions of the deep inelastic scattering in the semi-inclusive limit $x\to 1$. As explained above, the correlator ${\rm EC}_{J}(y)$ is obtained by integrating the differential cross-section $d\sigma_{2\to q_1+X}$ with the weight factor $E_1^J \delta(\Omega_{\vec q_1}-\Omega_{\vec n_1})$. The energy of the detected particle can be written as $E_1=E-\omega$ where $0<\omega<E$. The key observation is that for $J\to \infty$ the dominant contribution to the integral over $E_1$ comes from the maximal value of $E_1$ or equivalently from small values of $\omega$. 

In this limit, we can replace the energy weight factor with
\begin{align}
    E_1^J = E^J (1-\omega/E)^J \sim E^J e^{-\omega J/E}\,.
\end{align}
This allows us to simplify the correlator ${\rm EC}_{J}(y)$ for $J\to\infty$ as 
\begin{align}\label{EC-J-int}
    {\rm EC}_{J}(y) = E^J \sum_X \int d\sigma_{2\to q_1+X} \, e^{-\omega J/E}
    \delta(\Omega_{\vec q_1}-\Omega_{\vec n_1})\,.
\end{align}
In the Born approximation, the final state $X$ consists of a single particle that moves back-to-back to $q_1$ and the differential cross section is proportional to $\delta(\omega)$. As a consequence, the tree level contribution ${\rm EC}^{(0)}_{J}(y)$ is independent of $J$ and is given by \re{EC-B1} and \re{EC-B2}. 

At loop order, the total invariant mass of the final state  $p_X^2=(p_1+p_2-q_1)^2 = 4E \omega$ vanishes for $\omega\to 0$. This suggests that for $J\to\infty$, the final state $X$ consists of a fast particle with momentum $q_2=E(1,-\vec n_1)$ accompanied by soft graviton radiation.\footnote{Due to the absence of collinear divergences in gravity, jet-like configurations do not produce the dominant contribution at $J\to\infty$.} This implies that in the large $J$ limit, the differential cross section in \re{EC-J-int} can be computed using the eikonal approximation \re{eik}. Repeating the above analysis we find that
\begin{align}\label{ds-x0} 
\sum_X \int d\sigma_{2\to q_1+X}  \delta(E-E_1-\tilde \omega)
    \delta(\Omega_{\vec q_1}-\Omega_{\vec n_1})
= H^2(E) \int_{-\infty}^\infty dx_0 \, e^{-ix_0 \tilde \omega} W(x_0)\,,
\end{align}
where the delta functions on the left-hand side fix the momentum of the detected particle. Here
the eikonal factor $W(x^0)$ is given  by \re{M} with the important difference that the
spatial transverse vector $(0,\vec x_\perp)$ is replaced with the time-like vector $(x^0,\vec 0)$. We recall that in the back-to-back region, the relation
\re{ds1} resums the contribution of the soft graviton radiation carrying the total transverse momentum $\vec k_{X,\perp}= 2E \vec \ell_\perp$. In the large $J$ limit, the relation \re{ds-x0} resums the contribution of the soft graviton radiation carrying the total energy $k_{X,0}=\omega$. 

Combining the last two relations we find 
\begin{align}\label{EC-pre}
{\rm EC}_J(y) = E^J  H^2(E)W(x_0= iJ/E)\,,
\end{align}
where the function $W(x_0)$ is evaluated for a pure imaginary argument. We have seen that $W(x_\perp)$ is free from infrared divergences but it has ultraviolet divergences. The same is true for the function $W(x_0)$. Repeating the calculation of $W(x_0)$ and absorbing its UV divergences into the renormalized hard function, we find
\begin{align}
W(x_0)= \exp\lr{- \frac12(\kappa E)^2 \gamma(y) \log (-E^2 x_0^2)}
\end{align}
where $y=y_1$.
This expression can be obtained from the virtual corrections by replacing the IR cutoff with $-1/x_0^2$. Substituting this relation in \re{EC-pre} we obtain
\begin{align} \label{eq:large-J-all-loops}
    {\rm EC}_J(y)\sim E^J \exp\lr{-(\kappa E)^2 \gamma(y) \log J} \,.
\end{align}
Expanding this relation in powers of $\kappa E$ we reproduce the one-loop results \re{eq:EJClargeJ} and \re{eq:EC+-largeJ}.

\section{Discussion}

Here we discuss some additional aspects of gravitational energy correlators that go beyond the scope of the main text, and we list a few possible open directions.

\subsection{Initial state singularity}

In the paper, we considered an initial state consisting of a pair of gravitons with definite momenta. Such a plane-wave initial state is known to yield an infinite total cross section, which in turn makes the one-point energy correlator non-integrable over the celestial sphere.
A plane-wave state is not physical by itself, since it has infinite norm. Rather, we view it as an idealization of a normalizable wave packet, which we can take to be
\be
\label{eq:statewavepacketUndressed}
| \psi \rangle = \int {d^3 \vec p_1 \over (2 \pi)^{3} (2 |\vec p_1|)} {d^3 \vec p_2 \over (2 \pi)^3 (2 |\vec p_2|)} \psi(\vec p_1, \vec p_2) a_{h_1}^\dagger(\vec p_1) a_{h_2}^\dagger(\vec p_2) | 0 \rangle ,
\ee
where $[a_h(\vec p), a_{h'}^\dagger(\vec q)] = \delta_{h,h'} (2 \pi)^3 2|\vec p| \delta^3(\vec p' - \vec q)$ are the graviton annihilation/creation operators and $h_i$ stands for helicity. The wave funciton $\psi(\vec p_1, \vec p_2)$ characterizes the shape of the wave packets for the incoming gravitons.

We can then normalize this state, $\langle \psi | \psi \rangle =1$, i.e.
\be
 \int {d^3 \vec p_1 \over (2 \pi)^{3} (2 |\vec p_1|)} {d^3 \vec p_2 \over (2 \pi)^3 (2 |\vec p_2|)} |\psi(\vec p_1, \vec p_2)|^2 = 1 \ .
\ee
Assuming that $\psi(\vec p_1, \vec p_2)$ is peaked around certain values, we can approximate the narrow wave packets by plane waves. We might think of our calculation as capturing correctly the physics of the narrow wave packets. However, as we review below, this simple intuition is not correct.

The reason is that the normalizability of the state \eqref{eq:statewavepacketUndressed} depends on the interference between states with different ingoing momenta $\vec p_1$ and $\vec p_2$. For simplicity, we can restrict the discussion to the center-of-mass frame and set $\vec p_1 + \vec p_2 = 0$. However, the BMS symmetry of gravitational scattering implies that there is no nontrivial interference between such states \cite{Carney:2018ygh}. 
The reason is that, for different momenta we have for the supertranslation charges,
\be
Q_{\text{BMS}}(\vec p_1 , -\vec p_1) \neq Q_{\text{BMS}}(\vec p_{1'}, -\vec p_{1'}) \ ,
\ee
which implies that in four dimensions any inclusive cross section $\langle \vec p_{1'} , -\vec p_{1'} | X \rangle \langle X | \vec p_1, - \vec p_1 \rangle$ is zero. It is indeed straightforward to generalize the analysis of Weinberg \cite{Weinberg:1965nx} to this case, to see that the IR divergences do not cancel among such non-diagonal initial states.

To rectify the problem, we can consider instead a family of dressed states, 
\begin{align}
\label{eq:statewavepacketMain}
| \psi \rangle_{\text{dressed}} &= W_{\text{in}}^\dagger \int {d^3 \vec k_1 \over (2 \pi)^{3} (2 |\vec k_1|)} {d^3 \vec k_2 \over (2 \pi)^3 (2 |\vec k_2|)} \psi(\vec k_1, \vec k_2) W(\vec k_1, \vec k_2)a_{h_1}^\dagger(\vec k_1) a_{h_2}^\dagger(\vec k_2) | 0 \rangle\, ,
\end{align}
where $W(\vec k_1, \vec k_2)$ stands for the Faddeev-Kulish gravitational Wilson line dressing. We have chosen a \emph{fixed} $W_{\text{in}} \equiv W(\vec p_1, \vec p_2)$, such that for a \emph{given} set of momenta $\vec k_1 = \vec p_1$ and $\vec k_2 = \vec p_2$, the dressing is absent. It has been argued in  
\cite{Carney:2018ygh}, and it is easy to see it explicitly by repeating the calculation of Section~\ref{sec:backtoback}, that such states do exhibit nontrivial interference. 

If we  choose the wave function $\psi(\vec k_1, \vec k_2)$ to be narrowly peaked around $(\vec p_1, -\vec p_1)$, we effectively return to the calculations performed in the main body of the paper.
Our proposal, therefore, is that the standard plane-wave calculation of collider observables in $4d$ gravity is a good approximation to the calculation with the \emph{dressed} narrow wave packets away from the forward peak. 

Near the forward peak, we expect the plane-wave approximation to break down, and the details of how normalized states are defined to become important, rendering the energy correlators integrable over the celestial sphere. We have not carried out a nonperturbative resolution of the forward peaks in this paper,\footnote{Close to the forward limit, an eikonal resummation of the amplitude is necessary; see e.g. \cite{tHooft:1987vrq} and the related discussion in \cite{Lippstreu:2025jit}.} and we leave this interesting problem for future work.

\subsection{Extra scales}

In the paper, we restricted our analysis to the case in which both the initial and final states are massless. We also examined the effects of stringy modes on the low-energy energy correlators. It would be interesting, however, to study the situations in which physical scales are present more broadly and to understand their imprint on the energy correlators.\footnote{In a cosmological context, this has been explored in \cite{Arkani-Hamed:2015bza}.}

Perhaps the simplest situation arises when the energy of the state crosses a physical production threshold. A natural setting for this is gravity coupled to matter, e.g. the Standard Model. In that case, once the center-of-mass energy satisfies $s > (m_a + m_{\bar a})^2$, a new production channel should open, and its onset should leave a visible imprint on the energy correlators.\footnote{Here $a$ could be a neutrino, an electron, or any other stable particle.}
In the context of the present paper, if we view ${\cal N}=8$ SG as the low-energy limit of string theory compactified on $T^6$, there are two obvious physical scales: the string scale and the KK scale. In this work, we focus on energies below both thresholds, so that neither string nor KK modes are produced in the final state.
It would be very interesting to quantify how these scales affect the energy correlator once these channels become accessible.

More broadly, it is interesting to ask what happens as we increase the energy in a gravitational collider experiment. As discussed in \cite{Amati:1987wq,Amati:1987uf,Amati:1988tn}, see also \cite{Giddings:2009gj,Haring:2022cyf}, we expect gravitational nonlinearities to become more important. This leads, in particular, to the expectation that gravitational radiation becomes relevant and eventually a black hole is formed in the collision, see Figure~\ref{eq:eventshapekinematics}. In Appendix \ref{app:BHD}, we discuss the asymptotic ($s \to \infty$) form of the gravitational energy correlators, assuming that they are dominated by black hole production, and we conclude that the resulting distribution is homogeneous on the celestial sphere. 

\subsection{Bootstrap}

It is natural to ask whether constraints on gravitational theories can be derived from the consistency of energy correlators, which must be nonnegative for all angles in all states and satisfy the energy-momentum conservation Ward identities.
A less obvious constraint is the associativity of the OPE between the gravitational detectors.

In the context of AdS/CFT, the leading stringy correction to the strong-coupling result computed in \cite{Hofman:2008ar} directly probes the first higher-derivative correction to the AdS gravitational effective action and must be sign-definite because of the energy correlators multipole positivity \cite{Fox:1978vu}, see also \cite{Mecaj:2025ecl,Dempsey:2025yiv}.

In flat space the situation appears to be more complicated. For example, in the context of the present paper, we could aim to constrain the sign of the leading stringy correction to the energy–energy correlator.
This correction is related to the first higher-derivative correction to the gravitational effective action in flat space. If we could find a state $|\psi \rangle$ for which the leading supergravity contribution to the energy correlator is zero, the constraint would follow. 

Regarding the OPE of detector operators, we found that it takes the simple form \eqref{eq:OPE}. It would be very interesting to determine whether this persists at higher orders in perturbation theory, or even at finite coupling, and whether such a simple OPE structure imposes nontrivial constraints on the three- and higher-point energy correlators in gravity, see \cite{MuratKologluSeminar2024}.

\section*{Acknowledgments}
We thank Daniel Carney, Alexandre Fouquet, Murat Kologlu and Ian Moult for helpful discussions. AZ is grateful to the \emph{Simons Collaboration on Celestial Holography} for its hospitality and for the opportunity to present and discuss the results of this work at the 2025 Annual Meeting. This project has received funding from the European Research Council (ERC) under the European Union’s Horizon 2020 research and innovation program (grant agreement number 949077). The work of DC and GK was supported
by the French National Agency for Research grant ``Observables'' (ANR-24-CE31-7996).

\appendix

\section{Phase space integrals and real emission corrections} 
\label{app:PSint}

In this Appendix we specify the phase space integrals that arise in the calculation of the generalized energy correlators at one loop.

The phase space measure for the $L$-particle final state of the process $p_1 + p_2 \to q_1+ q_2 + \ldots + q_L$, takes the usual form in dimensional regularization with $d=4-2\epsilon$ and $\epsilon <0$:
\begin{align}
d\text{PS}_{L} = (2\pi)^d \delta^d \left(p_1+p_2-\sum_{i=1}^{L} q_i\right)\prod_{i=1}^{L} \frac{d^d q_i}{(2\pi)^{d-1}} \delta_+(q_i^2)\,. \label{eq:PS}   
\end{align}

\subsection*{One-point energy correlators}

Let us start with the one-point correlator, and consider its perturbative expansion \p{eq:ECpertexp},
\begin{align}
\text{EC}_{J} &= E^J \left({\kappa E \over 2} \right)^4  \left(
\text{EC}_{J}^{(0)} + \left({\kappa E \over 2} \right)^{2} \text{EC}_{J}^{(1)} + \ldots \right) . \label{eq:ECJpert}
\end{align}
At LO, the EC receives only a two-particle tree-level contribution,
\begin{align}
\text{EC}_{J}^{(0)} = \frac{E^{-2\epsilon}}{8(2\pi)^{2-2\epsilon}} \mathbb{M}^{(0)}_{2 \to 2} (q_1,q_2)\biggr|_{\begin{subarray}{l} q_1 = E(1,\vec n_1) \\ q_2 = E(1,-\vec n_1) \end{subarray}} \,.
\label{eq:EJC2}
\end{align}
The NLO correction is a sum 
the virtual and real corrections 
\begin{align}\label{B.4}
\text{EC}_{J}^{(1)} = 
\text{EC}_{J}^{\text{virt}} + \text{EC}_{J}^{\text{real}} \,,
\end{align}
which are given by the integrals over the two- and three-particle 
phase spaces, respectively,  
\begin{align}   
& \text{EC}_{J}^{\text{virt}} \equiv \frac{1}{E^{6+J}}\int\limits_{q_1,q_2} d\text{PS}_{2}(q_1,q_2) \, 
\mathbb{M}_{2 \to 2}^{(1)}(q_1,q_2)\,E_{1}^J\,\delta(\Omega_{\vec{q}_{1}}-\Omega_{\vec{n}_{1}})\,, \label{B.5} \\  
& \text{EC}_{J}^{\text{real}} \equiv \frac{1}{2!}\frac{1}{ E^{6+J}}\int\limits_{q_1,q_2,q_3} d\text{PS}_{3}(q_1,q_2,q_3)\, 
\mathbb{M}_{2 \to 3}^{(0)}(q_1,q_2,q_3)\,E_{1}^J\,\delta(\Omega_{\vec{q}_{1}}-\Omega_{\vec{n}_{1}})\,. \label{eq:ECreal}
\end{align}
The symmetry factor \(1/2!\) in the last relation arises as follows. The amplitude \(\mathbb{M}^{(0)}_{2\to 3}\) involves three identical particles in the final state, which yields the usual factor \(1/3!\)  This is compensated by a factor of \(3\), reflecting the fact that any one of the three particles can be detected by the calorimeter. Let us note that for an $L$-particle final state with $\mathbb{M}^{(0)}_{2\to L}$ in \eqref{eq:ECreal}, the symmetry factor would be $\frac{1}{(L-1)!}$.

Both the virtual and real contributions are IR divergent. The integral over the two-particle phase space in \re{B.5} is localized by the calorimeter angular delta function together with the momentum conservation. As a result, the two-particle contribution takes the following form:
\begin{align}
\text{EC}_{J}^{\text{virt}} = \frac{E^{-2\epsilon}}{8(2\pi)^{2-2\epsilon}} \mathbb{M}^{(1)}_{2 \to 2} (q_1,q_2)\biggr|_{\begin{subarray}{l} q_1 = E(1,\vec n_1) \\ q_2 = E(1,-\vec n_1) \end{subarray}} \,.
\label{eq:EJC2virt}
\end{align}
In contrast, the three-particle phase-space integration in the real contribution is nontrivial. After taking momentum conservation into account, there remain the integrations over a solid angle and an energy fraction; we do it following \cite{Bork:2009nc}. The idea is to split 
$d{\rm PS}_3 = d{\rm PS}_1 \times d{\rm PS}_2$ by relabeling the final state momenta, $p_1 + p_2 \to q_1 +k + (q-k)$, where $q_1^\mu = xE(1,\vec n_1)$, and by first doing the integration over $d{\rm PS}_2(k,q-k)$. We have
\begin{align}
& \text{EC}_{J}^{\text{real}}= \frac{1}{4}(2\pi)^{-2d+3} E^{-2\ep} \int^1_0 dx \, x^{J+1-2\ep} \, d\sigma(x,y) \,,  \label{splitPS} \\
&  d\sigma(x,y) \equiv \frac{1}{E^4}\int d^d q d^d k\, \delta(k^2) \delta((q-k)^2) \delta^{d}(q_1+q-p_1-p_2) \, \mathbb{M}^{(0)}_{2\to 3}(q_1,k,q-k) \, , \notag
\end{align} 
where $y \equiv y_1$ is the angular variable \p{ys}. After performing a partial-fraction decomposition of the squared matrix element $ \mathbb{M}^{(0)}_{2 \to 3}$, we rewrite the phase-space integral as a linear combination of standard van Neerven integrals \cite{vanNeerven:1985xr} with rational coefficients $a_{i,j},\, b_{i,j}$,
\begin{align}
d\sigma(x,y)  =  \sum_{i \neq j} \left[ a_{i,j}(x,y) \int d^d k \frac{\delta(k^2) \delta((q-k)^2)}{(k l_i)(k l_j)} + b_{i,j}(x,y) \int d^d k \frac{\delta(k^2) \delta((q-k)^2)}{(k l_i)((q-k)l_j)} \right] \,.
\end{align}
Here we use a uniform notation for the momenta, $(l_1,l_2,l_3)\equiv(p_1,p_2,q_1)$. In this way,  $d\sigma(x,y)$ evaluates in terms of the hypergeometric functions ${}_2 F_1$. After doing the remaining integral over the energy fraction $x$ in \p{splitPS}, we extract the IR pole $1/\ep$ originaing from the region $x\sim 1$, which corresponds to the emission of a soft graviton.

\subsection*{Two-point energy correlators}

Let us consider the EEC defined in  \p{eq:ECpertexp},
\begin{align} \label{eq:EECJ1J2pert}
\text{EEC}_{J_1,J_2} &= E^{J_1+J_2} \left({\kappa E \over 2} \right)^4  \left(
\text{EEC}_{J_1,J_2}^{(0)} + \left({\kappa E \over 2} \right)^{2} \text{EEC}_{J_1,J_2}^{(1)} + \ldots \right) .
\end{align}
At LO, there is only a two-particle contribution, and the phase-space integration is trivial. Taking into account that $\vec{q}_1 + \vec{q}_2 = 0$, see \eqref{eq:EJC2}, we obtain
\begin{align}
\text{EEC}_{J_1,J_2}^{(0)} = 
\left( \delta(\Omega_{\vec{n}_{1}}+\Omega_{\vec{n}_{2}}) + \delta(\Omega_{\vec{n}_{1}}-\Omega_{\vec{n}_{2}}) \right) \text{EC}_{J_1+J_2}^{(0)} \,. \label{eq:EEC_LO}
\end{align}
The NLO correction is a sum of several terms,
\begin{align}
\text{EEC}_{J_1,J_2}^{(1)} =  \delta(\Omega_{\vec{n}_{1}}+\Omega_{\vec{n}_{2}}) \text{EC}_{J_1+J_2}^{\text{virt}}   + \text{EEC}_{J_1,J_2}^{\text{real}} + \delta(\Omega_{\vec{n}_{1}}-\Omega_{\vec{n}_{2}})\text{EC}_{J_1+J_2}^{(1)} \,. \label{eq:EEC_NLO}
\end{align}
The last term is the diagonal contribution, which is equivalent to the NLO correction to the EC. 
The first two terms are off-diagonal, and they correspond to a virtual and a real contributions. The two-particle virtual contribution is given by \eqref{eq:EJC2virt}. The off-diagonal three-particle real contribution is
\begin{align}
\text{EEC}_{J_1,J_2}^{\text{real}} \equiv  \frac{1}{E^{J_1+J_2+6}} \int\limits_{q_1,q_2,q_3}  d\text{PS}_{3}\, 
\mathbb{M}_{2 \to 3}^{(0)}   E_{1}^{J_1}E_{2}^{J_2}\,
\delta(\Omega_{\vec{q}_{1}}-\Omega_{\vec{n}_{1}})
\delta(\Omega_{\vec{q}_{2}}-\Omega_{\vec{n}_{2}}) \,. \label{eq:EECreal}
\end{align}
For a final state involving $\mathbb{M}_{2 \to L}^{(0)}$, the symmetry factor is ${1 \over (L-2)!}$. In the formula above we have $L=3$, so this factor is trivial. It arises as follows. We have the usual factor ${1 \over L!}$ for $L$ identical particles in the final state of $\mathbb{M}_{2\to L}$, and we have an additional factor $L(L-1)$, which is the number of different ordered pairs of particles detected by two calorimeters.

If we rewrite the latter equation in terms of the calorimeter variables, see \p{M5simpl}, the real contribution takes the form of a univariate integral over the energy fraction,
\begin{align}
\text{EEC}_{J_1,J_2}^{\text{real}}(z,y_1,y_2)  = \frac{E^{-4-4\ep}}{16(2\pi)^{5-4\ep}} \int\limits^1_0 dx \, \frac{x^{J_1+1-2\ep}(1-x)^{J_2+1-2\ep}}{(1-zx)^{J_2+2-2\ep}} \mathbb{M}_{2 \to 3}^{(0)}(x,z,y_1,y_2)   \,.\label{eq:EECintx}
\end{align}
The latter is finite for $0<z<1$, so in this case we can set $\ep =0$. However, the integral diverges as $z \to 1$ that corresponds to a $1/\ep$ pole in the back-to-back contact term, see Appendix~\ref{Gelfand}.  Thus, the real contribution is finite for a generic configuration of the calorimeters and it contains an IR-divergent back-to-back contact term $\sim \delta(\Omega_{\vec{n}_{1}}+\Omega_{\vec{n}_{2}}) $, which is expected to cancel the IR divergence of the virtual contribution.

\section{Energy correlators with arbitrary energy weight}
\label{app:EJC}

We find convenient to combine  the NLO corrections to   $\text{EC}_{J}^{(1)}$ with $J \geq 1$ (see \p{eq:ECJpert}) in ${\cal N}=8$ SG, into a generating function,  
\begin{align}
&\sum_{J \geq 0} t^J \text{EC}_{J}^{(1)} (y) \notag  \\
& = \frac{1}{2 \pi^4} \biggl\{ \frac{t}{(y-\bar{y})y(\bar{y}-t)} \biggl( \bar{y} \log^2(y) + \left( -\bar{y}+t y + (1-t)\bar{y}^2(1+\bar{y}) \right) \frac{2{\rm Li}_2(y) }{(1-t)\bar{y}^2} \biggr)
\notag\\
& -\frac{t}{(y-\bar{y})\bar{y}(y-t)} \biggl( y \log^2(\bar{y}) + \left(-y+t\bar{y} + (1-t)y^2(1+y)\right) \frac{2  {\rm Li}_2(\bar{y}) }{(1-t)y^2} \biggr) 
\notag\\
& + \frac{t^3}{(1-t)y \bar{y} (y-t)(\bar{y}-t)} \biggl( -2 \log(1-t) ( y \log(y) + \bar{y} \log(\bar{y})) +(1-t)\log^2(1-t)  \notag\\
& -2 t {\rm Li}_2(t) + \frac{\pi^2}{3 t^2} \left( - \frac{t(1-t)}{y(1-y)} + (1-t)^3 + t (1+t)+ 2 (1-t) y(1-y) \right) \biggr) \biggr\} \label{eq:GenFunEJC}\ ,
\end{align} 
where $t$ is an auxiliary parameter and $\bar{y} \equiv 1- y$. The crossing symmetry relation
\begin{align}
 \text{EC}_{J}^{(1)} (y) =  \text{EC}_{J}^{(1)} (1-y)
\end{align} 
 is manifest. Series-expanding the previous equation in powers of $t$, we recover $\text{EC}^{(1)}$ given in Eq.~\eqref{first corr}, and $\text{EC}_{J=2}^{(1)}$ from Eq.~\eqref{eq:E2C1L}. Both expressions are of maximal transcendentality  two. For $J \geq 4$, the lower transcendentality terms $\log(y)$ and $\log(1-y)$ appear; for $J \geq 5$ also  rational terms are present.

The generating function allows us to calculate the collinear beam-calorimeter limit $y\to 0$,
\begin{align}
\label{eq:EJCasypmy0}
\sum_{J \geq 0} t^J \text{EC}_{J}^{(1)} (y) & = \frac{1}{2 \pi^4} \frac{1}{y} \biggl( -\frac{t}{1-t}\log^2(y)  -\frac{2t}{1-t}\log(y)  \\ 
&+ \frac{\pi^2}{3} \frac{t(2-4t+t^2)}{(1-t)^2}
+  \frac{t}{1-t} (2-t\log^2(1-t)) +\frac{2t^3 {\rm Li}_2(t)}{(1-t)^2} \biggr) + O(\log^2(y))\,, \notag
\end{align}
which generalizes  \eqref{eq:ECy0}. We also use the generating function \eqref{eq:GenFunEJC} to calculate the asymptotics of $\text{EC}_{J}^{(1)}$ at large $J$. Indeed, it behaves as $O\left( \frac{\log(1-t)}{(1-t)}\right)$ at $t \sim 1$, implying a logarithmic $\log(J)$  asymptotics, see  \eqref{eq:EJClargeJ}.

In the ancillary files we also provide the generating function of NLO energy correlators in pure gravity.

\section{Origin of the contact term  $\delta(1-z)$}
\label{Gelfand}

In this Appendix, we explain the origin and normalization of the contact term
$\delta(1-z)$ appearing in \eqref{contact}. The discussion is purely
geometric and is related to a degeneracy of the angular variables in the back-to-back limit.

We parametrize the relative geometry of the unit vectors
$\vec n,\vec n_1,\vec n_2$ by the variables  $y_1,z$ (see \eqref{ys}),
together with the angle $\beta\in[0,\pi]$ between the unoriented planes
spanned by $(\vec n_1,\vec n_2)$ and $(\vec n,\vec n_1)$, defined in
\eqref{beta}.    Keeping the vectors $\vec n$ and $\vec n_1$ fixed in a non-(anti)collinear configuration ($0<y_1<1$), the third vector $\vec n_2$ sweeps a $(2-2\epsilon)$-dimensional unit sphere $S^{2-2\epsilon}_{\vec n_2}$.  So, we fix  $y_1$  and parametrize the vector $\vec n_2$ by the angles
$(\theta,\beta)$, where $\cos\theta=1-2z$ and $\theta\in[0,\pi]$, see Figure~\ref{fig:angles-sphere}. Thus, $(\theta,\beta)$ may be viewed as the polar and azimuthal coordinates on the sphere.  At the poles $\theta=0,\pi$
(equivalently $z=0,1$), the azimuthal angle $\beta$ becomes arbitrary.

The solid-angle measure may be written as (with $\epsilon<0$)
\begin{equation}
\int d\Omega
=
\frac{2^{2-2\epsilon}\pi^{\frac12-\epsilon}}{\Gamma(\frac12-\epsilon)}
\int_0^\pi d\beta\,(\sin\beta)^{-2\epsilon}
\int_0^1 dz\,(z(1-z))^{-\epsilon} \, .
\label{eq:OmegaMeasureApp}
\end{equation}
In the  limit $\epsilon\to0$ this convention implies
\begin{equation}
\left.
\int d\Omega \right|_{\epsilon=0}=4\int_0^1 dz\int_0^\pi  d\beta=4\pi\,.
\end{equation}

We now consider the singular distribution $f(\beta)/(1-z)$ and extract its contact term by integrating against a smooth test function
$\varphi(z,\beta)$,
\begin{equation}
\bigl(f(\beta)(1-z)^{-1},\varphi\bigr)
=
\int d\Omega\,
\frac{f(\beta)}{1-z}\,\varphi(z,\beta) \, .
\end{equation}
Smoothness on the sphere implies that, at the pole $z=1$, the test function
becomes independent of the degenerate angle,
$\varphi(z,\beta)\to\varphi(1) \equiv {\rm const}$. Next, we write the identity
\begin{align}
\bigl(f(\beta)(1-z)^{-1},\varphi\bigr)
&=
\int d\Omega\,
\bigl[\varphi(z,\beta)-\varphi(1)\bigr]\frac{f(\beta)}{1-z}
+\int d\Omega\,\frac{f(\beta)}{1-z}\, \varphi(1) \,.
\label{eq:splitApp}
\end{align}

The first term on the right-hand side  is finite as $\epsilon\to0$ and defines the plus distribution
$f(\beta)/(1-z)_+$ on the sphere. In the second term we can do the $z-$integral,
\begin{align}
&\int d\Omega\,\frac{f(\beta)}{1-z}\, \varphi(1) =
\frac{4 \pi ^{1-\epsilon } \Gamma (-\epsilon )}{\Gamma \left(\frac{1}{2}-\epsilon \right)^2} \int_0^\pi d\beta\, (\sin\beta)^{-2\epsilon}f(\beta) \, \varphi(1) \,. 
\end{align}
Further, we denote by $\Omega_0=(1,\beta)$ the pole of the sphere corresponding to $\theta=\pi$ and we write (see \re{omega})
\begin{align}
&\varphi(1) =\int d\Omega\, \delta(\Omega-\Omega_0)\,  \varphi(\Omega) = \frac1{4\pi}  \int_0^1 \delta(1-z)  \varphi(z,\beta) \,.
\end{align}
Expanding in $\epsilon$ and removing the test function $\varphi(z,\beta)$, we obtain a
distributional identity on the $(2-2\epsilon)$-dimensional unit sphere,
\begin{align}
\left.\frac{f(\beta)}{1-z}\right|_{S^{2-2\epsilon}}
&=
\delta(1-z)
\int_0^\pi \frac{d\beta}{\pi}\,f(\beta)\, 
\left[
-\frac{1}{\epsilon}
+\gamma_E
+\log(4\pi)
+2\log\!\bigl(2\sin\beta\bigr)
\right]
\nonumber\\
&\quad
+\frac{f(\beta)}{(1-z)_+}
+O(\epsilon) \,,
\label{eq:distribSphereFinalApp}
\end{align}
which coincides with the $\ep-$expanded version of Eq.~\eqref{contact}.
We have thus shown that the contact term $\delta(1-z)$ originates from the spherical pole at
$z=1$ and necessarily involves an averaging  over the degenerate angle $\beta$. 

We conclude with a comment on the origin of the contact term generated by the
singular distribution $(1-z)^{-1}$. A standard construction is to consider the
distribution $(1-z)^\lambda$, which is regular for $\Re\lambda>-1$, and to
analytically continue it in $\lambda\in\mathbb C$. This continuation develops a simple pole,
$(1-z)^\lambda \sim (1+\lambda)^{-1}\delta(1-z)$.
A  related phenomenon occurs when $\mathbb R^n$ is parametrized by
spherical coordinates $(r,\vec\omega_{n-1})$ \cite{GelfandShilov1964}. At the origin $r=0$, smooth test
functions satisfy $\varphi(r\vec\omega)=\varphi(0)$, and one finds
$r^\lambda \stackrel{\lambda \to -n}{\longrightarrow} {C_n \over  n +\lambda}\delta^n(x)$.
In our case, the regulator $\epsilon$ originates from the dimension of
the space parametrized by $(z,\beta)$ rather than from a deformation of the
power of $(1-z)^{-1}$. Nevertheless, the mechanism producing the contact term is
analogous: at $z=1$ the coordinate $\beta$ becomes degenerate, and smoothness on
the sphere enforces an averaging over this angle, in direct analogy with the
angular independence of test functions at $r=0$ in spherical coordinates.

\section{The square of the superamplitude summed over the final states}
\label{sec:SFStates}

In this Appendix, we explain how to compute the square of the five-point superamplitude ${\cal M}_{2\to 3}$, summed over all three-particle final states. The supermultiplet contains $2^{\cal N}$ helicity states, and we use on-shell superspace to carry out the sum.

\subsection*{R-symmetry $SU({\cal N})$} \label{app:SU_N}

Let us consider an $n$-point superamplitude in a theory with chiral supersymmetry charges, which transform under the fundamental irrep of the R-symmetry  $SU({\cal N})$. We introduce anticommuting variables $\eta^A_i$, where $A=1,\ldots,{\cal N}$ and $i=1,\ldots,n$, to parametrize the on-shell supermultiplets of scattered states.  There are $2 {\cal N}$  chiral (complex) supercharges
$Q^{\alpha A} = \sum_i \la^{\alpha}_i \eta^A_i$, where $\la^{\alpha}_i$ are the chiral helicity two-component spinor variables defined through the relation $p^{\alpha\dot\alpha}_i = \la^{\alpha}_i \tilde\la^{\dot\alpha}_i$. The supercharge conservation is imposed via a Grassmann delta function $\delta^{2\cal N}(Q)$.\footnote{The anti-chiral supercharges $\bar Q^{\dot\alpha}_A$ are realized as differential operators in $\eta$; we do not need them here. } The superamplitude involves $n-3$ bosonic functions of the kinematical variables. They are the coefficients in the sum of $n-3$ nilpotent supersymmetry invariants of degree $(2+k)\,{\cal N}$, with $k=0,\ldots,n-4$, which correspond to the N$^{k}$MHV helicity sectors. Since the ${\cal N}=8$ SG supermultiplet is self-conjugate, the N$^{k}$MHV and N$^{n-4-k}$MHV sectors are related by  charge conjugation.

In the five-point case, $n=5$, the superamplitude of the process $2 \to 3$ contains only an MHV and an NMHV sectors. In terms of the holomorphic odd variables  $\eta^A$, the superamplitude takes the following form: 
\begin{align}
{\cal A}_{2 \to 3}(\eta,\la,\tilde\la) = \delta^{2\cal N}(Q) A(\la,\tilde\la) + {\bf F}_{\eta}[\delta^{2\cal N}(\bar Q)] A^* (\la,\tilde\la) \,. \label{eq:A5super}
\end{align}
The bosonic functions $A(\la,\tilde\la)$ and $A^* (\la,\tilde\la)$ are related by complex conjugation, which swaps the chiral $\la$ and anti-chiral $\tilde\la$  Lorentz spinors. To represent the NMHV sector in terms of  $\eta^A$, we Fourier transform the anti-holomorphic variables $\bar\eta_{i\,A}$ of the conjugate $\overline{\rm MHV}$ sector,
\begin{align}
{\bf F}_{\eta}[\delta^{2\cal N}(\bar Q)] = \int d\bar \eta \, e^{\sum \eta \bar{\eta}} \delta^{2\cal N} (\bar Q) \,,\label{eq:Fourier}
\end{align}
where $\bar Q^{\dot\alpha}_{A} = \sum_i \tilde\la_i^{\dot\alpha} \bar\eta_{i\,A}$, and $d\bar \eta \equiv \prod_{i} d^{\cal N} \bar \eta_i$.

The charge-conjugate amplitude is naturally written in the anti-chiral variables,
\begin{align}
{\cal A}^{c}_{2 \to 3}(\bar \eta,\la,\tilde\la) = \delta^{2\cal N}(
\bar{Q}) A^*(\la,\tilde\la) + {\bf F}_{\bar\eta}[\delta^{2\cal N}(Q)] A(\la,\tilde\la) \,.
\end{align}
After the Fourier transform to the holomorphic  variables, it coincides with the initial amplitude,  since the MHV and NMHV sectors are exchanged by charge conjugation, 
\begin{align}
{\cal A}_{2 \to 3}(\eta,\la,\tilde\la) = {\bf F}_{\eta}[ {\cal A}^c_{2 \to 3}] (\eta,\la,\tilde\la) \,. \label{eq:A5c}
\end{align}

In view of calculating the squared matrix element of the process $2 \to 3$, we split the odd variables, $\eta = (\eta_I,\eta_F)$, where $I=\{ 1,2 \}$ are the initial states and $F=\{ 3,4,5 \}$ are the final states. To calculate the square of the amplitude summed over the final states, we multiply the amplitude by its conjugate, both written in the holomorphic odd variables $\eta$ and $\xi$, respectively, and integrate out the odd variables of the final states $\eta_F = \xi_F$, keeping the initial state variables $\eta_I$ and $\xi_I$,
\begin{align}
\mathbb{M}_{2 \to 3}(\eta_I,\xi_I) = \int {\cal A}_{2 \to 3}(\eta,\la,\tilde\la) \, {\bf F}_{\xi} [ {\cal A}^c_{2 \to 3}] (\xi,\la,\tilde\la) \, \delta(\eta_F- \xi_F) \, d\eta_F d\xi_F \,, \label{eq:M3super}
\end{align}
where $d\xi_F \equiv \prod_{i\in F} d^{\cal N}\xi_i $, $d\eta_F \equiv \prod_{i\in F} d^{\cal N}\eta_i $ and $\delta(\eta_F - \xi_F) \equiv \prod_{i \in F}\delta^{\cal N} (\eta_i - \xi_i)$. Substituting the superamplitude \p{eq:A5super} and its conjugate \p{eq:A5c} in the previous relation, we obtain four terms. Two of them vanish, 
\begin{align}
& \int  \delta^{2\cal N}(Q_\eta) \, \delta^{2\cal N}(Q_\xi) \, \delta(\eta_F- \xi_F) \, d\eta_F d\xi_F =  0 \,, \notag\\
& \int {\bf F}_{\eta}[\delta^{2\cal N}(\bar{Q})] \, {\bf F}_{\xi}[\delta^{2\cal N}(\bar{Q})]  \, \delta(\eta_F- \xi_F) \, d\eta_F d\xi_F =  0  \,,
\end{align}
and the remaining two cross-terms are identical upon the exchange of $\eta$ and $\xi$. They are easy to calculate. Indeed, we substitute the definition of the Fourier transform \p{eq:Fourier} and rewrite $\delta^{2\cal N}(\bar{Q})$ in exponential form (Grassmann Fourier transform) by introducing an auxiliary integration variable $\bar \theta^{A}_{\dot \alpha}$, 
\begin{align}
& \int  \delta^{2\cal N}(Q_\eta)  \, {\bf F}_{\xi}[\delta^{2\cal N}(\bar{Q})] \, \delta(\eta_F- \xi_F)\, d\eta_F d\xi_F \notag\\
& = \int \delta^{2\cal N}(Q_{\eta}) \exp\left[ {\sum(\bar\theta \cdot \tilde\la_i  + \xi_i) \bar\xi_i}\right] \delta(\eta_F- \xi_F)\, d\eta_F d\xi_F d\bar \xi d^{2{\cal N}}\bar\theta \notag\\
& = \int \delta^{2\cal N}(Q_{\eta_I} + Q_{\xi_F}) \, \delta(\bar\theta \tilde\la_F + \xi_F) \, \delta(\bar\theta \tilde\la_I + \xi_I) \, d\xi_F d^{2{\cal N}}\bar\theta \notag\\
& = \int  \delta^{2\cal N}(Q_{\eta_I} + \bar\theta p_I) \, \delta(\bar\theta \tilde\la_I + \xi_I) \, d^{2{\cal N}}\bar\theta  = s_{12}^{\cal N} \,  \delta(\eta_I - \xi_I) \,. \label{eq:intQFQ}
\end{align}
Here $p_I \equiv p_1 + p_2$, and $p_I^2 =s_{12}$ arises upon integration as a Jacobian factor. Summing the four terms in \p{eq:M3super}, we obtain
\begin{align}
\mathbb{M}_{2 \to 3}(\eta_I,\xi_I) =  2 s_{12}^{\cal N}|A(\la,\tilde\la) |^2 \,  \delta(\eta_I - \xi_I)\,. \label{eq:M3susy}
\end{align}
Due to the delta function in the previous equation, the initial states of the amplitude and its conjugates are correctly matched, $\eta_I =\xi_I$. In particular, the result does not depend on the choice of the initial state, so all two-particle initial states have the same squared matrix element.

In the four-point case, $n=4$, the superamplitude is of the MHV type. It coincides with its conjugate,  
\begin{align}
{\cal A}_{2 \to 2}(\eta,\la,\tilde\la) = \delta^{2\cal N}(Q) A(\la,\tilde\la) \,, \quad
{\cal A}^c_{2 \to 2}(\bar \eta,\la,\tilde\la) = \delta^{2\cal N}(\bar Q) A^*(\la,\tilde\la) \,,
\end{align}
upon the Grassmann Fourier transform. The squared matrix element summed over the  final states $F=\{3,4\}$,
\begin{align}
\mathbb{M}_{2 \to 2}(\eta_I,\xi_I) = \int {\cal A}_{2 \to 2}(\eta,\la,\tilde\la) \, {\bf F}_{\xi}[ {\cal A}^c_{2 \to 2}] (\xi,\la,\tilde\la) \, \delta(\eta_F- \xi_F) \, d\eta_F d\xi_F \,, \label{eq:M2super}
\end{align}
receives contributions only from \p{eq:intQFQ}, so 
\begin{align}
\mathbb{M}_{2 \to 2}(\eta_I,\xi_I) = s_{12}^{\cal N}|A(\la,\tilde\la) |^2 \,  \delta(\eta_I - \xi_I)\,.  \label{eq:M2susy} 
\end{align}

In the case of ${\cal N} = 8$ SG, we obtain from \p{eq:M3susy} and \p{eq:M2susy} the expressions \p{eq:MMtree2pt} with the bosonic helicity function $A(\la,\tilde\la)$ given by $M^{(0)}$ in \p{eq:Mtree5}.

Note that \p{eq:M3susy} also gives the MHV and N${}^{\text{max}}$MHV contributions to the square of any  $n$-point amplitude. However, for $n>5$ other N${}^k$MHV components have to be taken into account as well.

\subsection*{Closed superstring with R-symmetry $SU(4) \times SU(4)$}
\label{app:SU4xSU4}

Compared to ${\cal N}=8$ SG, the R-symmetry of the closed superstring amplitude is broken, $SU(8) \to SU(4) \times SU(4)$. Consequently, the simple relation \eqref{eq:A5super} among the helicity amplitudes is modified, and the summation over the supermultiplet in Eq.~\eqref{eq:M3susy} has to be adjusted accordingly.   

The KLT relations in their supersymmetric form \cite{Elvang:2010kc} provide the following expression for the five-point tree-level closed-string superamplitude, 
\begin{align}
{\cal M}^{\text{string}}(12345) =  &   g_1 {\cal A}_L(12345) {\cal A}_R(21435) + g_2 {\cal A}_L(13245) {\cal A}_R(31425) \,, \label{eq:M5super}
\end{align}
where the coefficient functions $g_1$ and $g_2$ are defined in \p{eq:defg1g2}. The tree-level open-string superamplitudes ${\cal A}_L$ and ${\cal A}_R$ correspond to the left- and right-moving modes of the closed string. Both have explicit ${\cal N}=4$ supersymmetry and R-symmetry $SU(4)$ for the toroidal compactification. We present these five-point amplitudes as a sum \p{eq:A5super} of MHV and NMHV nilpotent invariants,
\begin{align}
& {\cal A}_L = \delta^{8}(Q_\eta) A (\la,\tilde\la) + {\bf F}_{\eta}[\delta^{8}(\bar Q)] A^* (\la,\tilde\la) \,, \notag \\
& {\cal A}_R = \delta^{8}(Q_{\hat{\eta}}) A (\la,\tilde\la) + {\bf F}_{\hat{\eta}}[\delta^{8}(\bar Q)] A^* (\la,\tilde\la) \,, \label{eq:ALAR}
\end{align} 
where $\eta_i^{A}$ with $A=1,\ldots,4$ are the holomorphic odd variables of ${\cal A}_L$, and $\hat{\eta}_i^{A}$ with $A=5,\ldots,8$ are those of ${\cal A}_R$. The bosonic function $A(\la,\tilde\la)$ (and its complex conjugate $A^*(\la,\tilde\la)$) is the same for ${\cal A}_L$ and ${\cal A}_R$. Its explicit expression is given in \p{AYM5}. The supercharges of ${\cal A}_L$ and ${\cal A}_R$,
\begin{align}
Q_\eta = \sum_i \la_i \eta_i \,, \quad
Q_{\hat{\eta}} = \sum_i \la_i \hat{\eta}_i \,,
\end{align}
combine into the ${\cal N}=8$ supercharges of the closed string. However, compared to ${\cal N}=8$ SG, the R-symmetry of the compactified closed string is $SU(4) \times SU(4)$ \cite{Elvang:2010kc}.
As a consequence, the helicity sectors of the closed string amplitude are classified by a pair of indices; there are four sectors in the five-point amplitude, i.e. N${}^{(k_1,k_2)}$MHV where $k_1,k_2=0,1$. Substituting \p{eq:ALAR} into \p{eq:M5super}, we easily identify these four supersymmetric sectors.

In order to calculate the squared amplitude summed over the final states, we proceed as before, this time integrating over the odd variables of ${\cal A}_L$ and ${\cal A}_R$,
\begin{align}
\mathbb{M}_{ 2 \to 3}(\eta_I,\hat\eta_I,\xi_I,\hat\xi_I) = \int {\cal M}(\eta,\hat\eta,\la,\tilde\la) {\bf F}[ {\cal M}^c] (\xi,\hat\xi,\la,\tilde\la) \, \delta(\eta_F- \xi_F) \delta(\hat\eta_F- \hat\xi_F)  \, d\eta_F d\hat\eta_F d\xi_F d\hat\xi_F  \,. \label{eq:M5susysum}
\end{align}
 In the product ${\cal M} {\bf F}[ {\cal M}^c]$ there are four terms of the following form, 
\begin{align}
{\cal A}_L(\eta) {\cal A}'_R(\hat\eta) {\bf F}_{\xi}[{\cal A}^{''c}_L ] {\bf F}_{\hat\xi}[{\cal A}^{'''c}_R ].
\end{align}
For each of these terms, the odd integrations of \p{eq:M5susysum} factorize  into holomorphic and anti-holomorphic odd  variables, so we can utilize  the previous results: 
\begin{align}
& \int {\cal A}_L(\eta) {\bf F}_{\xi}[{\cal A}^{''c}_L ] \, \delta(\eta_F- \xi_F) \, d\eta_F d\xi_F  \int {\cal A}'_R(\hat\eta) {\bf F}_{\hat\xi}[{\cal A}^{'''c}_R ] \, \delta(\hat\eta_F- \hat\xi_F) \, d\hat\eta_F d\hat\xi_F \notag \\
& = s_{12}^8 \left( A A^{''*} + A^{*} A^{''}\right) \left( A^{'} A^{'''*} + A^{'*} A^{'''}\right) \delta(\eta_I - \xi_I) \delta(\hat\eta_I - \hat\xi_I)\, .
\end{align}
Another novelty of the current situation is that there are several independent bosonic functions in the expression for the closed string superamplitude \p{eq:M5super}. Finally, we obtain
\begin{align}
& \mathbb{M}_{2\to 3}(\eta_I,\hat\eta_I,\xi_I,\hat\xi_I) = 2 s_{12}^8 \Bigl( 2g_1^2 |A_{12345}|^2 |A_{21435}|^2 + 2g_2^2 |A_{13245}|^2 |A_{31425}|^2 \\
& + g_1 g_2 \left( A_{13245} A_{12345}^* + A_{12345} A_{13245}^* \right) \left( A_{31425} A_{21435}^* + A_{21435} A_{31425}^* \right) \Bigr)\delta(\eta_I - \xi_I) \delta(\hat\eta_I - \hat\xi_I) \,, \notag
\end{align}
where $A_{ijklm} \equiv A(ijklm)$.
Like the supergravity case, the squared matrix element of the closed string does not depend on the choice of the initial state.

\section{Relation to the cusp anomalous dimension}\label{App:cusp}

According to the relations \eqref{EE-fin1} and \eqref{F-eq}, the same function \eqref{gamma} governs the behavior of the energy correlators for $z\to 1$ and the specific ultraviolet divergence of the eikonal function \re{M1}. This property is yet another manifestation of the  relationship between the infrared asymptotics of  the 
observables and the ultraviolet (cusp) singularities of the eikonal integrals \cite{Korchemsky:1985xj,Korchemsky:1991zp}.  

\subsection*{Lightlike gravitational cusp anomalous dimension}
 
The ultraviolet divergence of the eikonal function \eqref{M1} originates from the virtual corrections described by the functions $G_{\rm v}(0)$ and its conjugate $\bar G_{\rm v}(0)$ defined in \eqref{GG}. Using the representation \re{WL} of the eikonal phase, this function can be written in an equivalent form as
\begin{align}\label{WL'}
  e^{-{\kappa^2\over 8}G_{\rm v}(0)} = \left\langle\exp\lr{{i\kappa\over 2} \int_C dt\, \dot x^\mu(t)\dot x^\nu(t) h_{\mu\nu}(x(t)) }\right\rangle_h \,,
\end{align}
where the integration path $C$ consists of four semi-infinite lines running along the momenta of the incoming and outgoing particles and meeting at the same point. This path can be thought of as describing the world lines of the particles in the $2\to 2$ scattering amplitude. The relation~\eqref{WL'} is the gravitational counterpart of the eikonal phase in (nonabelian) gauge theory in terms of Wilson loops.

It is convenient to flip the momenta of the outgoing particles
$p_3=-q_1$ and $p_4=-q_2$ and treat $p_i$ (with $i=1,\dots,4$) as incoming momenta satisfying $\sum_i p_i=0$. Then, the path in \eqref{WL'}
can be parametrized as the union of four semi-infinite lines $x_i^\mu(t)=p_i^\mu t$ with $-\infty< t \le 0$. These lines meet at the origin and each pair of them forms a lightlike cusp. 

Performing the Gaussian averaging in \eqref{WL'} we obtain
\begin{align} \label{G-sum-I}
  {}&  G_{\rm v}(0) = \sum_{i, j } \int\limits_{-\infty}^0 dt \int\limits_{-\infty}^0dt' \, p_i^\mu p_i^\nu p_j^{\mu'} p_j^{\nu'} D_{\mu\nu,\mu'\nu'}(p_it-p_j t') \, e^{-i\lambda^2(t+t')}\,, 
\end{align}
where $D_{\mu\nu,\mu'\nu'}(x)$ is the graviton propagator %
in  configuration space and the exponential factor regularizes the infrared divergence. The relation \re{G-sum-I} is gauge invariant, which allows us to replace the propagator by its expression in the de Donder gauge (see \re{prop}).
Introducing  UV dimensional regularization, %
the double integral in \re{G-sum-I} can be evaluated as  
\begin{align}\label{I-cusp}
 \mu^{2\epsilon} {\Gamma(1-\epsilon)\over 4\pi^{2-\epsilon}}\int\limits_{-\infty}^0  
    {dt\, dt'\, (p_ip_j)^2\, e^{-i\lambda^2(t+t')}\over [-(p_it -p_jt')^2+i0]^{1-\epsilon}}
 = s_{ij} {\Gamma^2(\epsilon)\Gamma(1-\epsilon)\over 16\pi^{2-\epsilon}} \lr{-{(s_{ij}+i0)\mu^2\over \lambda^4}}^{\epsilon} \,, 
\end{align}
where $s_{ij}=2(p_i p_j)$ and $\lambda^2$ is the IR cutoff. As expected, the eikonal integral develops both ultraviolet and infrared divergences. Moreover, the dependence on the corresponding cutoffs enters through the ratio $\mu^2/\lambda^4$ and, as a consequence, the infrared and ultraviolet divergences of $G_{\rm v}(0)$ are in one-to-one correspondence. 

The ultraviolet divergence of \eqref{I-cusp} is due to the nonzero angles between the vectors $p_i$ and $p_j$. In fact, since these vectors are lightlike, the hyperbolic angles are infinite, hence the integral \re{I-cusp} develops a double pole $1/\epsilon^2$. Substituting \eqref{I-cusp} into \eqref{G-sum-I}, we find that the coefficient of the double pole $1/\epsilon^2$ is proportional to the total invariant mass $\sum_{i,j}s_{ij}$ and it vanishes. As a consequence, $G_{\rm v}(0)$ only contains a single pole $1/\epsilon$ and satisfies the evolution equation
\begin{align}\label{G0-eq}
    \mu {\partial\over\partial\mu} \lr{{\kappa^2\over 8}G_{\rm v}(0)} = \gamma_{\rm cusp}\,,
\end{align}
where the additional factor $\kappa^2/8$ comes from the exponent of \re{WL'} and  $\gamma_{\rm cusp}$ is the lightlike gravitational cusp anomalous dimension,
\begin{align}\label{ll}
    \gamma_{\rm cusp}={\kappa^2 \over 32\pi^2}\sum_{ i<j } s_{ij}\log(-s_{ij}-i0) \,.
\end{align}
Depending on the sign of $s_{ij}$, some terms in the sum develop an imaginary part. 

The relation \eqref{M1} involves the sum $G_{\rm v}(0)+\bar G_{\rm v}(0)$. It satisfies the evolution equation \eqref{G0-eq}, with the cusp anomalous dimension replaced by twice  its real part,
\begin{align}\notag\label{Re-gamma}
    2\,{\rm Re} \, \gamma_{\rm cusp} {}&= {\kappa^2 \over 8\pi^2} \Big[(p_1p_2) \log (p_1p_2)+(q_1q_2) \log (q_1q_2)- (p_1q_1) \log (p_1q_1)
\\[1.5mm]    
   {}& \qquad\quad - (p_1q_2) \log (p_1q_2)- (p_2q_1) \log (p_2q_1)- (p_2q_2) \log (p_2q_2) \Big] \,,
\end{align}
where we substituted $p_3=-q_1$ and $p_4=-q_2$. Using \eqref{kin0}, we find for $E=E_1=E_2$
\begin{align}
   2 {\rm Re} \, \gamma_{\rm cusp} %
   = 2 (\kappa E)^2\gamma(y_1) \,,
\end{align}
where the function $\gamma(y_1)$ is defined in \re{gamma} and $y_1=(1-(\vec n\vec n_1))/2$, see \p{ys}. 

Comparing the last relation with \eqref{EE-fin1} and \eqref{gamma}, we conclude that the asymptotic behavior of the energy--energy correlator in the back-to-back region is indeed governed by the lightlike gravitational cusp anomalous dimension. This property is very general and it also holds in four-dimensional gauge theories including QCD
\cite{COLLINS1981381,KODAIRA198266,Sterman:1986aj,Korchemsky:1993uz,deFlorian:2004mp,Moult:2018jzp,Korchemsky:2019nzm}.

\subsection*{Relation to the gravitational Bremsstrahlung function }

It is well known \cite{Weinberg:1965nx} that the infrared divergences in the differential cross section of the (elastic) $2\to 2$ scattering process can be factorized into an exponential factor, 
\begin{align}
   \exp\lr{-\frac12 B_{\rm gr} \log(\mu^2/\lambda^2)}\,,
\end{align}
where $\lambda$ and $\mu$ are  IR and UV cutoffs on the soft gravitons momenta, respectively.   The Bremsstrahlung function is given by
\begin{align}\label{B-gr}
    B_{\rm gr} = {G_N \over 2\pi}\sum_{i,j}  m_i m_j{1+\beta_{ij}^2\over \beta_{ij}(1-\beta_{ij}^2)^{1/2}}\log{1+\beta_{ij}\over 1-\beta_{ij}}\,,
\end{align}
where   all particles with on-shell momenta $p_i^2=m_i^2$ are considered incoming, and $\beta_{ij}$ are their relative velocities, 
\begin{align}
\beta_{ij} = \sqrt{1-{m_i^2m_j^2\over (p_ip_j)^2} } = \tanh \gamma_{ij}\,.
\end{align}
Here $\gamma_{ij}$ is the relative angle between the momenta of particle, $\cosh\gamma_{ij} = (p_i p_j)/ (m_i m_j)$.

It is interesting to compare the  Bremsstrahlung function \re{B-gr} with the  (non-lightlike) gravitational cusp anomalous dimension  given by \cite{Miller:2012an},
\begin{align}
    \Gamma_{\rm cusp}=-{\kappa^2\over 32\pi^2}\sum_{i<j} m_i m_j \left[ (i\pi -\gamma_{ij}) {\cosh (2 \gamma_{ij} )\over \sinh(\gamma_{ij} )}
      +\cosh\gamma_{ij}\right] \,,
\end{align}
where $\kappa^2=32\pi G_N$.
This relation holds for an arbitrary total momentum $\sum_i p_i$. The latter vanishes for the scattering amplitude and we obtain
\begin{align}
  2 {\rm Re}\, \Gamma_{\rm cusp}={\kappa^2\over 16\pi^2}\sum_{i<j}  m_i m_j  
    \gamma_{ij} {\cosh (2 \gamma_{ij} )\over \sinh(\gamma_{ij} )}=B_{\rm gr}\,.
\end{align}
 
In the massless limit, for $m_i=m$ and $m\to 0$, we find that $\beta_i=1+O(m^4)$ and hence the  individual terms in the sum \re{B-gr} diverge logarithmically with $m$. However, due to momentum conservation, 
$\sum_i p_i =0$, the divergent terms cancel in the sum \re{B-gr} leading to
\begin{align} 
\lim_{m\to 0} B_{\rm gr} = {2G_N \over  \pi}\sum_{i,j} (p_i p_j) \log (p_ip_j) \,.
\end{align}
Comparing this relation with  \eqref{Re-gamma} and identifying the momenta $p_3=-q_1$ and $p_4=-q_2$, we arrive at
\begin{align}
\lim_{m\to 0} B_{\rm gr} = 2\,{\rm Re} \, \gamma_{\rm cusp} \,,
\end{align}
where $\gamma_{\rm cusp}$ is the lightlike cusp anomalous dimension defined in \re{ll}.

\section{Black-hole dominance}
\label{app:BHD}

Gravitational scattering at high energies and fixed impact parameters is universal: it produces a black hole \cite{Eardley:2002re,Yoshino:2002tx,Yoshino:2005hi}. It is interesting to ask how this universality manifests itself at the level of the gravitational energy correlators studied in the paper. We do not have a rigorous way to address this question; instead, in this Appendix we consider a scenario that seems natural.

The answer is not immediate for the reason that we are scattering plane waves. These are not localized in the space of the impact parameter $b$.
For a given collision energy $2E$, part of the wave function such that $b \lesssim R_{\text{Sch}}(2E)$, where   $R_{\text{Sch}}(E) = 2 G_N E$ is the Schwarzschild radius, produces a black hole. On the other hand, for the part of the wave function for which $b \gtrsim R_{\text{Sch}}(2E)$, we expect that the final state consists of deflected hard gravitons accompanied by gravitational Bremsstrahlung. To apply this intuition, we assume that the energy of the process is larger than the species scale $R_{\text{Sch}}(2E) > L_{\text{sp}}$, such that a semiclassical treatment of the collision applies.

If a semi-classical black hole is formed, it will decay through Hawking radiation \cite{Hawking:1975vcx}. If we consider a normalized spherically-symmetric black hole state of mass $M$, we expect that, to leading order in $G_N M^2 \gg 1$,
\be
\langle \text{BH} | {\cal E} (\vec n_1) ... {\cal E} (\vec n_k) | \text{BH} \rangle = \left( {M \over 4 \pi} \right)^k + \dots \ . 
\ee
The total cross section of black hole production in the two-graviton collision is expected to grow like the area of the black disc of  Schwarzschild radius,
\be
\sigma_{tot}^{\text{BH}} \simeq  2 \pi R_{\text{Sch}}(2E)^2 \ . 
\ee
Let us next consider the contribution of the states with $b \gtrsim R_{\text{Sch}}(2E)$. We know that the total cross section of such processes is infinite, however the infinity arises from  elastic scattering in the forward region. By placing the energy calorimeters away from the beam, $\vec n_i \neq \pm \vec n$, we project out the infinite contribution, making the total cross section that contributes to the energy correlator finite. 

It is however difficult to quantitatively estimate the contribution of these non-BH processes, since two competing effects are at play:
on the one hand, we expect the total cross section for such processes to be infinite; on the other hand, the contribution of the underlying process to the energy correlator of interest decreases as we increase $b$. It would be very interesting to develop a quantitative estimate of this effect.

A simple possibility that allows us to make a universal prediction is that of \emph{black-hole dominance}. %
That is, let us \emph{assume} that, when properly taken into account, the non-BH processes produce a subleading contribution to the off-beam energy correlators at high energies.
In this case, we expect the following universal formula for the leading energy correlator at high energies in gravitational theories:
\be
\textbf{Black-hole dominance:} ~~~
\lim_{\kappa E \to \infty} \langle {\cal E} (\vec n_1) ... {\cal E} (\vec n_k) \rangle \propto  {(\kappa E)^4 \over 4 \pi} \left( {2 E \over 4 \pi} \right)^k , ~~~ \vec n_i \neq \pm n \ ,  
\label{eq:BHD}
\ee
where we used that $2 s \sigma_{tot}^{\text{BH}} =  {(\kappa E)^4 \over 4 \pi}$ with $2s$ being the flux factor in the standard definition of the total cross section for the two-particle initial state. The simple, factorized form of the energy correlators in
\eqref{eq:BHD} is not specific to gravity; it is a general feature of
final states with many particles
\cite{Chicherin:2023gxt,Firat:2023lbp}. The off-beam assumption is important because for $\vec n_i \to \vec n$ we expect that forward elastic physics dominates. It would be interesting to understand whether the simple physical picture expressed by the formula above is actually realized.

In the context of celestial holography, it is interesting to consider the Mellin transform of the EEC. Introducing the dimensionless parameter $\omega=(\sqrt{G_N}E)^2$, one can consider the Mellin transform $\int_0^{\infty} d\omega \ \omega^{\Delta-1}{1 \over E^2}\langle {\cal E}(\vec n_1){\cal E}(\vec n_2)\rangle$. For $\omega\ll 1$ (the perturbative regime) the correlator scales as $\omega^{3}$, while for $\omega\gg 1$ (the black-hole regime, see \eqref{eq:BHD}) it grows as $\omega^{2}$. This suggests that the Mellin transform exists and is analytic for $-3<\mathrm{Re}(\Delta)<-2$.

\bibliographystyle{JHEP} 
\bibliography{main}

\end{document}